\documentclass[aps,groupedaddress,nofootinbib]{revtex4-1}

\usepackage{amsmath}
\usepackage{bm}
\usepackage{float}
\usepackage{graphicx}
\usepackage{lineno,hyperref}
\usepackage{mathrsfs}
\usepackage{natbib}
\usepackage{subfigure}
\usepackage{wrapfig}
\usepackage{xcolor}

\newcommand{\ave}[1]{\left\langle #1 \right\rangle}
\newcommand{\abs}[1]{\left| #1 \right|}
\newcommand{\be}{\begin{equation}}
\newcommand{\ee}{\end{equation}}
\newcommand{\bfx}{\ensuremath{{\bf x}}}
\newcommand{\bld}[1]{{\bf #1}} 
\newcommand{\bracet}[1]{\left\{ #1 \right\}}
\newcommand{\bracket}[1]{\left[ #1 \right]}
 \newcommand{\bx}{\mathbf {x}} 
\newcommand{\bv}{\mathbf v}
\newcommand{\dxdt}[1]{\frac{\partial #1}{\partial t}}

\newcommand{\fn}[1]{\left( #1 \right)}
\newcommand{\gradx}[1] {{\boldsymbol \nabla}_{#1}}
\newcommand{\measvol}{\ensuremath \mathscr{V}_m}
\newcommand{\mrm}[1]{\mathrm {#1} }

\newcommand{\pdxpdt}[1]{\frac{\partial #1}{\partial t}}
\newcommand{\phaseindex}{\ensuremath{\beta}}
\newcommand{\LEQ}[1]{\label{eq:#1}}
\newcommand{\regA}{\ensuremath{\mathscr{V}_1}}
\newcommand{\regB}{\ensuremath{\mathscr{V}_2}}
\newcommand{\regR}{\ensuremath{\mathscr{V}_R}}
\newcommand{\regr}{\ensuremath{\mathscr{V}_r}}
\newcommand{\rhof}{\rho_{{f}}}
\newcommand{\volinf}{\ensuremath \mathscr{V}_{\infty}}

\begin{document}


\title{Multiphase Flows: Rich Physics, Challenging Theory, and Big Simulations}


\author{Shankar Subramaniam}
\email[]{shankar@iastate.edu}
\affiliation{Center for Multiphase Flow Research \& Education, Department of Mechanical Engineering, Iowa State University}


\date{\today}

\begin{abstract}
Understanding multiphase flows is vital to addressing some of our most pressing human needs: clean air, clean water and the sustainable production of food and energy. This article focuses on a subset of multiphase flows called particle--laden suspensions involving non-deforming particles in a carrier fluid.
The hydrodynamic interactions in these flows result in rich multiscale physics, such as clustering and pseudo-turbulence, with important practical implications. Theoretical formulations to represent, explain and predict these phenomena encounter peculiar challenges that multiphase flows pose for classical statistical mechanics. A critical analysis of existing approaches leads to the identification of key desirable characteristics that a formulation must possess in order to be successful at representing these physical phenomena. The need to build accurate closure models for unclosed terms that arise in statistical theories has motivated the development of particle-resolved direct numerical simulations (PR--DNS) for model--free simulation at the microscale. A critical perspective on outstanding questions and potential limitations of PR--DNS for model development is provided. Selected highlights of recent progress using PR-DNS to discover new multiphase flow physics and develop models are reviewed. Alternative theoretical formulations and extensions to current formulations are outlined as promising future research directions. The article concludes with a summary perspective on the importance of integrating theoretical, modeling, computational, and experimental efforts at different scales.
\end{abstract}


\maketitle

\section{Relevance and Motivation\label{sec:intro}}
Multiphase flows involve the flow of matter in two or more thermodynamic phases, and include gas-solid flows, sprays and bubbly flows. They permeate practically every aspect of human life and an improved understanding of multiphase flows can play an important role in ensuring three vital needs of humans---clean air, clean water and sustainable food production---that are foundational to quality of life. Improved understanding of multiphase flows leads to better prediction of key phenomena in the natural environment and in man-made systems. Accurate prediction, in turn, opens the door to control of multiphase flows in order to achieve desired outcomes. Traditionally, multiphase flow control has been explored in the context of industrial devices~\citep{chu2009cfd,chu2009cfd-dem,mcgovern2008multiphase} but there is also recent interest in climate engineering~\citep{latham1990control,latham2012marine,cooper2014}.
Of particular topical interest in the midst of the COVID-19 pandemic is the accurate prediction and control of droplet dispersion caused by exhalation of human breath, and its implications for transmission of respiratory diseases in various settings. 

Broadly speaking, multiphase flows have applications in energy, environmentally sustainable technologies, chemical processing, critical infrastructure, healthcare and biological applications including pharmaceuticals, the design of materials and advanced manufacturing processes~\citep{comfre}.
In the following we consider representative examples of multiphase flow applications involving solid particles, droplets and bubbles to understand those questions that design and production engineers encounter, which require improved understanding of multiphase flows. These examples will illustrate how multiphase flow science can help with the design and scale--up of processes and devices in multiphase flow applications.

Fluidized beds in which solid particles are suspended by a fluid are commonly used as multiphase reactors. Here we describe two applications of fluidized beds that pertain to sustainable energy generation: (i) fast pyrolysis of biomass for bio-oil production, and (ii) CO$_2$ capture using dry sorbents.

Thermochemical processing of biomass to produce bio-oil involves heating biomass in an inert environment to extract the volatile components which are subsequently filtered and condensed into bio-oil. Biomass is injected into a bed of sand particles which is heated by heat transfer from the reactor walls and fluidized by the up-flow of hot nitrogen gas.
Process heat both adds to the cost of producing bio-oil and heating of the reactor surface area is a bottleneck to scaling up the reactor and limits biomass throughput. Autothermal pyrolyis is a promising modification of traditional biomass pyrolysis that promises, on the basis of laboratory scale studies, to reduce the cost of bio-oil from \$~3.27 per gallon to \$~2.58 per gallon by quintupling the reactor feed rate and halving the reactor cost per tonne of biomass processed~\citep{polin2019conventional,polin2019process,polin2019thesis}. In autothermal pyrolysis, the heat needed to devolatilize biomass is provided volumetrically by exothermic gas-phase reactions that are initiated by introducing small amounts of oxygen into the inert atmosphere, without drastically affecting product yield. The challenge is how to successfully scale this reacting multiphase flow process from laboratory scale, with biomass throughput of milligrams per hour, to pilot scale (kg/hour) to commercial plant scale (tons/hour), while recognizing that the hydrodynamics of a fluidized bed change with scale~\citep{knowlton2005scale}. There is also a complex interaction between the chemical reaction kinetics (as yet, poorly characterized in the low temperature range of interest) and the hydrodynamics, that can also vary with scale, and which can affect the trade-off between efficiency of volumetric heat generation while not compromising the yield of bio-oil. For example, an important design question that engineers face is where to place the biomass injection inlet in order to maximize bio-oil yield and biomass throughput?

The second example is the capture of carbon dioxide from flue gas by flowing it past dry sorbent particles in a circulating fluidized bed. 
In such gas--solid reactors it is important to be able to accurately predict the conversion rate given the high cost of the sorbent particles and the need to reduce the level of effluent contaminants~\citep{breault2009mass}. Existing models for interphase mass transfer are often specific experimentally--based empirical correlations for the reactor system under consideration because correlations for the mass transfer coefficient between the solid and gas phases in the literature differ by up to 7 orders of magnitude. The same is true for gas--solid heat transfer coefficients as well~\citep{sun2015modeling}. The wide variation in reported mass transfer coefficients in the existing literature is attributed to flow regime differences~\citep{breault2009mass}. Clustering of particles inhibits gas-solid contacting and diffusion of CO$_2$ from the gas stream to the particle surface, which reduces the overall efficiency of the device. An important question confronting process designers is the prediction of clusters and heat and mass transfer in this application to maximize conversion and throughput.

\subsubsection{Multiscale nature of multiphase flow}
\begin{figure}
\begin{centering}
\includegraphics[scale=0.4]{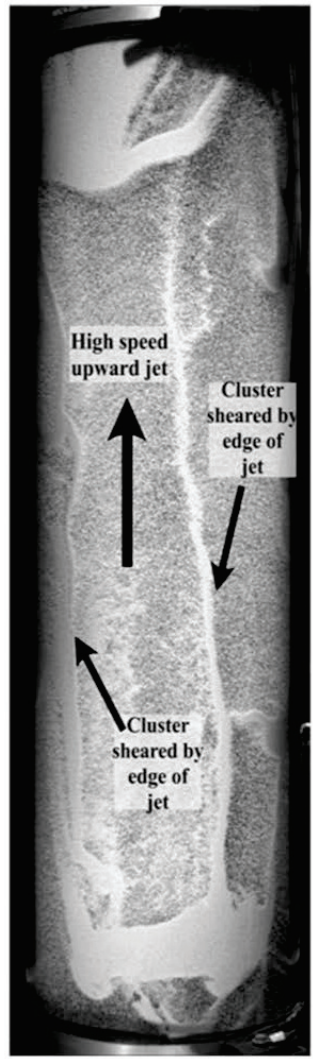}
\caption{An image from high-speed video of a riser flow showing the complex hydrodynamics and multiscale features of the particle--laden suspension. The superficial gas velocity is 6.6 m/s and the solids flux is 20 kg/(m$^2$-s). Reprinted from {\em Powder Technology}, vol. 242, F. Shaffer, B. Gopalan, R. W. Breault, R. Cocco, S. R. Karri, R. Hays, and T. Knowlton, "High speed imaging of particle
flow fields in CFB risers," 86, Copyright (2013), with permission from Elsevier.}
\label{fig:multiscale}
\end{centering}
\end{figure}

High-speed videography experiments of gas-solid flows~\citep{shaffer2013high} (see Fig.~\ref{fig:multiscale}) reveal a complex multiscale structure, and it is useful to classify these scales into three ranges: (i) the {\em microscale} representing structures with a characteristic length scale in the range $O(1-10 d_p$), where $d_p$ is the particle diameter, (ii) the {\em mesoscale} representing structures with a characteristic length scale in the range $O(10-100 d_p)$, and (iii) the {\em macroscale} representing structures on the scale of the device. This image is at the boundary of the flow obtained through a transparent riser but owing to the high solid loading in these flows it is difficult to obtain such non-intrusive measurements in the interior of such flows. The macroscale behavior can change quite dramatically when the problem parameters such as the inlet gas velocity and solids flux are varied~\citep{mcmillan2013particle}, and it is yet unknown to what extent the structures at different scales influence each other. The level of simulation fidelity at each scale that is needed to predict macroscale behavior is also not known. There are indications that mesoscale structures affect macroscale behavior, and that microscale interactions can in turn affect mesoscale structures, but a definitive quantification of these influences is still an outstanding problem.

Sprays and droplet-laden flows are another example of multiphase flows that also exhibit a similar range of multiscale phenomena as particle--laden suspensions arising from the hydrodynamic interaction between the liquid and gas phases. This is a harder problem than particle-laden suspensions because the deformation of droplets and their coalescence and breakup are additional physical phenomena that have to be considered. Sprays have been studied extensively in the context of internal combustion and gas turbine engines. They also find wide applications in agriculture for spraying pesticides on crops, in advanced manufacturing, coating applications, pharmaceutical and healthcare applications and other processing operations such as drying. Of late, droplet-laden flows have come into prominence because of the
COVID-19 pandemic and questions concerning the transport of the SARS-n-CoV-2 virus, and other similar pathogens. Prediction of droplet dispersion from human breath and its interaction with masks and face shields as well as airflow in groups of people in different settings, such as a restaurant, airplane or schoolroom, can lead to more effective control and mitigation strategies.

When the density of the dispersed phase is lower than that of the carrier fluid, such as for bubbles and buoyant particles, the buoyant and added mass forces play an important role in multiphase hydrodynamics. Multiphase gas-liquid flows are frequently encountered in the chemical and nuclear industry~\citep{jakobsen2008chemical,anglart1996cfd,serizawa1997experiment}. Multiple chemical processes are performed in bubble column reactors, which are vessels filled with liquid reactants, where the gas-phase is fed by means of a distribution system, specifically studied to create a bubble size distribution that maximizes the bubble surface area. Heat and mass transfer phenomena are key elements of the processes in these reactors due to the high exchange efficiency that characterizes these systems~\citep{jakobsen2008chemical,kantarci2005bubble}. Important examples are the Fischer-Tropsch process~\citep{haghtalab2012kinetic}, polymerization reactions, and oxidation and hydrogenation~\citep{storm1992modelling,winterbottom1993bubble,ferrer1985homogeneous,debellefontaine2000wet} processes, and biochemical processes (fermentation~\citep{kantarci2005bubble,debellefontaine2000wet}, wastewater treatment, production of antibiotics, proteins and enzymes~\citep{kantarci2005bubble,ranjbar2008high}, algae growth~\citep{kosseva1991biotransformation,nauha2013modeling,posten2009design}). Bubbly flows are also particularly relevant to the nuclear industry~\citep{anglart1996cfd,munoz2012simulation}, where safety of design and operation motivated some of the early developments in multiphase flow modeling. Typical operating conditions of these devices involve gas phase volume fractions between 5 and 40\%. Engineers rely on models to predict the behavior of these flows in large-scale systems. Bubbly flows also exhibit a multiscale structure and diverse phenomena over the range of operating parameters~\citep{tryggvason2013multiscale}.

Multiphase hydrodynamics can significantly influence heat and mass transfer in flows with bubbles or buoyant particles. Recent experiments by~\citet{gvozdic2018experimental} show that the introduction of small air bubbles (average bubble diameter $\sim 2.5$ mm) at fairly low concentrations (gas volume fraction 0.9\%) into a flow in a vertical channel with one heated vertical wall, enhances the heat transfer relative to its baseline natural convection value by a factor of 20. \citet{wang2019self} have extended this approach by adding a minute concentration ($\sim 1$\%) of a heavy liquid (hydrofluoroether) to a water-based turbulent convection system resulting in heat transfer that supersedes turbulent heat transport by up to 500\%. In both studies the enhancement in heat transfer is attributed to the generation of pseudo-turbulent temperature fluctuations in the liquid by the motion of bubbles.

\subsection{Context and road map}
Even this brief introduction should convince the reader that multiphase flow is such a broad topic that a comprehensive review is not feasible within the constraints of a journal article. Therefore, although the multiphase flows in the aforementioned applications are often non-isothermal, turbulent reacting flows involving bubbles, drops or solid particles, the remainder of this article will focus on the hydrodynamics of isothermal, non-reacting particle-laden flows. Brevity demands that even this overview be selective. Consequently, this article represents my own, admittedly limited, perspective on the hydrodynamics of particle-laden flows.
This article is based on an invited talk given at the 2019 Annual Meeting of the American Physical Society's Division of Fluid Dynamics in Seattle, WA~\citep{subramaniam2019multiphase}. 
\section{Particle-laden suspensions}
Dilute particle suspensions in canonical turbulent flows involving particles of diameter $d_p$ smaller than the Kolmogorov scale $\eta$ have been studied extensively over the past several decades (see~\citet{bala_eaton_2010} for an edifying review). In these multiphase flows the particles are usually modeled as point particles, representing point sources of mass, momentum and heat transfer. 
Here we focus on particle-laden suspensions with inertial solid particles with sizes in the range of 100 to 500 microns that can be larger than the smallest scales of fluid motion, which in the case of turbulent flow is the Kolmogorov scale. In this regime the point particle approximation is no longer valid and a particle-resolved description of the flow field is needed. The interaction of finite size particles with flows where the mean motion of the fluid with respect to the particles is non-zero leads to interesting and important new physics arising from interacting wakes, the details of which are neglected in earlier studies of point particles in homogeneous, isotropic turbulence. The multiphase flow applications described earlier involve solid volume fractions that range from very dilute to close-packed, and so the range of solid volume fraction considered here must far exceed that in earlier studies of dilute particle suspensions. Higher solid volume fractions bring into play additional particle-fluid interactions which are detailed in Sec.~\ref{subsubsec:pfinter}. These make the problem considerably more challenging than dilute suspensions. In order to establish a precise definition of the problem, we first examine a mathematical description of a particle-laden suspension.
\subsubsection{Complete description of a particle-laden suspension}
\label{subsubsec:compdesc}

A complete description of particle-laden suspension specifies information about
the state of every particle and fluid point at every time instant, and
it completely determines the future time evolution of the gas--solid
flow. The state of a particle-laden suspension $S(t)$ at any instant $t$, can be
represented in terms of the state of the particles $S_{\mrm{p}}(t)$
and that of the fluid $S_{\mrm{f}}(t)$. As a specific example we consider a
simple gas--solid flow with smooth, monodisperse spheres but complete
descriptions of more complex gas--solid flows are also possible. The
set of positions and velocities $\bracet{\bld{X}^{(i)}, \bld{V}^{(i)},
  i = 1,\dots,N(t)}$ of $N(t)$ monodisperse spherical particles characterizes the state of
the particles $S_{\mrm{p}}(t)$. The state of the fluid is
characterized by the knowledge of the fluid velocity field ${\bf
  u}({\bf x},t)$ and the pressure field $p({\bf x},t)$. This complete
description of the gas--solid flow at the scale of the particles is
denoted a microscale description.

The evolution of $S(t)$ is given by
\begin{equation}
  \frac{dS}{dt} = \mathcal{L} \bracket{S(t)},
\label{eq:stateevol}
\end{equation}
where $\mathcal{L}$ represents a nonlinear operator operating on the
state $S(t)$ of the gas--solid flow.  For incompressible flows, this operator becomes the mass and
momentum conservation equations for the fluid--phase:
\begin{equation} 
\boldsymbol\nabla \boldsymbol\cdot \mathbf{u}  = 0\mrm{,}
\label{eq:cont_fp}
\end{equation}
\begin{equation} 
  \rho_f \pdxpdt{ \mathbf{u}} + \rho_f \boldsymbol\nabla \cdot \left( \mathbf{u}\mathbf{u}\right) = \mathbf{b} +  \boldsymbol\nabla\cdot \boldsymbol\tau
\label{eq:nse_fp}
\end{equation}
subject to appropriate boundary conditions, 
and the position and velocity evolution equations for the solid phase:
\begin{align}
\frac{d \mathbf{X}^{(i)}(t)}{dt} &= \mathbf{V}^{(i)}(t), \\
m^{(i)} \frac{d \mathbf{V}^{(i)}(t)}{dt}&=
\mathbf{B}+\mathbf{F}_h^{(i)}(t)+\sum_{\substack{{j=1}\\{j \neq
i}}}^{N(t)} \mathbf{F}_\mrm{int}^{(i,j)}(t),
\label{eq:lag_vel}
\end{align}
where particle rotation has been omitted for simplicity.  
In Eq.~\ref{eq:nse_fp}, $\rhof$ is the thermodynamic density of the
fluid--phase, $\bld{b}$ represents body forces (e.g., hydrostatic
pressure gradient, acceleration due to gravity, etc) acting throughout
the volume of an infinitesimal fluid element, and $\boldsymbol{\tau}$
represents the surface stresses (both pressure and viscous stresses)
acting on the surface of an infinitesimal fluid
element. Equations~\ref{eq:cont_fp} and~\ref{eq:nse_fp} are subject to
no--slip and no--penetration velocity boundary conditions on the surfaces of the
particles and appropriate initial conditions for the fluid fields and particle configuration. In the velocity evolution equation for the particles
(Equation ~\ref{eq:lag_vel}), $m^{(i)}$ is the mass of the $i$--th
particle, $\mathbf{B}$ is any external body force,
$\mathbf{F}_h^{(i)}$ is the hydrodynamic force (due to the pressure $p$
and viscous stresses at the particle surface) given by
\begin{align}
\mathbf{F}_h^{(i)}=-\oint_{\partial \mathcal{V}_s^{(i)}} p\, \mathbf{n}\, dA + \mu_f \oint_{\partial \mathcal{V}_s^{(i)}} \boldsymbol\nabla \mathbf{u}\cdot d\mathbf{A} ,
\end{align}
and
$\mathbf{F}_\mrm{int}^{(i,j)}$ is any interaction force (eg. contact
force due to collisions, cohesion, electrostatics, etc) between particles $i$ and $j$. This
coupled set of equations together with the boundary conditions governs
the state of the gas--solid flow at the microscale. Particle--resolved direct numerical simulation (PR--DNS) is a computational approach to solve these governing equations with adequate resolution to represent the flow field over each individual particle, and to directly compute the hydrodynamic force on each particle surface from the solution without resorting to a drag model.

\subsubsection{Key particle-fluid interactions} 
\label{subsubsec:pfinter}
A logical starting point to understand flow past several particles is flow past a single isolated particle. Classical results for uniform steady incompressible flow past an isolated solid particle begin with the Stokes solution at zero Reynolds number (extended to fluid particles by Hadamard and Rybczynski) and extended to account for finite Reynolds number effects by Oseen and many other researchers (see~\citet{clift} for details). These studies provide expressions for the velocity and pressure fields from which the steady drag experienced by a spherical particle in flow is easily calculated, this being the principal hydrodynamic interaction between particle and fluid. The effects of flow non-uniformity are provided by Fax\'{e}n, and the effects of flow unsteadiness on drag are described by early works of Basset and Boussinesq. The Maxey-Riley-Gatignol (MRG) equation summarizes all these effects~\citep{maxey_riley_pf83}. These isolated particle studies established the foundations for the study of an ensemble of solid particles in fluid flow.

It is customary to distinguish particle-laden flows based on the nature of the key interactions between particles and fluid. Isolated single particle studies highlight the principal hydrodynamic effect that the fluid has on a particle, namely to exert a drag force on it. Newton's third law requires that the particle also exerts an equal and opposite force on the fluid. When a fluid interacts with an ensemble of particles, this force exerted by particles on the fluid can be neglected if the mass loading (or phase mass ratio~\citep{fox_cup_book}) of the particles is much less than unity. In these so-called dilute systems, the assumption of a one-way coupling is sufficient. In one-way coupling, the dynamical equation for the particle motion accounts for the force exerted by the fluid, but the force exerted by particles on the fluid is neglected. It should be noted that particles also displace fluid volume and it is expected that this volume-displacement effect can be neglected if the solid phase volume fraction is small (~\citep{apte2008accounting}). However, the averaged fluid--phase mass conservation equation reveals that an additional restriction arises from the gradient of particle volume fraction as well~\citep{shankar_pecs_2013}.

For mass loading (phase mass ratio) equal to, or greater than unity, two-way coupling of the momentum equation in both phases must be accounted for and the fluid phase momentum equation must now contain a term corresponding to the force (per unit volume) exerted by the particles on the fluid. Turning now to interactions between particles themselves, we can distinguish two types of interactions: (a) particle-particle interactions mediated by the presence of fluid, and (b) direct particle-particle interactions due to collisions on contact, cohesive forces, or electrostatics. Traditionally, the inclusion of particle-particle interactions due to collisions on contact has been termed four-way coupling~\citep{elgho_asr_94}. In the examples of fluidized bed applications described in Sec.~\ref{sec:intro} all these fluid-particle interactions are present and the flow is four-way coupled, which makes its modeling and prediction considerably more challenging. With these key interactions in mind, we can now turn to the question of understanding the behavior of several particles in a flow.

\section{Theoretical Approaches}
\label{sec:theoapp}
The fundamental question concerning particle-laden suspensions is whether we can understand and predict macroscale phenomena arising from a collection of particles in fluid flow from the microscale description detailed in Sec.~\ref{subsubsec:compdesc}, and if we can do this for the case of finite size particles interacting with a laminar or turbulent flow over a wide range of solid volume fraction? This is properly the domain of statistical mechanics.
Theoretical descriptions of particle-laden suspensions draw from two principal branches of physics: (i) statistical physics of many-particle systems~\citep{mcquarrie2000statistical}, in particular the kinetic theory of gases~\citep{liboff_book}, and (ii) statistical fluid mechanics of single--phase turbulent flow~\citep{monin_yaglom2,popebook}. It is worthwhile to reflect at this point that insofar as particle-laden suspensions is concerned, the success of early theoretical descriptions~\citep{brenner1972suspension,hinch1977averaged,koch_pof_1990} in answering the fundamental statistical mechanics question as it pertains to dilute suspensions in Stokes flow has been slow to be generalized. In other words, unlike the remarkable success of the kinetic theory of gases in predicting the transport properties of simple gases (see e.g. the diffusivity of CO$_2$ in~\citet{liboff_book}), there are fewer definitive analytical results of broad applicability in particle-laden suspensions. In part, this is because multiphase flows involve many parameters, such as the volume fraction and density ratio to name a few, and the principal physical interactions change significantly across the parameter space, thereby making it difficult to develop a general theory with broad applicability. Another reason is that much of the information needed to complete these theories, which involves developing closure models for unclosed terms, is intractable analytically and must therefore be obtained from computational studies~\citep{shankar2014arfm}. In this regard, careful use of machine learning tools~\citep{jofre2020data,moore2019hybrid} can help map out the form of these unclosed terms over the entire parameter space and help complete these theories. It seems appropriate at this point to recall an appropriately reworded statement~\citep{lumley1990whither} by the eminent fluid mechanician, the late John Lumley (1930--2015), that underpins my expectation of theoretical approaches to multiphase flows: "I believe in the ultimate possibility of developing general computational {\em statistical mechanical} procedures based on first principles; and under certain circumstances I believe that it is possible to do this rationally."~\footnote{Lumley's original quote referred to computational procedures applied to single--phase turbulent flows.}

\subsection{Statistical Mechanics of Particle-laden Suspensions}
Statistical approaches to particle-laden suspensions are usually classified on the basis of the frame of reference (Eulerian or Lagrangian) in which the final equations corresponding to each phase are presented, but this classification actually obscures the mathematical basis of each approach and impedes a systematic analysis  of their inter-relationships. It is much more instructive to trace the mathematical basis of these statistical approaches from their origins in particle physics and turbulent flow, which, as we shall see here, results in valuable insights.

The classical approach to the statistical description of many--particle systems in physics is founded on the mathematical theory of stochastic point processes (SPP).
The state of particles in a particle--laden suspension as given in Sec by $S_{\mrm{p}}(t;\omega)= \bracet{\bld{X}^{(i)}(t;\omega), \bld{V}^{(i)}(t;\omega), i = 1,\dots,N(t;\omega)}$ on a given realization $\omega$ in the event space $\Omega$ is also a stochastic point process. Statistical representations of gas--solid flows based on the SPP
theory have been described by several researchers and they are
reviewed in~\cite{shankar_pecs_2013}. The computational implementation of the SPP approach leads naturally to the simulation of particles in a Lagrangian reference frame as they interact with an Eulerian representation of the carrier fluid flow, giving rise to the Lagrangian-Eulerian (also called Euler-Lagrange) simulation methods for particle-laden suspensions~\cite{shankar_pecs_2013}.

In statistical fluid mechanics of single--phase turbulent flow the starting point is the representation of the flow velocity and pressure fields as random fields~\citep{monin_yaglom2,popebook,adlerbook,panchev}. Theoretical approaches to particle-laden suspensions have also been considered by representing both the dispersed and carrier phases as random fields (RF) in the Eulerian frame of reference~\citep{drew_arfm,pai_ss_jfm09}. Prior to the rigorous analysis of Drew~\cite{drew_arfm} based on ensemble--averaging, conservation equations in the Eulerian frame of reference were also derived on the basis of volume--averaging by~\citet{And_jack} and time--averaging by \citet{ishii_tft}. Historically there has been some controversy about the validity of, and inter-relationship between, these approaches~\citep{joseph_lundgren_etal_ijmf90}. Since the equations obtained from ensemble--averaging and volume--averaging look superficially similar they are used interchangeably in the literature, but they are fundamentally different. Later in this section we will see that the RF basis yields a formal relationship between the deterministic ensemble--averaged equations and their stochastic volume--averaged counterparts, which opens the door for modeling based on stochastic partial differential equations. 

The volume--averaging approach of ~\citet{And_jack} also involved filtering the fluid and particle fields. This seemingly minor variation to straightforward volume averaging is in fact deserving of a separate classification as a third approach to modeling and simulation of particle-laden suspensions, whose origins also lie in single-phase fluid turbulence. Spatial filtering of turbulence fields leads to  large-eddy simulation (LES) ~\citep{smagorinsky1963general,lilly1967proceedings,popebook} methods, which have proved to be very successful in computation of turbulent single--phase flow~\citep{moin1982numerical,smagorinsky1993large,goc2020wall}. \citet{olivier_les_jcp_2013} extended the spatial filtering in Anderson \& Jackson's~\cite{And_jack} volume--averaging approach to particle-laden suspensions and formulated the volume--filtered Euler-Lagrange (VFEL) method. Although VFEL and LES share a common basis in spatial filtering of the governing equations, the differences in multiscale nature of fluid turbulence and particle-laden suspensions inform the nature of the approximations and models that are needed. A brief summary of these approaches is provided below with a view to assessing their adequacy in capturing the rich physics described in the preceding section. 


\subsection{Stochastic point process approach}
Whereas a complete
characterization of all multiparticle events requires consideration of
the Liouville PDF~\citep{pprocess}, there is a hierarchy of equations
analogous to the BBGKY hierarchy found in the classical kinetic theory
of molecular gases~\citep{liboff_book} that leads to the multiphase
version of the one--particle (Boltzmann--Enskog) density
equation~\citep{fox_arfm2011}, 
\begin{equation}
  \dxdt{f} + \gradx{\bf x}\cdot \fn {{\bf v} f} + \gradx{\bf v} \cdot
  {\fn {\ave { {\bf A}\, |\,\bx, \bv ;t}f}} =
  \dot{f}_{\mrm{coll}}\mathrm{,} 
  \label{eq:opdf} 
  \end{equation} where
$\gradx{\bf x}$ and $\gradx{\bf v}$ denote the gradient operators in
the position and velocity space, respectively, angle brackets $\langle \,\,\, \rangle$ denote ensemble--averaging, and
$\dot{f}_{\mathrm{coll}}$ is the collisional term that depends on
higher--order statistics. Equally instructive is the derivation of the one--particle density equation using the Klimontovich approach, which defines the one--particle density as the ensemble average of $f_K'$, the sum of fine--grained densities $f_1'^{(i)}=\delta(\mathbf{x}-\mathbf{X}^{(i)}(t))\delta(\mathbf{v}-\mathbf{V}^{(i)}(t)), i = 1, \ldots, N$ corresponding to each of the $N$ particles:
\begin{equation}
f (\mathbf{x},\mathbf{v},t) \equiv \left\langle f_K' \right\rangle  = \left\langle \sum_{i=1}^{N(t)}f_1'^{(i)} \right\rangle= \left\langle \sum_{i=1}^{N(t)} \delta (\mathbf{x} - \mathbf{X}^{(i)}(t) ) \delta (\mathbf{v} - \mathbf{V}^{(i)}(t) )\right\rangle .
\end{equation}
This reveals important differences between this one--particle density and the PDF associated with a single particle: namely, that the one--particle density is an unnormalized, weighted sum of the PDFs associated with each particle, which are, in general, not independently or identically distributed~\cite{pprocess}.  
In addition to the model for the
collisional term, the one--particle density equation for gas--solid flow requires a closure model for the conditional
particle acceleration term $\ave { {\bf A}\,|\,\bx, \bv ;t}$, which
represents the average hydrodynamic force experienced by a particle
in the suspension at location $\mathbf{x}$ with velocity
$\mathbf{v}$ due to fluid pressure and velocity gradient fields at
the particle surface. Note that in particle--laden suspensions the conditional 
particle acceleration depends on particle velocity through slip with 
respect to the fluid, and therefore appears inside the 
velocity derivative in the velocity transport term. This equation is also known as the number density function (NDF) equation, or Williams' equation for the droplet distribution function in the context of spray droplets, and has been used extensively as the starting point for many analyses of particle-laden suspensions~\citep{koch_pof_1990,fox_arfm2011}.

As noted in~\citet{pprocess}, the key differences between the SPP approach in the classical kinetic theory molecular gases and particle-laden suspensions are: (i) the lack of scale separation, (ii) the importance of fluctuations in number (and hence, volume fraction or bulk density~\footnote{The bulk density is simply the product of the material density of the solid particles \protect$\rho_p$ and the average volume fraction $\phi$}), and (iii) the challenges in representing the effect of the fluid phase. We now examine these key differences to gain an appreciation of the challenges in using the SPP approach to represent the physics of particle-laden suspensions. 

The Boltzmann--Enskog closure of the collision integral on the right hand side of Eq.~\ref{eq:opdf} which writes the two--particle PDF as a product of one--particle PDFs is based on the assumption of binary, instantaneous collisions which relies on a separation of scales~\citep{liboff_book}: that the time between collisions should be much larger than the duration of a collision. In this sense, the NDF equation with a BE closure for the collision term is already time-averaged over a time scale larger than the duration of a collision but smaller than the time between collisions~\citep{tenneti2016stochastic}, and this point will become relevant when we consider numerical simulations of Eqs.(~\ref{eq:cont_fp}--~\ref{eq:lag_vel}). Simple scaling estimates of the terms representing interparticle collisions and
  fluid--particle drag in the NDF lead to the identification of the
  principal nondimensional groups~\citep{fox_cup_book}---the particle Knudsen number
  ($\mathrm{Kn_{p}}=\tau_c U_p/L$), the particle Mach number
  ($\mathrm{Ma} = U_p/\sqrt{\Theta_p}$), and the Stokes number
  ($\mathrm{St}= \tau_p/\tau_f$)---that characterize gas--solid flow. The particle Knudsen number characterizes the ratio of the time between collisions to the characteristic time that a particle takes to traverse a characteristic hydrodynamic length scale $L$ (over which mean and second moments vary spatially) with a characteristic velocity based on the particle granular temperature $\Theta_p$. The particle Mach number characterizes the ratio of a characteristic mean particle velocity scale $U_p$ to a characteristic velocity based on the particle granular temperature $\Theta_p$, while the Stokes number characterizes the ratio of a characteristic particle momentum response time $\tau_p$ to a characteristic fluid momentum response time $\tau_f$. This
  allows the delineation of important regimes of gas--solid flow in
  terms of these nondimensional groups~\citep{fox_cup_book}, resulting in: (a) flows
  dominated by collisions ($\mathrm{Kn_p} <0.1$) or convective transport; (b) low Mach number
  collisional flows and high Mach number flows characterized by shocks
  in the particle phase; and (c) low Stokes number and high Stokes number
  flows with the associated characteristic spatial distribution of
  particles in the flow field~\cite{eaton_fessler_ijmf94}.

Particle--laden suspensions that are dominated by collisions ($\mathrm{Kn_p} <0.1$) are amenable to further simplification using kinetic theory extensions of Chapman-Enskog-like approximations to gas-solid flow that expand the NDF about a homogeneous cooling base state~\citep{vgstsscmh_jfm_2012}. This permits the derivation of Navier--Stokes--Fourier equations for the particle phase that account for the effects of inelastic collisions and the influence of the fluid on particle acceleration in the transport coefficients that appear in constitutive relations with analytical expressions~\citep{vgstsscmh_jfm_2012}. However, since a particle--laden suspension may violate the particle Knudsen number restriction in some parts of the flow, these models are not uniformly valid. Quadrature--based moment methods have emerged as a powerful alternative to tackle regions dominated by convective transport by directly computing the collision integral through quadrature summation, thus eliminating the need to assume $\mathrm{Kn_p} <0.1$~\citep{kong2017solution,heylmun2019quadrature}. However, there is another ratio of length scales that is important in particle--laden suspensions, which is related to the fluctuations in number (volume fraction or bulk density) and has received less attention in the literature. 

It is well known from observations of particle--laden suspensions (see Fig.~\ref{fig:multiscale}) that particles can organize in spatial patterns that are called clusters. We also know from the statistical mechanics of dense gases and liquids that a {\em statistically homogeneous} fluid can manifest structure which is characterized by the relative probability of neighbors at different separation distances by the pair correlation function (PCF), which for isotropic systems is also denoted the radial distribution function (RDF). This information is not available from the one--particle PDF which only characterizes the variation of the average number of particles $\langle N \rangle = \displaystyle \int f (\mathbf{x},\mathbf{v},t)\, d\mathbf{x}\, d\mathbf{v}$ in space through the average number density
\begin{equation}
n(\mathbf{x},t)= \int f (\mathbf{x},\mathbf{v},t)\, d\mathbf{v},
\end{equation}
with associated length scale $\ell_n = n/|\nabla n|$.
On the other hand, the RDF is related to the second--order density characterizing the variance of the number of particles and contains information regarding particle number fluctuations. 

An important insight from the theory of stochastic point processes is that the counting measure $N(\mathscr{A})$ on a region $\mathscr{A}$ in physical space is a {\em random measure} (see Appendix~\ref{app:spp}). Unlike constant density single--phase  turbulent flows where fluid particles are space-filling with fluid (material) volume always equaling the geometric volume, multiphase flows must deal with the additional challenge that particle number and volume fraction (or bulk density) are {\em random measures}. Although the average of the counting measure $N(\mathscr{A})$ yields a first--order theory based on the average number $\langle N(\mathscr{A})\rangle$, and its density the NDF or one--particle density $f(\mathbf{x},\mathbf{v},t)$ (and from it the average number density $n(\mathbf{x},t) $), a {\em major limitation of the NDF equation is that it cannot account for fluctuations in number.} 

One challenge in modeling particle--laden suspensions is that we do not know {\em a priori} whether the spatial patterns in particles arise from an inhomogeneous mean number density field or from second--order structure. A systematic statistical analysis of experimental data is needed to determine the length scales corresponding to the mean number density $\ell_n$, and $\ell_g$, the characteristic length scale associated with the PCF/RDF, before this question can be definitively answered. However, VFEL simulations of particles settling under Stokes drag in a {\em statistically homogeneous} problem called cluster--induced turbulence (CIT) indicate that the length scale of clusters corresponding to second--order structure is $O(10-100) d_p$ (cite Jesse).
Therefore, it is clear that a representation of fluctuations in number is needed to characterize the rich physics of clustering observed in {\em statistically homogeneous} particle--laden suspensions. 

Fluctuations in number can be represented by extending the one--particle fine--grained density in the Klimontovich approach to its two--particle counterpart as follows~\cite{krall_trivelpiece}:
\begin{align}
  f'_1f'_2 =
  \sum_{i=1}^{N}f'^{(i)}_1 \sum_{\substack{j=1\\j\neq i}}^{N}f'^{(j)}_2=
\sum_{i=1}^{N}\delta(\mathbf{x}_1-\mathbf{X}^{(i)}(t))\delta(\mathbf{v}_1-\mathbf{V}^{(i)}(t))
\sum_{\substack{j=1\\j \neq i}}^{N}\delta(\mathbf{x}_2-\mathbf{X}^{(j)}(t))\delta(\mathbf{v}_2-\mathbf{V}^{(j)}(t)) ,
\label{eqn:fprime2}
\end{align}
where
$[\mathbf{x}_k,\mathbf{v}_k, k=1, 2]$ are the Eulerian coordinates of the
position--velocity phase space for the particle pair~\footnote{The summation
over distinct pairs $j\neq i$ is necessary for the definition 
of the two--particle density, whose integral is the second factorial
measure. If all pairs are included, an atomic contribution arises in
the second moment measure that does not have a
density~\cite{stoyan_stoyan,stoyan_kendall_mecke}.}
The ensemble average of the two--particle fine--grained density
function $f'_1f'_2$ is the two--particle density 
$\rho^{(2)}(\mathbf{x}_1,\mathbf{x}_2,\mathbf{v}_1,\mathbf{v}_2,t)$,
which is defined as
\begin{equation}
\rho^{(2)}(\mathbf{x}_1,\mathbf{x}_2,\mathbf{v}_1,\mathbf{v}_2,t) \equiv
\langle f'_1f'_2\rangle. 
\label{eqn:rho2}
\end{equation}
Integrating the two--particle density over the velocity spaces results
in the unnormalized pair--correlation function
\begin{equation}
\rho^{(2)}(\mathbf{x}_1,\mathbf{x}_2,t) = \int
\rho^{(2)}(\mathbf{x}_1,\mathbf{x}_2,\mathbf{v}_1,\mathbf{v}_2,t) \,
d\mathbf{v}_1 d\mathbf{v}_2 ,
\end{equation}
which in turn can be integrated over a region $\mathscr{A}$ in
physical space to obtain the second factorial moment of the counting measure $N(\mathscr{A})$ :
\begin{equation}
  \left\langle 
    N(\mathscr{A}) \, \left[ N(\mathscr{A})-1\right] \right\rangle = \int 
  \rho^{(2)}(\mathbf{x}_1,\mathbf{x}_2,t) 
  d\mathbf{x}_1 d\mathbf{x}_2.
\end{equation}

Although the two--particle density enables the representation of fluctuations in number, its evolution equation is not closed~\citep{sund_coll_jfm99,shankar_pecs_2013,markutsya2012coarse} and presents unique modeling challenges for unclosed terms which are a function of particle pair separation~\citep{rani2014stochastic}. Furthermore, the full solution of the two--particle density in a statistically {\em inhomogeneous} flow requires it to be represented in a higher dimensional space. The twin challenges of additional closure modeling and higher dimensionality have together contributed to slow development and limited adoption of this approach. 

Accurate coupling of the fluid phase with the SPP description of the particles is critical for prediction and modeling of the variety of phenomena described earlier. For ease of modeling and computational representation of boundary conditions, a solution approach based on Lagrangian particle methods is commonly used to indirectly solve the evolution of the NDF (cf. Eq.~\ref{eq:opdf}) in a computationally efficient manner~\citep{shankar_pecs_2013,kiva2,garg2007accurate,garg2009numerically}. The Lagrangian--Eulerian (LE) method corresponds to a closure of the SPP representation at the level of the
droplet distribution function or number density function (NDF), with
the carrier phase represented in an Eulerian frame through a
Reynolds--averaged Navier--Stokes (RANS) closure, LES or DNS. 
The influence of the dispersed phase on the carrier phase is represented by the addition of interphase coupling source terms to the usual fluid-phase RANS, LES or DNS equations. 
The overarching challenge common to all SPP modeling approaches, be it the one--particle/NDF equation or the transport equation for the two--particle density, is that there is no {\em joint statistical representation of fluid and particles in the SPP description}. 
This constitutes a fundamental limitation of the SPP approach. Given that the interaction between fluid and finite-size particles is fundamentally a two-point phenomenon~\cite{shankar_ddf,pprocess}, representing and modeling this coupling at the single-point level is challenging.
We will now examine the relative merits of the RF approach in capturing the rich physics of particle--laden suspensions.
  
\subsection{Random field approach}
In a particle--laden suspension there are two distinct
thermodynamic phases: a carrier fluid phase and a dispersed particle
phase.  In a
single realization $\omega$, which is an element of
the sample space $\Omega$ of all possible
realizations, the phases
can be distiguished using an indicator function
$I_{\phaseindex}(\mathbf{x},t;\omega)$ for the $\phaseindex$th phase, defined
as
\begin{equation}
  I_\beta(\bfx,t; \omega) =
  \begin{cases}
    1 & \text{if $\bfx$ is in phase $\beta$ at time $t$}\\
    0 & \text{if $\bfx$ is not in phase $\beta$ at time $t$.}
  \end{cases}
  \LEQ{ind_fn_defn}
\end{equation}
The phase indicator fields $I_{\phaseindex}(\mathbf{x},t;\omega)$ are random fields that satisfy, at each space--time location $(\mathbf{x},t)$, the relation
\be
\label{eq:indfnsumto1}
\sum_{\phaseindex=\{f,s\}} I_{\phaseindex}(\mathbf{x},t) = 1 , 
\ee
where $f$ represents the fluid phase and $s$ represents the
solid phase.
Statistical information at only a single space--time location
$(\mathbf{x},t)$ of the RF representation results in {\em single--point} Eulerian--Eulerian
(EE) theories. Just as the quest for tractable engineering models of single--phase turbulent flows based on a random--field statistical description at the single--point level led to the familiar Reynolds-averaged equations of turbulent flow, so too the RF approach to particle--laden suspensions
at the closure level of first moments leads to the Eulerian--Eulerian (EE)
two--fluid (TF) theory
~\citep{drew_arfm,drew_passman}. In the EE TF theory, the phases are distinguished by the phase
indicator function and phasic averages are defined as averages
conditional on the presence of the fluid or solid phase. Corresponding
to the complete description in Sec.~\ref{subsubsec:compdesc}, the TF
representation of the same system at the level of first moments would
be $\ave{S}_{TF}=\{\langle \phi\rangle,\langle \mathbf{u}^{(f)}\rangle,
\langle \mathbf{u}^{(s)}\rangle, \ave{p^{(f)}},\ave{p^{(s)}}\}$, where $\langle \phi (\mathbf{x},t)\rangle = \langle I_s(\mathbf{x},t)\rangle$ is the average solid volume fraction and $\langle
\mathbf{u}^{(f)}\rangle (\ave{p^{(f)}})$  and $\langle
\mathbf{u}^{(s)}\rangle (\ave{p^{(s)}}) $ are the average velocities (pressures) of the fluid and
solid phases, respectively. One of the advantages of the RF approach, as compared to the SPP approach, is that it a simultaneous statistical representation of both the fluid phase and the particle phase. Therefore, the sources of
randomness arising from different configurations of particles and from
different realizations of the velocity field are considered
simultaneously. However, as already noted previously, because the interaction between fluid and finite-size particles is fundamentally a two-point phenomenon~\cite{shankar_ddf,pprocess}, representing and modeling this coupling at the single-point level is difficult. Before discussing two--point RF representations, we first resolve a longstanding debate in the RF approach which is later shown to lead to insights for improved modeling in the VFEL approach.

\subsubsection{Ensemble--averaging and volume--averaging}

There has been a longstanding debate between three approaches to deriving the equations in the two--fluid theory---namely, (i) a time--averaging approach~\citep{ishii_tft,kataoka_ijmf}, (ii) a volume--averaging approach~\citep{jackson1997locally,And_jack}, and (iii) the ensemble--averaging approach~\citep{drew_arfm,drew_passman}--- leading to a controversy about the validity of, and inter-relationship between, these approaches~\citep{joseph_lundgren_etal_ijmf90}. Here we resolve the differences between the volume average and ensemble average by means of a simple example based on the mass conservation equation . 

\paragraph{Ensemble--averaging approach}
The ensemble--averaged equations have been derived by several
researchers~\citep{drew_arfm,pai_ss_jfm09}. Starting from the mass conservation equation 
\begin{equation}
\frac{\partial \rho}{\partial t} + \boldsymbol \nabla \cdot ( \rho \mathbf{u} ) = 0,
\end{equation}
which is trivially satisfied inside constant density, rigid particles,
and multiplying it by the indicator function and taking the ensemble
average yields
\begin{equation}
\frac{\partial \langle \phi \rangle}{\partial t}  + \boldsymbol \nabla \cdot ( \langle \phi \rangle \langle \mathbf{u}^{(p)} \rangle ) = 0 ,
\label{eq:ensavmasscons}
\end{equation}
where $\langle \tilde{\mathbf{u}}^{(p)} \rangle$ is the phasic average velocity of the particles (see Appendix~\ref{app:ensav} for details).
\paragraph{Volume--averaging approach}
Let us consider an arbitrary measurement region (or observation window) $\measvol$ with volume
$V_m$ in a gas-solid flow and consider the total volume occupied by
the particles in $\measvol$ at time $t$ corresponding to realization
$\omega$:
\begin{equation}
\label{eq:Phip_def}
\Phi_p (\measvol; t, \omega) = \int_{\measvol} I_p (\mathbf{x},t; \omega)\, d\mathbf{x}.
\end{equation}
Note that $\Phi_p$ does not have a smooth
density in $\mathbf{x}$ because $I_p(\mathbf{x},t; \omega)$ is a
generalized Heaviside function. In mathematical terminology, $\Phi_p
(\measvol; t, \omega)$ is a set function on the set
$\measvol$. Since this set function always yields a non-negative real value, $\Phi_p$ is a measure. Its dependence on the realization $\omega$ reminds us that it is a {\it random measure}. 

A key relationship we establish here is obtained by expressing this volumetric quantity $\Phi_p$ in terms of the
ensemble-averaged solid volume fraction field $\langle \phi \rangle
(\mathbf{x},t)$ as:
\begin{equation}
\label{eq:Phip_and_ensavg}
\Phi_p (\measvol; t, \omega) = \int_{\measvol} \langle \phi \rangle (\mathbf{x},t)\, d\mathbf{x} + 
\tilde{\Phi}_p(\measvol; t, \omega),
\end{equation}
where $\tilde{\Phi}_p(\measvol; t, \omega)$ is a fluctuation that represents the departure of $\Phi_p$ from its ensemble average (first term on right hand side) on realization $\omega$.
Note that $\tilde{\Phi}_p (\measvol; t, \omega)$ also does not have a
smooth density in $\mathbf{x}$ because by definition it involves a
generalized Heaviside function:
\begin{equation}
\tilde{\Phi}_p(\measvol; t, \omega) = \int_{\measvol}  \tilde{I}_p(\mathbf{x},t; \omega) \, d\mathbf{x} ,
\end{equation}
where
\begin{equation}
\label{eq:indfunfluc}
\tilde{I}_p(\mathbf{x},t; \omega) \equiv I_p(\mathbf{x},t; \omega) - \langle \phi \rangle (\mathbf{x},t) .
\end{equation}

For the statistically homogeneous case, we can think of
$\tilde{\Phi}_p(\measvol; t, \omega)$ as the statistical variability
in the estimate of
\begin{equation}
\left\langle  \Phi_p (\measvol; t) \right\rangle
=
\int_{\measvol} \langle \phi \rangle (\mathbf{x},t)\, d\mathbf{x} ,
\end{equation}
arising from a finite number of samples in the measurement region
$\measvol$. For the statistically homogeneous case,
$\tilde{\Phi}_p(\measvol; t, \omega) \rightarrow 0$ as
$1/\sqrt{N({\measvol})}$. But for all finite $V_m$ it is
non-zero and it is a random quantity since it depends on $\omega$.

We can write a conservation equation for $\Phi_p (\measvol; t,
\omega)$ by considering its temporal evolution
\begin{equation}
\frac{\partial \Phi_p (\measvol; t, \omega)}{\partial t} = \int_{\measvol} \frac{\partial I_p (\mathbf{x},t; \omega)}{\partial t} \, d\mathbf{x} ,
\end{equation}
to obtain (see Appendix~\ref{app:volav} for details):
\begin{equation}
\label{eq:volavgmasscons}
\overbrace{\frac{\partial }{\partial t} \int_{\measvol} \langle \phi \rangle (\mathbf{x},t)\, d\mathbf{x}}^{(1)}+ 
\overbrace{\frac{\partial \tilde{\Phi}_p}{\partial t}(\measvol; t, \omega) }^{(2)}
 =   \underbrace{- \int_{\partial \measvol} \langle \phi \rangle  \langle \mathbf{u}_p\rangle \boldsymbol \cdot \mathbf{n} \, dA}_{(3)} -
 \underbrace{\int_{\partial \measvol} \mathbf{J}^{\prime} \boldsymbol \cdot \mathbf{n} \, dA }_{(4)}.
\end{equation}
Note that the above equation is a stochastic equation with explicit dependence on the realization $\omega$.
Clearly, terms (1) and (3) in the above equation represent the ensemble--averaged mass conservation (cf. Eq.~\ref{eq:ensavmasscons}) for particles with constant material density, which is a deterministic equation. Terms (2) and (4) on the other hand, represent the {\em departure of the stochastic volume--averaged mass conservation equation from its deterministic ensemble--averaged counterpart}. 

To summarize, we have formally established an important relationship between the volume average and the ensemble average for multiphase flows. It is shown that the volume average is a random quantity that differs from the non-random ensemble average by a stochastic fluctuation. For statistically homogeneous flows it is shown that volume--averaged particle properties such as the particle geometric volume converge to their ensemble--averaged counterparts as $1/\sqrt{N(\measvol)}$. In the case of statistically inhomogeneous flows, a separation of scales is needed to ensure a similar relationship between the volume average and the local ensemble average and this criterion is explicitly established.

The volume average of particle geometric volume is used to derive the weak form of mass conservation for the particle phase. This equation can be interpreted as the ensemble-average mass conservation equation representing a balance between the temporal derivative of ensemble-averaged particle volume and average flux of the same quantity at the system boundary, augmented by the temporal derivative of the stochastic fluctuation which is balanced by a flux fluctuation which varies with each realization of the multiphase flow. The stochastic fluctuation term contains important information about second-order fluctuations in particle number and volume which is absent in first-order ensemble-averaged quantities such as the ensemble-averaged particle number density and volume fraction. 
\subsubsection{Second--moment and PDF closures}
Similar to the unclosed Reynolds stress term in the Reynolds--averaged equations for single--phase turbulent flow, the ensemble--averaged TF momentum equations
  for the fluid and solid phases also contain unclosed terms 
  corresponding to stresses arising from fluid velocity
  fluctuations and particle velocity fluctuations,
  respectively. 
Attempts to improve the single--point statistical description of single--phase turbulent flow by resorting to higher levels of closure have led from the Reynolds-averaged equations  to Reynolds-stress transport equations involving second moments, and culminated in the transport equation for the single--point probability density function~\cite{popebook}. In the case of riser flows involving particle--laden suspensions (see Fig.~\ref{fig:multiscale}), \citet{Hrenya_aiche97} 
showed that an additional transport equation for the particle velocity variance is necessary to predict phenomena such as the core-annular structure. Similar efforts to extend the EE TF theory beyond average quantities have led to second--moment closures~\citep{simonin_95,ahmadi_ma_1990a,xu2006multiscale} as well as the single--point PDF approach~\citep{pai_ss_jfm09}. {\em It is worth noting that these single--point second--moment and PDF closures are still at the level of first-order ensemble-averaged quantities such as the ensemble-averaged particle number density and volume fraction, and do not contain information concerning fluctuations in particle number (or volume).} 
~\citet{pai_ss_jfm09} showed that the ensemble--averaged EE TF equations of ~\citet{drew_arfm} can be derived from the single--point PDF approach and the consistency conditions for the correspondence of mean and second--moment equations between the RF and SPP approaches were established. ~\citet{fox2014multiphase} derived a multiphase turbulence theory based on a sequential phase and Reynolds averaging procedure starting from a mesoscale description based on a continuous volume fraction field. Although this approach is useful at the macroscale in incorporating mesoscale fluctuations, it does not reconcile with the microscale picture derived here.

The advantages of the single--point PDF approach, as explained in~\citet{pai_ss_jfm09}, include allowing for a
  distribution of solid and fluid velocities at the same
  location, and a clear identification of events and their probabilities in the single--point EE TF formulation. The latter are useful in distinguishing a key difference between the RF approach in multiphase flow and constant-density single--phase turbulent flow which is overlooked in some theoretical treatments. In constant-density single--phase flow, two--point statistics such as the Eulerian velocity autocovariance reduce to the single--point Reynolds stress tensor in the limit of zero pair separation distance because the underlying measure is not random, so $\lim_{\mathbf{r} \rightarrow 0} R_{ij}(\mathbf{x},\mathbf{x} + \mathbf{r}) = R_{ij}(\mathbf{x})$. The same is not true for multiphase flows where certain two--point statistics such as fluid--particle velocity covariance vanish at zero separation, e.g. $\lim_{\mathbf{r} \rightarrow 0} R^{pf}_{ij}(\mathbf{x},\mathbf{x} + \mathbf{r}) = 0$, because particle and fluid cannot coexist at the same physical location and the two--point statistic is based on a different measure~\citep{sund_coll_jfm99}.  The analysis of events and probabilities in the single--point RF approach reveals that correlation between gas-- and solid--phase motions does not manifest at the single--point level of closure, because for
   particles of finite size, every space--time location can be occupied
   only by either solid or gas~\citep{xu_ss_pf06}. However, these correlations
   do appear in the two--point representation.

The unclosed terms that need to be modeled in all the single--point RF approaches discussed thus far (including the average interphase momentum transfer, the particle velocity covariance, the fluid velocity covariance, and the conditional expectation of acceleration in each phase that arise in the PDF approach) represent motions
  at all scales (micro, meso and macro). However, the mechanisms that
  produce fluid velocity fluctuations, for instance, at the microscale are different
  from the interaction of particle clusters with fluid flow at the
  mesoscale, or the interphase interaction at the macroscale.  Since
  the single--point RF approaches do not contain scale information they cannot
  distinguish between these different scale--dependent mechanisms that
  contribute to the transport of fluid velocity fluctuations.  Just as two--point statistics and spectral closures were introduced in single--phase turbulent flow to account for the multiscale nature of turbulence, similar attempts~\citep{sund_coll_jfm99} have been made to account for the multiscale nature of particle--fluid interactions in particle--laden suspensions.  

\subsubsection{Two--point closure}
Higher order representations such as a two--point statistical description of gas--solid flow based on the RF representation can be
found in~\citet{sund_coll_jfm99}.  These are based on the second moment of the phase indicator random field $C_{\beta \gamma}(\mathbf{x},\mathbf{x} + \mathbf{r}) = \langle I_{\beta} (\mathbf{x})I_{\gamma}(\mathbf{x} + \mathbf{r})\rangle, \beta, \gamma \in \{f,p\}$, which appears as a natural two--point extension of $\langle I_{\beta} (\mathbf{x})\rangle$. Just as fluctuations in a random variable are described by the variance, fluctuations in a random field are described by the covariance field. 
The covariance of the random field $I_p(\mathbf{x},t; \omega)$ (see p. 203 of~\citet{stoyan_kendall_mecke}) in the statistically homogeneous, isotropic case is simply the second moment of $I_p$ and is given by:
\begin{equation}
C_{pp}(\mathbf{r}) = \langle I_p(\mathbf{x},t)I_p(\mathbf{x}+\mathbf{r},t) \rangle ,
\end{equation}
which is similar to the two-point single-time Eulerian autocovariance in constant density single-phase turbulence. 
The covariance of the particle indicator
function field represents the expected value of the event where two spatial locations separated by $\mathbf{r}$ are simultaneously occupied by the
particle phase. In the limit of very large separation $\mathbf{r}\rightarrow \infty$ and $C_{pp}(\infty)=\langle \phi\rangle^2$ corresponding to zero correlation between the occurrence of particles at points separated far apart. 
In the limit of zero separation $\mathbf{r}\rightarrow 0$, $C_{pp}(0)=\langle\phi \rangle$, which confirms that {\em fluctuations in volume fraction do not show up in the EE TF single--point theory}, just as the single--point NDF in the SPP approach contains no information concerning number fluctuations.

Note that the covariance $C_{pp}(\mathbf{r})$  is not centered. If we consider the second central moment, it is called the {\it covariance function} $k(\mathbf{r})$ and is defined in terms of $\tilde{I}_p \equiv I_p - \langle \phi \rangle$ by
\begin{equation}
k(\mathbf{r}) = \langle (\tilde{I}_p(\mathbf{x},t) \tilde{I}_p(\mathbf{x}+\mathbf{r},t) \rangle =  C_{pp}(\mathbf{r}) - \langle \phi \rangle^2.
\end{equation} 
The single-point limit of the covariance function at $\mathbf{r}\rightarrow 0$ is $k(0)=\langle\phi \rangle (1 - \langle\phi \rangle)$, while for $\mathbf{r}\rightarrow \infty$, $k(\infty)=0$ corresponding to zero correlation between the occurrence of particles at points at far separations. These results have also been reported by~\citet{sund_coll_ijmf1}.

There is a close connection between the random field $I_p(\mathbf{x},t; \omega)$ 
 and the distribution of the centers of the particles which can be described by a stochastic point process. So there is intrinsically a connection between the statistics of the random field $I_p(\mathbf{x},t; \omega)$ and those of 
the underlying stochastic point process. In the case of monodisperse spheres, this connection is easy to write down analytically for model systems. One such system is overlapping spheres with centers distributed according to a Poisson point process, although this is not a good model for many multiphase systems. Systems of non-overlapping spheres are a better model for particle-laden flows and it is easy to see in this case that the covariance $C_{pp}(\mathbf{r})$ contains information about clustering of particles, which is not contained in the mean solid volume fraction. 

This connection between the statistics of the random field and the underlying point process yields a very useful relation in the case of spherical particles, which can be used to relate the SPP and RF approaches at the level of two--point statistics.
The covariance function $k(\mathbf{r})$ is related to the pair correlation function which is a number-based statistic associated with the underlying point process. This relationship has been derived previously by other researchers for both monodisperse (\citet{torquato1982microstructure}, see also Eq. 10 of ~\citet{sund_coll_ijmf1}) and polydisperse spheres~\citep{sund_coll_ijmf2}. For a system of monodisperse non--overlapping spheres that are distributed corresponding to a homogeneous number density $n$ and a pair correlation function $g(\mathbf{r})$, it can be shown that
\begin{equation}
k(\mathbf{r}) =  n I(\mathbf{r}) + n^2
\int h(\mathbf{z})I(\mathbf{r}-\mathbf{z}) \, d\mathbf{z}. 
\end{equation}
In this expression $h(\mathbf{z}) = g(\mathbf{z}) - 1$ is the total correlation function and $I(\mathbf{r})$ is the volume of intersection of two spheres whose centers are separated by a vector $\mathbf{r}$:
\begin{equation}
I(r) = V_p \left\{ 
\begin{array}{cc}
\displaystyle 1 - \frac{3}{2}\frac{r}{r_p} + \frac{1}{2} \left( \frac{r}{r_p}\right)^3 & \displaystyle   \frac{r}{r_p} \leq 1 \\
& \\
\displaystyle 0 &   \displaystyle  \frac{r}{r_p} > 1 
\end{array}
\right.
\end{equation}
These relations show that just as ~\citet{pai_ss_jfm09} established consistency conditions for the correspondence of mean and second--moment equations between the RF and SPP approaches at the single--point level, the same is possible at the two--point level also. 

There is an important connection between how {\em estimates} of particle volume fraction fluctuations are computed from point fields such as PR--DNS or LE/VFEL particle data, and the representation of these fluctuations in the volume--averaging approach that merits discussion here. ~\citet{lu1990local} showed that the {\em estimates} of particle volume fraction fluctuations in a statistically homogeneous distribution of spheres obtained by volume-averaging over a region $\mathscr{V}_m$ actually depend on the volume of that region and the two--point density $k(\mathbf{r}$ (details are given in Appendix~\ref{app:repflucvolavg}). This is a very important result which shows that volume fraction fluctuations cannot be written as a single--point density! In Appendix~\ref{app:repflucvolavg} we also make the connection between the representation of volume fraction fluctuations in the volume--averaging approach and this result, thereby providing a route to rigorously incorporating volume fraction fluctuations into extensions of current single--point theories.

The two--point theory needs to be extended to statistically inhomogeneous flows before it can be applied to realistic problems, and even in the homogeneous case the resulting two--point equations lead to additional unclosed terms that require closure models.
Just as in in two--particle SPP approach, here also the challenges of closure modeling and higher dimensionality have hampered the development and adoption of this approach.

But perhaps the most important drawback of both the SPP and RF statistical approaches is that {\em the averaging process obscures the emergence of scale--dependent structure from instabilities that develop as a consequence of nonlinear interactions between particle configuration and flow conditions specific to each realization which are not represented in the ensemble average}. For instance, the stability theory developed by ~\citet{koch_pof_1990} analyzes the stability of a suspension in terms of the average solid volume fraction $\langle \phi \rangle$, and stability limits are derived on the basis of perturbations to $\langle \phi \rangle$ about a base state. However, it is clear from both simulations and experiment that there is a natural statistical variability to the particle volume $\mathbf{\Phi}_p$ from one realization to another, and it is probable that the origin of instabilities in particle--laden suspensions lies in the microscale interactions related to volume fraction fluctuations, rather than mean perturbations. Therefore, a more realistic stability theory must arrive at stability limits that account for the variance of volume fraction fluctuations which are related to the second--order structure. Whether this can be captured by a two--point theory, or if ensemble--averaging obscures the interaction between the two phases on each realization has not been definitively established. However, current analysis of the two--point theory~\citep{murphy2017analysis} and VFEL simulations~\citep{capecelatro2014numerical} indicate that a two--point theory may not be adequate, or too complicated, to provide such stability limits. Just as the importance of capturing coherent structures led to the development of large-eddy simulation approaches in single--phase turbulent flow, the importance of capturing scale--dependent structure in particle--laden suspensions has prompted the development of spatial filtering approaches leading to the VFEL method.

\subsection{Spatial filtering approach}

The volume averaging approach described earlier is useful to gain a
formal understanding of conventional approaches to multiphase flows
and to highlight its differences from the ensemble-averaging
approach. However, a limitation of Eq.~\ref{eq:volavgmasscons} is that
$\tilde{\Phi}_p$ does not have a smooth density (nor does
$\mathbf{J}^{\prime}$) and so a differential form cannot be
derived. It could be interesting to explore whether in a finite volume
context, the stochastic form of Eq.~\ref{eq:volavgmasscons} could be directly modeled and
solved. An alternative is to consider
spatial filtering of the indicator function $I_p$ and derive
conservation equations based on filtered quantities. This was the
approach pursued by~\citet{And_jack} (AJ67) and~\citet{jackson1997locally} (J97).

Here we derive the mass conservation equation in the spatial filtering
approach and compare it with the results of AJ67 and J97. Spatial
filtering of the indicator function $I_p$ by a smooth, infinitely
differentiable, homogeneous kernel~\footnote{A {\em homogeneous} kernel function~\cite{popebook} is a special case of the general form $\mathscr{G}(\mathbf{r},\mathbf{x})$ when the kernel is independent of $\mathbf{x}$ and depends only on the separation $\mathbf{r} = \mathbf{x} - \mathbf{y}$.} function $\mathscr{G}(\mathbf{r}=\mathbf{x} - \mathbf{y}; \Delta)$ leads to a
filtered instantaneous particle volume fraction field $\bar{\phi}$
\begin{equation}
\label{eq:phibar}
\bar{\phi} (\mathbf{x}, t; \omega) \equiv \int_{\volinf} I_p (\mathbf{y},t) \mathscr{G}(\mathbf{x} - \mathbf{y}; \Delta) d\mathbf{y}  ,
\end{equation}
which is also a smooth, infinitely differentiable {\it random}
field. In the filtering operation the kernel is integrated over the entire domain $\volinf$. 
Taking the time derivative of $\bar{\phi}$ defined by Eq.~\ref{eq:phibar} and substituting the topological equation governing the evolution of the indicator function~\cite{drew_arfm} 
\[
\frac{\partial I_p (\mathbf{y},t)}{\partial t}  = - \nabla_{\mathbf{y}} \boldsymbol \cdot (\mathbf{u} I_p) 
\]
results in the spatially filtered mass conservation equation (see Appendix~\ref{app:massconsspatfilt})
\begin{equation}
\frac{\partial \bar{\phi} }{\partial t} + \nabla_{\mathbf{x}} \boldsymbol \cdot ( \bar{\phi} \,\bar{\mathbf{v}} ) = - \int_{\partial V_{\infty}} \mathbf{n} \boldsymbol\cdot
\left[ \mathbf{u} I_p \mathscr{G}(\mathbf{x} - \mathbf{y}; \Delta) \right] dA.
\label{eq:genspatfiltmasscons}
\end{equation}
For kernels with compact support, the right hand side term vanishes at interior points sufficiently far from the boundary, resulting in a familiar equation
\begin{equation}
\frac{\partial \bar{\phi} }{\partial t} + \nabla_{\mathbf{x}} \boldsymbol \cdot ( \bar{\phi} \,\bar{\mathbf{v}} ) = 0,
\label{eq:spatfiltmasscons}
\end{equation}
which also holds everywhere for periodic boundary conditions. Although this equation {\em looks} exactly like the ensemble--averaged mass conservation equation Eq.~\ref{eq:ensavmasscons}, they are significantly different. For one, Eq.~\ref{eq:spatfiltmasscons} applies to smooth random fields whereas Eq.~\ref{eq:ensavmasscons} applies to the deterministic average volume fraction field. Many computational implementations of the EE TF model such as the widely--used MFIX code~\citep{mfix_theory} ascribe their origins to AJ67 and J97, but in fact they are based on ensemble--averaging because they rely on deterministic closure models.

The spatial filtering approach forms the basis for the volume--filtered Euler--Lagrange (VFEL) simulation approach of~\citet{olivier_les_jcp_2013}. It is noteworthy that VFEL simulations represent a realization of a particle--laden suspension and the solution will in general depend on initial conditions, filter width and choice of kernel function. The biggest advantage of VFEL simulations is that it opens the door to performing simulations of significantly larger domains and discovering mesoscale physics hitherto inaccessible to PR--DNS of the equations governing the complete description given in Sec.~\ref{subsubsec:compdesc}. VFEL simulations of cluster--induced turbulence~\citep{capecelatro2014numerical} were performed on a $2048 \times 512 \times 512$ mesh with $55 \times 10^6$ particles. Equally noteworthy is that models for the unclosed terms in VFEL will be different from their ensemble--averaged counterparts. Broadly speaking, spatial filtering confers the potential advantage of capturing scale--dependent fluctuations, but this promise is realized only insofar as the accuracy of models for the unclosed terms.

\subsection{Summary perspective of theoretical approaches}
\label{subsec:sumpersptheoapp}
In this section we have seen that the rich physics observed in particle--laden suspensions poses significant challenges to classical statistical mechanics approaches. The interaction of fluid with particles generates stress at the particle surface which manifests as a hydrodynamic force that accelerates the particle center of mass. Furthermore, particles can organize in structures spanning ten to hundred particle diameters due to interaction with the flow and form clusters. Joint two--point statistical representation of particles and fluid is needed to capture these interactions, but the multiscale nature of these interactions poses unique challenges. Considerable progress has been made in single--point statistical theory and consistent theories have been developed using SPP and RF approaches, but single--point theories in both approaches are incapable of representing fluctuations in particle number or volume. In this work it is also shown how volume--averaging in the RF approach differs from ensemble--averaging by 
a stochastic fluctuation term that contains important information about second-order fluctuations in particle number and volume, which is absent in first-order ensemble-averaged quantities such as the ensemble-averaged particle number density and volume fraction.
This leads naturally to the consideration of spatial filtering, which results in smooth fields describing a realization of a particle--laden suspension with the potential to more faithfully capture the emergence of scale--dependent structure from instabilities~\citep{kumaran2003stability} that develop as a consequence of nonlinear interactions between particle configuration and flow conditions specific to each realization, which are not represented in the ensemble average.
There have been attempts to produce single--point theories~\citep{fox2014multiphase} that represent fluctuations in particle volume fraction starting from a mesoscale description involving smooth fields, but as noted earlier this does not reconcile with the microscale picture derived here. It is desirable that both theoretical approaches and simulation methods--PR--DNS at the microscale, VFEL at the mesoscale, and EE TF or LE/QBMM at the macroscale--reconcile across scales, and are validated by experimental data wherever possible. 

In Sec.~\ref{sec:futdir} we propose a new class of formulations, including one involving spatial filtering of the Klimontovich density, as a path forward to rigorously incorporating second--order statistical information characterizing particle number (and volume) fluctuations by explicitly modeling it in a one--particle/single--point theory. These formulations would be reconcilable across micro, meso and macroscales and consistent counterparts could be identified in the SPP and RF approaches. We now discuss the model--free PR--DNS method to solve the governing equations of the complete description in Sec.~\ref{subsubsec:compdesc} that can be used to quantify and model the unclosed terms appearing in the governing equations of the theoretical approaches discussed in this section.
\section{Particle--Resolved Direct Numerical Simulation}
Particle--resolved direct numerical simulation of the complete description of a particle-laden suspension given in Sec.~\ref{subsubsec:compdesc} is a useful approach to discover and explain the rich physics in these flows and to develop closure models in the various modeling approaches by quantifying unclosed terms that appear in them~\citep {shankar2014arfm}. PR-DNS involves solving the governing equations (cf. Eqs~\ref{eq:cont_fp}--~\ref{eq:lag_vel}) described in Sec.~\ref{subsubsec:compdesc} with sufficient resolution to resolve the flow around each particle in the suspension. PR-DNS has been successfully used to discover and quantify flow phenomena such as pseudo-turbulence~\citep{mehrabadi2015pseudo} and pseudo-turbulent heat flux~\citep{sun2016pseudo} as well as particle force fluctuations~\citep{akiki2016force}. 
There are many computational methods to perform PR--DNS~\citep{hu_jcp2001,nomura_hughes,bagchi_balachandar_03,bagchi_balachandar_04,burton_eaton,patankaretal_ijmf2000,joseph_jcp2001,ladd_verberg_lbm,sharma_patankar_05,apte_etal,peskin_81,jamalphd,
taira_col_ibm_jcp2007,prosperetti_oguz_jcp01,Uhlmann:2005aa,rgarg_thesis,
kim_choi_06, lucci_etal_2010,zaleski_whole_domain,wylie_koch_jfm_03,tenneti_pt_2010,kriebitzsch_2013, zhou_2014,luo_2016,tang_free_2016,rubinstein_2016,rubinstein_2017, zaidi_2018} ( see~\citet{garg_book}
and~\citet{tenneti_ijmf_2011} for brief summaries, and ~\citet{prosperetti_trygg_book_2007} for an authoritative compendium of PR--DNS
approaches for a wide class of multiphase flows)
, but an exhaustive discussion of the numerical details and the tradeoff between accuracy and computational efficiency is neither feasible nor relevant to this discussion. Instead, here we will examine some common outstanding questions concerning PR--DNS, and identify key assumptions pertinent to developing closure models for the statistical mechanical theories described earlier. 

PR--DNS is computationally expensive and ~\citet{xu2010effect} estimated that the computational cost in terms of the number of grid points for a uniform grid scales as $R_{\lambda}^{3/2} \sqrt{Re_p}$, where $R_{\lambda}$ is the Taylor--scale Reynolds number of turbulence and $Re_p$ is the Reynolds number based on the mean particle slip velocity and diameter. Therefore, problem size is limited by available computational resources. It is not feasible with current computational resources to simulate macroscale particle--laden suspensions in industrial devices using PR--DNS. 
Even when these limitations are lifted, such simulations may not necessarily be useful, for the simple reason that more detailed information does not automatically translate into more insight that is useful in informing decision-making. A more useful approach, already effectively deployed in single--phase turbulent flow, is to use PR--DNS on problem sizes that are currently accessible through available computational resources, and use that simulation data at the microscale to inform statistical theories that are intended to predict phenomena at the meso and macroscale. There are also similarities to the use of molecular dynamics to inform statistical mechanical theories of atomic and molecular systems. 

The particle--laden suspension flows described in the applications are {\em statistically inhomogeneous} flows which means that the statistics vary with spatial location, and we can use $\ell_n$ as a characteristic length scale for these variations. We can use PR--DNS of microscale dynamics in statistically homogeneous problems on domains of characteristic size $L$ to develop models for unclosed terms in these statistical theories provided local statistical homogeneity can be assumed. This assumes a separation of scales $\ell_n \gg L$. Periodic boundary conditions are usually imposed on such PR--DNS which corresponds to the approximation of an infinite statistically homogeneous suspension by repeating periodic images. The PR--DNS domain has to be large enough to ensure that all Eulerian two--point correlations in both phases have decayed to zero, which implies that $L \geq \ell_g$. An interesting observation by~\citet{sun2016modeling} is that the length scale of variation of average fluid phase quantities can be significantly affected by interphase transfer terms. For instance, in a particle--laden suspension with heat transfer, fluid heating/cooling by heat exchange with particles in the entrance region can significantly change the length scale of variation of the mean fluid temperature, and this length scale depends on the solid volume fraction and Reynolds number based on the mean slip velocity between particles and fluid. For slow flow through dense particle suspensions, the length scale of variation of mean fluid temperature can be just a few particle diameters in the entrance region. This implies that local closure models may not always be adequate, and {\em non--local} closure models might be needed in some multiphase flow problems.

Another issue that has not been considered carefully is the choice of ensemble in the PR--DNS. In classical statistical mechanics, the canonical and microcanonical ensembles are fixed $N$ ensembles, whereas the grand canonical ensemble is a variable $N$ ensemble~\citep{allen_tildesley_cslbook}. Of course these classical ensembles refer to equilibrium thermodynamic states, whereas PR--DNS of canonical homogeneous problems such as steady flow through a fixed bed of particles established by a constant mean fluid pressure gradient pertain to a non--equilibrium steady state. There are also differences in how different PR--DNS studies initialize a statistically homogeneous particle configuration corresponding to a specified average solid volume fraction. Some researchers have used configurations with a PCF/RDF obtained from the steady solution to a granular gas undergoing elastic collisions at that volume fraction~\citep{tenneti_ijmf_2011}. 
The suitability of a specific computational approach to PR--DNS also depends on other issues such as accuracy and numerical convergence characteristics. A recent rigorous study of the numerical characteristics of the widely--used immersed boundary method (IBM) by~\citet{zhou2018investigation} provides precisely this kind of insight, and similar studies on other methods are needed to assess the suitability of each computational approach and provide guidance to the research community on their effective usage.

\section{Future Directions} 
\label{sec:futdir}
In Section~\ref{subsec:sumpersptheoapp} we established the need for the inclusion of fluctuations in particle number (or volume fraction) to capture multiscale interactions that underlie the emergence of scale--dependent structure from instabilities that develop between a particle configuration and flow conditions specific to each realization. We also noted that it is desirable that the formulation be reconcilable across micro, meso and macroscales so that PR--DNS can be used for modeling the unclosed terms at the microscale. For a statistically homogeneous suspension, ergodicity requires that the residual terms in VFEL in the limit of sufficiently large filter width should equal the ensemble--average from PR--DNS. VFEL simulations that satisfy this requirement can be used to model unclosed terms at the mesoscale. The formulations proposed in the following also provide a theoretical basis to relate VFEL simulation data at the mesoscale to EE TF or LE/QBMM simulation data at the macroscale.  Given the equivalence between statistical quantities in the SPP and RF approaches, a formulation in one approach should imply a consistent counterpart in the other. 

There are several promising approaches that could meet this need. The first one involves a Lagrangian particle representation of the fluctuation (second term on the right hand side of Eq.~\ref{eq:Phip_and_ensavg}) as either physical  particles
\begin{equation}
\nonumber
\tilde{\Phi}_p(\measvol; t, \omega) \approx \int_{\measvol} \sum_{i=1}^{N} V_p^{(i)} \delta \left(\mathbf{x} - \mathbf{X}^{(i)} \right)\, d\mathbf{x},
\end{equation}
or computational particles
\begin{equation}
\nonumber
\tilde{\Phi}_p(\measvol; t, \omega) \approx \int_{\measvol} \sum_{i=1}^{N_c} w^{(i)} V_p^{(i)} \delta \left(\mathbf{x} - \mathbf{X}^{(i)} \right)\, d\mathbf{x},
\end{equation}
where $V_p^{(i)}$ is the geometric volume associated with the $i$th particle, and $w^{(i)}$ is its statistical weight in the ensemble of $N_c$ computational particles. Substituting this representation into the volume--averaged mass conservation equation (Eq.~\ref{eq:volavgmasscons}) naturally extends the ensemble--averaged EE TF theory to account for fluctuations. A new set of additional unclosed terms (e.g. term (4) in Eq.~\ref{eq:volavgmasscons}, and corresponding terms in the momentum conservation equation) representing fluctuations are introduced, that require modeling. This EE Lagrangian fluctuation (EE--LF) approach has the advantage of building on existing finite--volume--based code structures in EE TF implementations of multiphase computational fluid dynamics (mCFD) such as the widely used open source OpenFOAM ~\citep{weller2018openfoam} and MFIX~\citep{mfix_theory} codes as well as commercial software such as ANSYS Fluent, CFX and Star--CCM+. This formulation appears closely related to traditional LE method but its computational cost should scale far more favorably since {\em only the fluctuation in number} is represented by Lagrangian particles. Several closure modeling questions need to be addressed pertaining to the interaction of fluctuations with the ensemble--averaged mean in both the mass and momentum conservation equations.

The second approach involves an Eulerian representation of fluctuations. Since the fluctuation term $\tilde{\Phi}_p(\measvol; t, \omega)$ in the volume--averaged approach does not have a smooth density, we must resort to spatial filtering. Here there are two choices concerning how we represent fluctuations. Filtering the expression for the fluctuation in the indicator function field $\tilde{I}_p(\mathbf{x},t; \omega)$ from its ensemble--average (cf. Eq.~\ref{eq:indfunfluc}) according to Eq.~\ref{eq:phibar}
yields the fluctuation in the spatially filtered volume fraction field from the spatially filtered ensemble average volume fraction field:
\begin{equation} 
\check{\phi}(\mathbf{x},t;\omega) \equiv \bar{\phi}(\mathbf{x},t;\omega) - \overline{\langle \phi \rangle} (\mathbf{x},t) .
\end{equation}
This is a true fluctuation in the sense that its ensemble average is zero. Although $\check{\phi}$ contains locally filtered information and so reflects the level of clustering over a length scale $\Delta$, the second moment of this field needs to be related to the two--point density (PCF/RDF) in order to quantify it. 

The advantage of this spatially filtered EE with Eulerian fluctuation (SFEE-EF) approach is that the deterministic part ties directly to the filtered two--fluid model of Sundaresan and co--workers~\citep{igci2008filtered}. The filtered two--fluid formulation has been adopted by mCFD users in industry as well as commercial mCFD developers for facilitating simulations of industrial devices on large domains using relatively coarse grids that would otherwise result in under--resolved simulations using traditional EE TF codes~\citep{ozarkar2015validation}. Enhancing that formulation by adding information concerning fluctuations and placing it on a rigorous footing are desirable outcomes. The filter width in this approach can be chosen to be much larger than typical values used in VFEL (where it is usually $\sim 10 d_p$) because it needs to only resolve macroscale or mesoscale variations.

Another Eulerian fluctuation field can be generated by simply taking the difference between the filtered field $\bar{\phi} (\mathbf{x}, t; \omega)$ and its ensemble--averaged counterpart $\langle \phi \rangle (\mathbf{x},t)$:
\begin{equation} 
\hat{\phi}(\mathbf{x},t;\omega) \equiv \bar{\phi}(\mathbf{x},t;\omega) - \langle \phi \rangle (\mathbf{x},t) .
\end{equation}
The ensemble average of this field is not zero and in that sense it is not a true fluctuation. However, it is worth exploring its relationship to the mesoscale volume fraction fluctuations in the multiphase turbulence theory of ~\citet{fox2014multiphase}.
Here also the key question is how to relate the second moment of this field to the two--point density (PCF/RDF). 

The relations between these fluctuations and the two--point density will be useful for both quantifying the fluctuations and as well as initializing them from specified two--point statistics. Using an approach similar to generalized polynomial chaos expansions (GPCE)~\citep{xiu2003modeling,soize2009reduced} to represent a realization of a random field, we can generate synthetic fields with specified single--point and two--point statistics. For instance, ~\citet{oztireli2012analysis} report an algorithm to generate a synthetic point field that corresponds to a given PCF. 

The third approach involves directly filtering the fine--grained (and hence, non-smooth) Klimontovich density $f_K'$ to obtain a filtered Klimontovich density function (FKDF) $\overline{f_K'}$ as follows:
\begin{equation}
\overline{f_K'} (\mathbf{x},\mathbf{v},t) = \int_{\mathbf{y}}\int_{\mathbf{w}} f_K' (\mathbf{y},\mathbf{w},t) 
\mathscr{G}(\abs{\mathbf{x} - \mathbf{y}}, \abs{\mathbf{v} - \mathbf{w}}; \Delta_x, \Delta_v)\, d\mathbf{y} \, d\mathbf{w} ,
\end{equation}
where the filter is defined in the product space to generate a smooth density.
The filter could be written as a separable product of filters in both physical space and velocity space. This filtered Klimontovich density function (FKDF) approach generates a filtered density function similar to the FDF approach in turbulent reacting flows~\citep{colucci1998filtered,fox_book}, but it is different because the underlying PDF in turbulent reacting flows is smooth and differentiable, whereas $f_K'$ is not. 

This observation opens up the possibility of developing a {\em fluctuation hydrodynamics} approach~\citep{klimontovich_2012} analogous to one which has been pursued in the context of granular gases~\citep{van1997mesoscopic}. Our starting point is to decompose the fine--grained Klimontovich density $f_K'$ in the same way as Eq.~\ref{eq:Phip_and_ensavg}  to express it as the sum of the NDF and a volume--dependent stochastic fluctuation in number $\widetilde{N}$:
\begin{equation}
\label{eq:fKprime_NDF}
N (\measvol; t, \omega) = \int_{\measvol} f_K' (\mathbf{x},\mathbf{v},t)\, d\mathbf{x} \, d\mathbf{v}  =  \int_{\measvol} f (\mathbf{x},\mathbf{v},t)\, d\mathbf{x} \, d\mathbf{v} + 
\widetilde{N}(\measvol; t, \omega),
\end{equation}
where now $\measvol$ denotes a region in the $\{\mathbf{x},\mathbf{v}\}$ space. Taking the time derivative of $N (\measvol; t, \omega)$ reveals that the evolution equation of the fine--grained Klimontovich density is a stochastic partial differential equation with a stochastic forcing of the standard NDF equation~\citep{klimontovich_2012} (cf. Eq.~\ref{eq:opdf}). In fluctuating hydrodynamics this stochastic term is directly modeled. An alternative approach that is amenable to simulation by modifying existing NDF solution approaches is proposed here.

Again a Lagrangian particle representation of the fluctuation $\widetilde{N}$ is possible in terms of physical  particles
\begin{equation}
\nonumber
\widetilde{N}(\measvol; t, \omega)
 \approx \int_{\measvol} \sum_{i=1}^{N} \delta \left(\mathbf{x} - \mathbf{X}^{(i)} \right) \delta \left(\mathbf{v} - \mathbf{V}^{(i)} \right) \, d\mathbf{x}, \, d\mathbf{v},
\end{equation}
or computational particles
\begin{equation}
\nonumber
\widetilde{N}(\measvol; t, \omega)
\approx \int_{\measvol} \sum_{i=1}^{N_c} w^{(i)} \delta \left(\mathbf{x} - \mathbf{X}^{(i)} \right) \delta \left(\mathbf{v} - \mathbf{V}^{(i)} \right) \, d\mathbf{x}, \, d\mathbf{v} ,
\end{equation}
where $w^{(i)}$ is its statistical weight in the ensemble of $N_c$ computational particles. This representation can be combined with Eulerian approaches to solving moments of the NDF equation (including QBMM), while noting that $\widetilde{N}(\measvol; t, \omega)$ does not have a smooth density. 
Smooth and differentiable Eulerian fields representing number fluctuations about the NDF can be defined by spatial filtering, analogous to $\check{\phi}(\mathbf{x},t;\omega)$ and $\hat{\phi}(\mathbf{x},t;\omega)$.

All these approaches are promising extensions to current statistical mechanical approaches to particle--laden suspensions, and because they are reconcilable at all scales (micro, meso and macro), it is possible to use PR--DNS to quantify the unclosed terms in each set of equations at the microscale. It will be interesting to revisit stability analyses of suspensions using these formulations to ascertain the effect of volume fraction fluctuations and see how they affect  stability limits based on the average volume fraction~\citep{koch_pof_1990}. Nevertheless, it is worth noting that none of these formulations truly represents {\em joint} statistics of particles and fluid, since that information is available only in two--point theories.
\section{Summary}
Multiphase flows are relevant to many important problems concerning human life.
Predictive models of multiphase flow can help in informed technical decision-making in both man--made and natural settings.
Hydrodynamic interactions in particle--laden suspensions constitute a complex phenomenon with rich physics.
Theoretical approaches are challenged to explain intriguing phenomena that manifest at different scales. Current advances in adapting classical statistical mechanics approaches have opened up a wide playground for theoreticians, modelers and computational researchers to pursue various new approaches that promise to address critical needs identified in this work. A key feature of the theoretical approaches proposed in this work is that they reconcile with the mathematical description at the microscale, and are easily related to observables at the meso and macroscales.
PR--DNS, which has proved to be a useful model-free simulation method for understanding flow physics at the microscale and model development, can be used to further develop these new formulations by quantifying unclosed terms that arise in them and by clarifying the role of fluctuations. This work envisions that a computational statistical mechanics approach to multiphase flow will explore new frontiers by establishing these theoretical foundations for
predictive model development, which requires integration of simulations, validated by experiment, at different scales.

\appendix
\section{Statistical characterization of  stochastic point processes}
\label{app:spp}
In the Klimontovich approach~\cite{plasma_theory,klimontovich_1986,shankar_ddf}, the ensemble
of particles is 
characterized by a fine--grained density
function $f'_1$ that is defined in a six-dimensional position-velocity space
$[\mathbf{x},\mathbf{v}]$ as
\begin{equation}
f'_1(\mathbf{x},\mathbf{v},t) \equiv
\sum_{i=1}^{N}f_1'^{(i)}=\sum_{i=1}^{N}\delta(\mathbf{x}-\mathbf{X}^{(i)}(t))\delta(\mathbf{v}-\mathbf{V}^{(i)}(t)) ,
\label{eqn:f}
\end{equation}
where the shortened notation
\[
f_1'^{(i)}=\delta(\mathbf{x}-\mathbf{X}^{(i)}(t))\delta(\mathbf{v}-\mathbf{V}^{(i)}(t))
\] 
is used to represent the delta function associated with the $i$th
particle.  The number of particles in any region $\mathscr{B}$ in
$[\mathbf{x},\mathbf{v}]$ space can be obtained by integrating the
fine--grained density $f'_1$ as follows:
\begin{equation}
N (\mathscr{B}) = \int_{\mathscr{B}} f'_1 d\mathbf{x}\, d\mathbf{v} .
\end{equation}
The ensemble average of the Klimontovich fine--grained density
function $f'_1$ is the one--particle density function $f$, which is
written as
\begin{equation}
f(\mathbf{x}, \mathbf{v},t) = \langle f'_1 \rangle = \left\langle
\sum_{i=1}^{N}f'^{(i)}_1
\right\rangle = \left\langle \sum_{i=1}^{N}
\delta(\mathbf{x}-\mathbf{X}^{(i)}(t))\delta(\mathbf{v}-\mathbf{V}^{(i)}(t))
\right\rangle.
\label{eqn:oneparticledensity}
\end{equation}
Integrating the one--particle density over velocity space results
in the number density $n(\mathbf{x},t)$ that forms the basis for the
continuum hydrodynamic description
\begin{equation}
n(\mathbf{x},t) = \int f(\mathbf{x}, \mathbf{v},t)\, d\mathbf{v} ,
\end{equation}
which in turn can be integrated over physical space to obtain the
expected number of particles:
\begin{equation}
\langle N\rangle = \int n(\mathbf{x}, t)\, d\mathbf{x} .
\end{equation}

In order to characterize structural properties such as the pair
correlation function, we need to consider the two--particle density.
The one--point fine--grained density in the Klimontovich approach can
be extended to its two--particle counterpart as follows~\cite{krall_trivelpiece}:
\begin{align}
  f'_1f'_2 =
  \sum_{i=1}^{N}f'^{(i)}_1 \sum_{\substack{j=1\\j\neq i}}^{N}f'^{(j)}_2=
\sum_{i=1}^{N}\delta(\mathbf{x}_1-\mathbf{X}^{(i)}(t))\delta(\mathbf{v}_1-\mathbf{V}^{(i)}(t))\times 
\sum_{\substack{j=1\\j \neq i}}^{N}\delta(\mathbf{x}_2-\mathbf{X}^{(j)}(t))\delta(\mathbf{v}_2-\mathbf{V}^{(j)}(t)) 
\end{align}
where
$[\mathbf{x}_k,\mathbf{v}_k, k=1, 2]$ are the Eulerian coordinates of the
position--velocity phase space for the particle pair. (The summation
over distinct pairs $j\neq i$ is necessary for the definition 
of the two--particle density, whose integral is the second factorial
measure. If all pairs are included, an atomic contribution arises in
the second moment measure that does not have a
density~\cite{stoyan_stoyan,stoyan_kendall_mecke}.)
The ensemble average of the two--particle fine--grained density
function $f'_1f'_2$ is the two--particle density 
$\rho^{(2)}(\mathbf{x}_1,\mathbf{x}_2,\mathbf{v}_1,\mathbf{v}_2,t)$,
which is defined as
\begin{equation}
\rho^{(2)}(\mathbf{x}_1,\mathbf{x}_2,\mathbf{v}_1,\mathbf{v}_2,t) \equiv
\langle f'_1f'_2\rangle. 
\end{equation}
Integrating the two--particle density over the velocity spaces results
in the unnormalized pair--correlation function
\begin{equation}
\rho^{(2)}(\mathbf{x}_1,\mathbf{x}_2,t) = \int
\rho^{(2)}(\mathbf{x}_1,\mathbf{x}_2,\mathbf{v}_1,\mathbf{v}_2,t) \,
d\mathbf{v}_1 d\mathbf{v}_2 ,
\end{equation}
which in turn can be integrated over a region $\mathscr{B}$ in
physical space to obtain the second factorial moment measure:
\begin{equation}
  \left\langle 
    N(\mathscr{B}) \, \left[ N(\mathscr{B})-1\right] \right\rangle = \int 
  \rho^{(2)}(\mathbf{x}_1,\mathbf{x}_2,t) 
  d\mathbf{x}_1 d\mathbf{x}_2. 
\end{equation}

Here we derive the expression for the pair--correlation function
$g(r)$ in a system
that contains particles whose centers are
distributed as statistically homogeneous and isotropic point
fields. The {\em second factorial moment measure} of a point field
(see~\citet{stoyan_stoyan}) is generalized to a
binary system with two particle types as
\begin{equation}
  \mu^{(2)}(\regA \times \regB) = \left\langle N(\regA) \,
    \left[ N(\regB)-1\right]  \right\rangle, 
\label{Eq1}
\end{equation}
where \regA\ and \regB\ are sets in
physical space, $N(\regA)$ is the number of 
particles in region $\regA$, and $N(\regB)$ is
the number of particles in region $\regB$.
The second factorial moment measure $\mu^{(2)}( \regA \times
\regB )$ has a density
$\rho^{(2)}(\mathbf{x}_1,\mathbf{x}_2)$ such that it
can be written as an integral
\begin{equation}
  \mu^{(2)}(\regA \times \regB) = \int_{\regA}\int_{\regB}\rho^{(2)}(\mathbf{x}_1,\mathbf{x}_2) 
  d\mathbf{x}_1 d\mathbf{x}_2.
\label{Eq2}
\end{equation}
This \emph{second-order product density} $\rho^{(2)}(\mathbf{x}_1,\mathbf{x}_2)$ is the unnormalized pair
correlation function.  

For a statistically homogeneous point field the second-order product
density $\rho^{(2)}(\mathbf{x}_1,\mathbf{x}_2)$ depends
only on the pair separation $\mathbf{r} = \mathbf{x}_2-\mathbf{x}_1$.
It is then convenient to transform $\regA \times \regB$ to $\regR
\times \regr$ in $(\mathbf{R}, \mathbf{r})$ space with $\mathbf{R} =
(\mathbf{x}_1+\mathbf{x}_2)/2$ and
$\rho_{\alpha\beta}^{(2)}(\mathbf{R},\mathbf{r})J=
\rho_{\alpha\beta}^{(2)}(\mathbf{x}_1,\mathbf{x}_2)$, where the
Jacobian of the transformation $J=\left| \partial (\mathbf{x}_1,
  \mathbf{x}_2)/\partial (\mathbf{R},\mathbf{r})\right|$ is unity,
leading to
\begin{equation}
  \mu^{(2)}(\regA \times \regB) = 
  \mu^{(2)}(\regR \times \regr) = 
\int_{\regR}\int_{\regr}\rho^{(2)}(\mathbf{R},\mathbf{r}) 
  d\mathbf{R} d\mathbf{r}.
\label{eqn:rho2Rrspace}
\end{equation}

For homogeneous and isotropic point fields, the
second-order product density $\rho^{(2)}$ depends only on the scalar
separation distance $r = | \mathbf{r}|$, and can be written as
\begin{equation}
\label{eq:rho2togr}
\rho^{(2)}(r) = n_\alpha n_\beta g(r),
\end{equation}
where $n$ is the average number density of the
 particles. Substituting
this expression into \ref{eqn:rho2Rrspace}, we obtain
\begin{equation}
 \mu^{(2)}(\regR \times \regr) =
 \int_{\regR}\int_{r}
n^2 g(r) 4\pi r^2 
  d\mathbf{R} \, dr,
\label{eqn:rho2iso}
\end{equation}
where the integral over \regr\ has been simplified using a spherical
volume element $4 \pi r^2 dr$.  Noting that
\begin{equation}
\left\langle N(\regR)\right\rangle = \int_{\regR} n d\mathbf{R},
\end{equation}
and considering the case where \regr\ is a spherical shell with volume
$V(r,\Delta r) = 4 \pi r^2\Delta r$ we obtain
\begin{equation}
 \mu^{(2)}(\regR \times \regr) =
\left\langle N(\regR)\right\rangle  n g(r) 4\pi r^2 
  \Delta r ,
\label{eqn:rho2iso2}
\end{equation}
provided $\Delta r$ is smaller than the scale of
variation of $g(r)$.

\section{Details of ensemble--averaged mass conservation equation derivation}
\label{app:ensav}
The ensemble--averaged equations have been derived by several
researchers (cite Drew, Pai etc). Defining the average solid volume
fraction as
\begin{equation}
\langle \phi \rangle  \equiv \langle I_p(\mathbf{x},t)\rangle ,
\label{eq:avphi}
\end{equation}
and starting from the mass conservation equation 
\begin{equation}
\frac{\partial \rho}{\partial t} + \boldsymbol \nabla \cdot ( \rho \mathbf{u} ) = 0
\end{equation}
which is trivially satisfied inside constant density, rigid particles,
and multiplying it by the indicator function and taking the ensemble
average yields
\begin{equation}
  \frac{\partial }{\partial t} \langle I_p \rho \rangle + \boldsymbol \nabla \cdot ( \langle I_p \rho \rangle \langle \tilde{\mathbf{u}}^{(p)} \rangle ) = 0 .
\end{equation}
For constant density particles $\langle I_p \rho \rangle = \rho_p \langle I_p  \rangle = \rho_p \langle \phi \rangle$, and 
\[
\langle \tilde{\mathbf{u}}^{(p)} \rangle = \langle  I_p \rho \mathbf{u} \rangle / \langle I_p \rho \rangle ,
\]
which for constant $\rho_p$ simplifies to 
\[
\langle \tilde{\mathbf{u}}^{(p)} \rangle = \langle \mathbf{u}^{(p)} \rangle = \langle  I_p \mathbf{u} \rangle / \langle I_p \rangle ,
\]
which is the phasic average velocity.
Therefore, the average mass conservation equation in the ensemble--averaged approach is:
\begin{equation}
\frac{\partial \langle \phi \rangle}{\partial t}  + \boldsymbol \nabla \cdot ( \langle \phi \rangle \langle \mathbf{u}^{(p)} \rangle ) = 0 .
\end{equation}

\section{Volume--averaged mass conservation equation}
\label{app:volav}
We can write a conservation equation for $\Phi_p (\measvol; t,
\omega)$ by considering its temporal evolution
\begin{equation}
\frac{\partial \Phi_p (\measvol; t, \omega)}{\partial t} = \int_{\measvol} \frac{\partial I_p (\mathbf{x},t; \omega)}{\partial t} \, d\mathbf{x} ,
\end{equation}
where we have used the fact that $\measvol$ does not depend on
$t$. Using Drew's topological equation and simplifying we obtain:
\begin{equation}
\frac{\partial \Phi_p}{\partial t} =  \frac{\partial}{\partial t} \int_{\measvol} \langle \phi \rangle \, d\mathbf{x} + \frac{\partial \tilde{\Phi}_p}{\partial t} 
 =   - \int_{\partial \measvol} \mathbf{J}\boldsymbol \cdot \mathbf{n} \, dA 
 =   - \int_{\partial \measvol} \langle \phi \rangle  \langle \mathbf{u}_p\rangle \boldsymbol \cdot \mathbf{n} \, dA - \int_{\partial \measvol} \mathbf{J}^{\prime} \boldsymbol \cdot \mathbf{n} \, dA 
\end{equation}
Or rewriting, we obtain the volume--averaged mass conservation equation to be
\begin{equation}
\overbrace{\frac{\partial }{\partial t} \int_{\measvol} \langle \phi \rangle (\mathbf{x},t)\, d\mathbf{x}}^{(1)}+ 
\overbrace{\frac{\partial \tilde{\Phi}_p}{\partial t}(\measvol; t, \omega) }^{(2)}
 =   \underbrace{- \int_{\partial \measvol} \langle \phi \rangle  \langle \mathbf{u}_p\rangle \boldsymbol \cdot \mathbf{n} \, dA}_{(3)} -
 \underbrace{\int_{\partial \measvol} \mathbf{J}^{\prime} \boldsymbol \cdot \mathbf{n} \, dA }_{(4)}.
\end{equation}
Note that the above equation is a stochastic equation with explicit dependence on the realization $\omega$.
Clearly, terms (1) and (3) in the above equation represent the ensemble--averaged mass conservation (cf. Eq.~\ref{eq:ensavmasscons}) for particles with constant material density, which is a deterministic equation. Terms (2) and (4) on the other hand, represent the departure of the stochastic volume--averaged mass conservation equation from its deterministic ensemble--averaged counterpart. 

\section{Mass conservation in the spatially filtered approach}
\label{app:massconsspatfilt}

Taking the time derivative of $\bar{\phi}$ defined by Eq.~\ref{eq:phibar} results in 
\begin{equation}
\frac{\partial \bar{\phi} }{\partial t} = \int_{V_{\infty}} \frac{\partial I_p (\mathbf{y},t)}{\partial t} \mathscr{G}(\mathbf{x} - \mathbf{y}) dV_y  .
\end{equation}
Substituting
\[
\frac{\partial I_p (\mathbf{y},t)}{\partial t}  = - \nabla_{\mathbf{y}} \boldsymbol \cdot (\mathbf{u} I_p) 
\]
in the above equation results in 

\[
\frac{\partial \bar{\phi} }{\partial t} = - \int_{V_{\infty}} \nabla_{\mathbf{y}} \boldsymbol \cdot 
\left[ \mathbf{u} I_p) \mathscr{G}(\mathbf{x} - \mathbf{y} \right] dV_y + \int_{V_{\infty}} \mathbf{u} I_p
\nabla_{\mathbf{y}} \mathscr{G}(\mathbf{x} - \mathbf{y}) dV_y .
\]

Using Gauss' divergence theorem, we can rewrite the above equation as
\[
\frac{\partial \bar{\phi} }{\partial t} = - \int_{\partial V_{\infty}} \mathbf{n} \boldsymbol \cdot 
\mathbf{u} I_p(\mathbf{y},t;\omega) \mathscr{G}(\mathbf{x} - \mathbf{y} dA - \int_{V_{\infty}} \mathbf{u} I_p
\nabla_{\mathbf{x}} \mathscr{G}(\mathbf{x} - \mathbf{y}) dV_y ,
\]
because
\[
\nabla_{\mathbf{y}} \mathscr{G}(\mathbf{x} - \mathbf{y}) = - \nabla_{\mathbf{x}} \mathscr{G}(\mathbf{x} - \mathbf{y}) .
\]
Now defining
\begin{equation}
\bar{\mathbf{v}} \equiv \frac{1}{\bar{\phi}} \int_{V_{\infty}} I_p (\mathbf{y},t) \mathbf{u} (\mathbf{y},t) \mathscr{G}(\mathbf{x} - \mathbf{y}) dV_y ,
\end{equation}
noting that
\[
\nabla_{\mathbf{x}} \boldsymbol \cdot ( \bar{\phi} \bar{\mathbf{v}} ) = \int_{V_{\infty}} \mathbf{u} I_p
\nabla_{\mathbf{x}} \mathscr{G}(\mathbf{x} - \mathbf{y}) dV_y ,
\]
we can rewrite the last term 
in the preceding form of the mass conservation equation as
\[
\frac{\partial \bar{\phi} }{\partial t} + \nabla_{\mathbf{x}} \boldsymbol \cdot ( \bar{\phi} \bar{\mathbf{v}} ) = - \int_{V_{\infty}} \nabla_{\mathbf{y}} \boldsymbol \cdot 
\left[ \mathbf{u} I_p \mathscr{G}(\mathbf{x} - \mathbf{y}) \right] dV_y = - \int_{\partial V_{\infty}} \mathbf{n} \boldsymbol\cdot
\left[ \mathbf{u} I_p \mathscr{G}(\mathbf{x} - \mathbf{y}) \right] dA
\]

\section{Representation of fluctuations in the volume averaging approach}
\label{app:repflucvolavg}
The fluctuation $\tilde{\Phi}_p (\measvol; t, \omega)$ which arises in
the volume averaging approach can be related to the two--point density
that is used to characterize fluctuations in the ensemble-averaging
approach. This important relationship is established in the following.
In order to consider fluctuations we consider the quantity  $\Phi^2_p
(\mathscr{A} \times \mathscr{B}; t, \omega)$ which we expand using Eq.~\ref{eq:Phip_and_ensavg} to obtain: 
\begin{equation}
\Phi^2_p
(\mathscr{A} \times \mathscr{B}; t, \omega) = 
\left[ \int_{\mathscr{A}} \langle \phi \rangle \, d\mathbf{x}_1 + \tilde{\Phi}_p(\mathscr{A}; t, \omega) \right]
\left[ \int_{\mathscr{B}} \langle \phi \rangle \, d\mathbf{x}_2 + \tilde{\Phi}_p(\mathscr{B}; t, \omega) \right].
\end{equation}
The expected value of $\Phi^2_p
(\mathscr{A} \times \mathscr{B}; t, \omega)$ is the second moment of particle volume which simplifies to:
\begin{equation}
\langle \Phi^2_p
(\mathscr{A} \times \mathscr{B}; t)\rangle = 
\left[ \int_{\mathscr{A}} \langle \phi \rangle \, d\mathbf{x}_1 
\int_{\mathscr{B}} \langle \phi \rangle \, d\mathbf{x}_2 + \langle \tilde{\Phi}_p(\mathscr{A}; t)  \tilde{\Phi}_p(\mathscr{B}; t) \rangle \right].
\end{equation}
The statistics of this
quantity have a density only in the {\it product space}. This connects it
to the two-point density, which is in turn related to $g(r)$ (see Eq.~\ref{eq:rho2togr} in Appendix~\ref{app:spp}).

We consider volume fluctuations in statistically homogeneous gas-solid
flows.  Gas-solid flows are characterized by an intrinsic statistical
variability in quantities such as the number of particles, or the
volume occupied by the particles in any given region. Moreover,
formation of clusters can lead to a lack of separation of length
scales. Therefore, the inherent statistical variability present in the
volume occupied the particles in any region or "measurement region"
needs to be characterized in a statistically homogeneous suspension.

\begin{figure}
\begin{centering}
\includegraphics[scale=0.6]{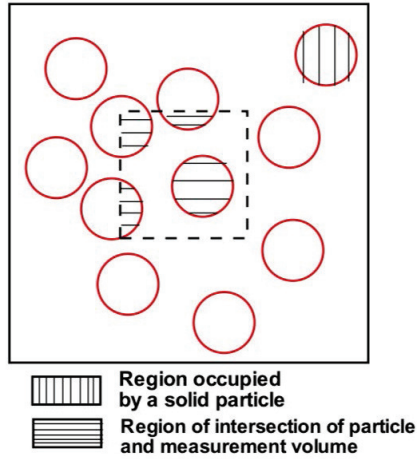}
\caption{Schematic showing the intersection of solid particles with the measurement region. The region of space occupied by the solids is hatched with vertical lines. The region of intersection of the solid particles with the measurement region is hatched with horizontal lines.}
\label{fig:schematic}
\end{centering}
\end{figure}

In this context we introduce the concept of measurement or observation region. A
measurement region is a region of arbitrary shape and fixed size in
the gas-solid flow domain. It can be thought of as an observation
window or frame in an experiment. We define statistical measures to
characterize the level of local volume fluctuations in a measurement
region. A schematic with the measurement region in the flow domain is
shown in Fig.~\ref{fig:schematic}. The solid phase is represented by
the indicator function $I_p (\mathbf{x})$ which is unity if the point
$\mathbf{x}$ lies in the solid phase and zero otherwise. The solid
phase volume fraction fluctuations in any measurement volume
$\measvol$ depend on the microstructure or the particle
configuration. It is useful to define the one--point and two--point
probability functions $S_1 (\mathbf{x}_1)$ and $C (\mathbf{x}_1,
\mathbf{x}_2)$ as follows:
\begin{eqnarray}
  S_1 (\mathbf{x}_1) & =&  \langle I_p (\mathbf{x}_1)\rangle \\
  C (\mathbf{x}_1, \mathbf{x}_2) & = & \langle I_p (\mathbf{x}_1) I_p (\mathbf{x}_2)
\end{eqnarray}
Here $S_1$ is the probability of finding a point in the solid
phase. For statistically homogeneous gas--solid suspnesions this
quantity is equal to the volume fraction of the solid phase. The
quantity $C (\mathbf{x}_1, \mathbf{x}_2)$ is the probability of
finding two points $\mathbf{x}_1$ and $\mathbf{x}_2$ simultaneously in
the solid phase (~\citet{torquato1982microstructure}; ~\citet{sund_coll_ijmf1}). For statistically homogeneous and isotropic gas--solid
suspensions, $C$ is only a function of the magnitude of the
separation between the points i.e., $C (\mathbf{x}_1, \mathbf{x}_2)
= C (|\mathbf{x}_1 - \mathbf{x}_2|)$. Two limits of this function
are of interest. When the separation between the points is very small,
$C \rightarrow \langle\phi \rangle$ and when the separation becomes very large $C
\rightarrow \langle \phi \rangle^2$.  The
inherent statistical variability in the volume fraction of the solid
phase can be characterized in terms of the variance of the local
volume fraction in a given measurement volume. In order to perform
this calculation, the local volume fraction (~\citet{lu1990local,quintanilla1997local}) in a measurement volume centered at location $\mathbf{x}$ is
defined as 
\begin{equation}
\varepsilon(\mathbf{x}) = \frac{1}{V_m} \int_{\measvol^{\mathbf{x}}}
I_p(\mathbf{z},t ; \omega )\, d\mathbf{z}.
\end{equation}
 When the measurement volume is large, i.e. $V_m \rightarrow \infty$ the local
volume fraction tends to the volume fraction of the solid phase. At
the other extreme when the measurement volume becomes very small, the
local volume fraction becomes the indicator function of the solid
phase at the point $\mathbf{x}$.  

It is useful to express the local particle volume in terms of volume
integrals over the entire domain. The physical domain is deterministic
and so the expectation operator and the integral operator commute. In
order to define the local volume fraction as an integral over the
whole domain, we define an indicator function for the measurement
region as follows:
\begin{equation}
I_{\measvol^{\mathbf{x}}} (\mathbf{y}) = \left\{ \begin{array}{cc} 1 & \mbox{if}\,\,\,\, y \in
    \measvol^{\mathbf{x}} \\
0 & \mbox{otherwise}.
\end{array}\right.
\end{equation}
From the definition it is clear that $I_{\measvol^\mathbf{x}}
(\mathbf{y}) = I_{\measvol^0} (\mathbf{y}-\mathbf{x}) $. Therefore,
the indicator function of a measurement region centered at any point
$\mathbf{x}$ is written in terms of the indicator function of the
measurement region centered at origin. With this definition, the local
particle volume in a measurement region centered at a point $\mathbf{x}$ can be
written as 
\begin{equation}
\varepsilon(\mathbf{x}) = \frac{1}{V_m} \int_{\measvol^{\mathbf{x}}}
I_p(\mathbf{y},t ; \omega )\,  I_{\measvol^0} (\mathbf{y}-\mathbf{x})
\, d\mathbf{y}.
\end{equation}
Note that in the above equation the integral is over all possible values of $\mathbf{y}$ and not over
$\measvol$.

It is clear that the expected value of the local particle volume is
nothing but the average particle volume of the solid phase. A measure of the
fluctuations in particle volume fluctuations can be obtained by examining the
ratio of the standard deviation of the local particle volume  to the
average solid phase volume. This ratio $k_{\phi} =
\sigma_{\varepsilon}/\langle \phi\rangle$ gives a measure of
the nonuniformity of the local particle volume. Here, the standard
deviation of the local particle volume is defined as
$\sigma_{\varepsilon}^2 = \langle \varepsilon^2\rangle - \langle \phi \rangle^2$.
It is instructive to examine the limits of the intensity of particle volume
fluctuations $k_{\phi}$. For large measurement volumes, $k_{\phi} \rightarrow 0$. For
very small measurement volumes $\sigma_{\varepsilon}^2 \rightarrow \langle \phi \rangle - \langle \phi \rangle^2$ and hence $\displaystyle k_{\phi} \rightarrow \sqrt{\frac{1 - \phi}{\phi}}$.

We now establish the relationship between the standard deviation of the particle volume fraction and the two--point correlation function $C$. The expected value of the square of the local
volume fraction fluctuations (for a homogeneous random field) is given by: 
\begin{eqnarray}
\nonumber
\langle \varepsilon^2\rangle & = & \frac{1}{V_m^2} \left\langle 
\int
I_p(\mathbf{y},t ; \omega )\,  I_{\measvol^0} (\mathbf{y}-\mathbf{x})
\, d\mathbf{y} \int
I_p(\mathbf{z},t ; \omega )\,  I_{\measvol^0} (\mathbf{z}-\mathbf{x})
\, d\mathbf{z}\right\rangle \\
\nonumber
& = & 
\frac{1}{V_m^2} 
\int \int
\langle 
I_p(\mathbf{y},t ; \omega )\,  I_p(\mathbf{z},t ; \omega )\rangle \,  
I_{\measvol^0} (\mathbf{y}-\mathbf{x}) I_{\measvol^0} (\mathbf{z}-\mathbf{x})
\, d\mathbf{y}
\, d\mathbf{z}\\
& = & 
\frac{1}{V_m^2} 
\int \int
C (\mathbf{y},\mathbf{z}) \,  
I_{\measvol^0} (\mathbf{y}-\mathbf{x}) I_{\measvol^0} (\mathbf{z}-\mathbf{x})
\, d\mathbf{y}
\, d\mathbf{z}
\label{eq:eps2}
\end{eqnarray}

Since the gas-solid flow
is homogeneous, $C (\mathbf{y}, \mathbf{z}) = C (\mathbf{y} - \mathbf{z})$. Now invoking a transformation $\mathbf{r} = \mathbf{y} - \mathbf{z}$, we see that 
\begin{eqnarray}
\nonumber
\langle \varepsilon^2\rangle & = & \frac{1}{V_m^2} 
\int  C (\mathbf{r}) \int
\,  I_{\measvol^0} (\mathbf{y}-\mathbf{x}) I_{\measvol^0} (\mathbf{y}-(\mathbf{x}+\mathbf{r}))
\, d\mathbf{y}
\, d\mathbf{r}\\
& = & \frac{1}{V_m^2} 
\int C (\mathbf{r}) V_2^{\mbox{\small int}}(\mathbf{r})
\, d\mathbf{r},
\label{eq:eps2_final}
\end{eqnarray}
where all integrals are over the entire domain and the dependence on $\mathbf{x}$ is dropped on account of statistical homogeneity.

In the above
expression, $V_2^{\mbox{\small int}}(\mathbf{r})$ is the volume of intersection of two measurement
regions separated by $\mathbf{r}$. This is because $I_{\measvol^0} (\mathbf{y}-\mathbf{x})$, is the
indicator function of a measurement volume
centered at $\mathbf{x}$. Similarly, $I_{\measvol^0} (\mathbf{y}-(\mathbf{x}+\mathbf{r}))$ is the indicator function of a measurement region
centered at the point $\mathbf{x} + \mathbf{r}$. Hence the volume integral 
\begin{equation}
V_2^{\mbox{\small int}}(\mathbf{r}) = \int_{\measvol}
\,  I_{\measvol^0} (\mathbf{y}-\mathbf{x}) I_{\measvol^0} (\mathbf{y}-(\mathbf{x}+\mathbf{r}))C
\, d\mathbf{y}
\end{equation}
gives the volume of intersection of these two measurement regions, which
depends on the shape and size of the measurement regions but is independent of $\mathbf{x}$. From the definition of the intersection volume, it is clear that $\displaystyle \int_{\mathscr{V}_r} V_2^{\mbox{\small int}}(\mathbf{r}) \, d\mathbf{r} = V_m^2$. Therefore, the variance of the
fluctuations can now be written as
\begin{eqnarray}
\nonumber
\sigma_{\varepsilon}^2 & = & \frac{1}{V_m^2} 
\int C (\mathbf{r}) V_2^{\mbox{\small int}}(\mathbf{r})
\, d\mathbf{r} - \langle \phi \rangle^2 \\
\nonumber
& = & 
\frac{1}{V_m^2} 
\int  C (\mathbf{r}) V_2^{\mbox{\small int}}(\mathbf{r})
\, d\mathbf{r} - \langle \phi \rangle^2 
\frac{1}{V_m^2} 
\int V_2^{\mbox{\small int}}(\mathbf{r})
\, d\mathbf{r}
\\
& = & 
\frac{1}{V_m^2} 
\int  \left[ C (\mathbf{r}) - \langle \phi \rangle^2 \right] V_2^{\mbox{\small int}}(\mathbf{r})
\, d\mathbf{r} = \frac{1}{V_m^2} 
\int  k (\mathbf{r}) V_2^{\mbox{\small int}}(\mathbf{r})
\, d\mathbf{r}
\end{eqnarray}

Therefore, the final expression for the intensity of volume fraction fluctuations becomes
\begin{equation}
k_{\phi} = \frac{1}{\langle \phi \rangle V_m}\left[ \int  k(\mathbf{r})  V_2^{\mbox{\small int}}(\mathbf{r}) 
\, d\mathbf{r} \right]^{1/2}.
\end{equation}

In the volume averaged approach, although the fluctuation $\tilde{\Phi}_p (\measvol; t, \omega)$ also does not have a smooth density in $\mathbf{x}$, it could be {\it modeled} in a manner similar to GPCE~\citep{xiu2003modeling} exploiting Karhunen--Loeve expansions~\citep{soize2009reduced}. A model for this particle volume fluctuation must imply a model for the evolution of the two-point density. This can give insight into how its spatially filtered counterpart can be modeled, which in turn implies a model for the spatially filtered two--point density.
\begin{acknowledgments}
I would like to acknowledge S. "Bala" Balachandar for first prompting my interest in the spatially filtered approach as an interesting alternative to the statistical approach. I would like to thank Sudheer Tenneti for his contribution to Appendix~\ref{app:repflucvolavg} from his thesis work, and Eric Murphy for useful comments on an early draft of this manuscript. 

This material is based upon work supported by the National Science Foundation under Grant No. 1905017 (ENG--CBET--Thermal Transport Processes) and the U.S. Department of Energy, Office of Energy Efficiency and Renewable Energy (EERE), Advanced Manufacturing Office, Award Number DE-EE0008326. SS also wishes to acknowledge financial support from SERDP WP19-1163 under the direction of Dr. Robin Nissan through the NSWC-Crane Naval Innovative Science and Engineering Program.
\end{acknowledgments}

\bibliography{PRF_invited}

\begin{thebibliography}{149}%
\makeatletter
\providecommand \@ifxundefined [1]{%
 \@ifx{#1\undefined}
}%
\providecommand \@ifnum [1]{%
 \ifnum #1\expandafter \@firstoftwo
 \else \expandafter \@secondoftwo
 \fi
}%
\providecommand \@ifx [1]{%
 \ifx #1\expandafter \@firstoftwo
 \else \expandafter \@secondoftwo
 \fi
}%
\providecommand \natexlab [1]{#1}%
\providecommand \enquote  [1]{``#1''}%
\providecommand \bibnamefont  [1]{#1}%
\providecommand \bibfnamefont [1]{#1}%
\providecommand \citenamefont [1]{#1}%
\providecommand \href@noop [0]{\@secondoftwo}%
\providecommand \href [0]{\begingroup \@sanitize@url \@href}%
\providecommand \@href[1]{\@@startlink{#1}\@@href}%
\providecommand \@@href[1]{\endgroup#1\@@endlink}%
\providecommand \@sanitize@url [0]{\catcode `\\12\catcode `\$12\catcode
  `\&12\catcode `\#12\catcode `\^12\catcode `\_12\catcode `\%12\relax}%
\providecommand \@@startlink[1]{}%
\providecommand \@@endlink[0]{}%
\providecommand \url  [0]{\begingroup\@sanitize@url \@url }%
\providecommand \@url [1]{\endgroup\@href {#1}{\urlprefix }}%
\providecommand \urlprefix  [0]{URL }%
\providecommand \Eprint [0]{\href }%
\providecommand \doibase [0]{https://doi.org/}%
\providecommand \selectlanguage [0]{\@gobble}%
\providecommand \bibinfo  [0]{\@secondoftwo}%
\providecommand \bibfield  [0]{\@secondoftwo}%
\providecommand \translation [1]{[#1]}%
\providecommand \BibitemOpen [0]{}%
\providecommand \bibitemStop [0]{}%
\providecommand \bibitemNoStop [0]{.\EOS\space}%
\providecommand \EOS [0]{\spacefactor3000\relax}%
\providecommand \BibitemShut  [1]{\csname bibitem#1\endcsname}%
\let\auto@bib@innerbib\@empty
\bibitem [{\citenamefont {Chu}\ \emph {et~al.}(2009{\natexlab{a}})\citenamefont
  {Chu}, \citenamefont {Wang}, \citenamefont {Yu},\ and\ \citenamefont
  {Vince}}]{chu2009cfd}%
  \BibitemOpen
  \bibfield  {author} {\bibinfo {author} {\bibfnamefont {K.}~\bibnamefont
  {Chu}}, \bibinfo {author} {\bibfnamefont {B.}~\bibnamefont {Wang}}, \bibinfo
  {author} {\bibfnamefont {A.}~\bibnamefont {Yu}},\ and\ \bibinfo {author}
  {\bibfnamefont {A.}~\bibnamefont {Vince}},\ }\bibfield  {title} {\bibinfo
  {title} {{CFD-DEM modelling of multiphase flow in dense medium cyclones}},\
  }\href@noop {} {\bibfield  {journal} {\bibinfo  {journal} {Powder
  Technology}\ }\textbf {\bibinfo {volume} {193}},\ \bibinfo {pages} {235}
  (\bibinfo {year} {2009}{\natexlab{a}})}\BibitemShut {NoStop}%
\bibitem [{\citenamefont {Chu}\ \emph {et~al.}(2009{\natexlab{b}})\citenamefont
  {Chu}, \citenamefont {Wang}, \citenamefont {Yu}, \citenamefont {Vince},
  \citenamefont {Barnett},\ and\ \citenamefont {Barnett}}]{chu2009cfd-dem}%
  \BibitemOpen
  \bibfield  {author} {\bibinfo {author} {\bibfnamefont {K.}~\bibnamefont
  {Chu}}, \bibinfo {author} {\bibfnamefont {B.}~\bibnamefont {Wang}}, \bibinfo
  {author} {\bibfnamefont {A.}~\bibnamefont {Yu}}, \bibinfo {author}
  {\bibfnamefont {A.}~\bibnamefont {Vince}}, \bibinfo {author} {\bibfnamefont
  {G.}~\bibnamefont {Barnett}},\ and\ \bibinfo {author} {\bibfnamefont
  {P.}~\bibnamefont {Barnett}},\ }\bibfield  {title} {\bibinfo {title}
  {{CFD-DEM study of the effect of particle density distribution on the
  multiphase flow and performance of dense medium cyclone}},\ }\href
  {https://doi.org/https://doi.org/10.1016/j.mineng.2009.04.008} {\bibfield
  {journal} {\bibinfo  {journal} {Minerals Engineering}\ }\textbf {\bibinfo
  {volume} {22}},\ \bibinfo {pages} {893 } (\bibinfo {year}
  {2009}{\natexlab{b}})},\ \bibinfo {note} {special issue: Computational
  Modelling}\BibitemShut {NoStop}%
\bibitem [{\citenamefont {McGovern}\ \emph {et~al.}(2008)\citenamefont
  {McGovern}, \citenamefont {Harish}, \citenamefont {Pai}, \citenamefont
  {Mansfield}, \citenamefont {Taylor}, \citenamefont {Pau},\ and\ \citenamefont
  {Besser}}]{mcgovern2008multiphase}%
  \BibitemOpen
  \bibfield  {author} {\bibinfo {author} {\bibfnamefont {S.}~\bibnamefont
  {McGovern}}, \bibinfo {author} {\bibfnamefont {G.}~\bibnamefont {Harish}},
  \bibinfo {author} {\bibfnamefont {C.}~\bibnamefont {Pai}}, \bibinfo {author}
  {\bibfnamefont {W.}~\bibnamefont {Mansfield}}, \bibinfo {author}
  {\bibfnamefont {J.}~\bibnamefont {Taylor}}, \bibinfo {author} {\bibfnamefont
  {S.}~\bibnamefont {Pau}},\ and\ \bibinfo {author} {\bibfnamefont
  {R.}~\bibnamefont {Besser}},\ }\bibfield  {title} {\bibinfo {title}
  {Multiphase flow regimes for hydrogenation in a catalyst-trap microreactor},\
  }\href@noop {} {\bibfield  {journal} {\bibinfo  {journal} {Chemical
  Engineering Journal}\ }\textbf {\bibinfo {volume} {135}},\ \bibinfo {pages}
  {S229} (\bibinfo {year} {2008})}\BibitemShut {NoStop}%
\bibitem [{\citenamefont {Latham}(1990)}]{latham1990control}%
  \BibitemOpen
  \bibfield  {author} {\bibinfo {author} {\bibfnamefont {J.}~\bibnamefont
  {Latham}},\ }\bibfield  {title} {\bibinfo {title} {Control of global
  warming?},\ }\href@noop {} {\bibfield  {journal} {\bibinfo  {journal}
  {Nature}\ }\textbf {\bibinfo {volume} {347}},\ \bibinfo {pages} {339}
  (\bibinfo {year} {1990})}\BibitemShut {NoStop}%
\bibitem [{\citenamefont {Latham}\ \emph {et~al.}(2012)\citenamefont {Latham},
  \citenamefont {Bower}, \citenamefont {Choularton}, \citenamefont {Coe},
  \citenamefont {Connolly}, \citenamefont {Cooper}, \citenamefont {Craft},
  \citenamefont {Foster}, \citenamefont {Gadian}, \citenamefont {Galbraith}
  \emph {et~al.}}]{latham2012marine}%
  \BibitemOpen
  \bibfield  {author} {\bibinfo {author} {\bibfnamefont {J.}~\bibnamefont
  {Latham}}, \bibinfo {author} {\bibfnamefont {K.}~\bibnamefont {Bower}},
  \bibinfo {author} {\bibfnamefont {T.}~\bibnamefont {Choularton}}, \bibinfo
  {author} {\bibfnamefont {H.}~\bibnamefont {Coe}}, \bibinfo {author}
  {\bibfnamefont {P.}~\bibnamefont {Connolly}}, \bibinfo {author}
  {\bibfnamefont {G.}~\bibnamefont {Cooper}}, \bibinfo {author} {\bibfnamefont
  {T.}~\bibnamefont {Craft}}, \bibinfo {author} {\bibfnamefont
  {J.}~\bibnamefont {Foster}}, \bibinfo {author} {\bibfnamefont
  {A.}~\bibnamefont {Gadian}}, \bibinfo {author} {\bibfnamefont
  {L.}~\bibnamefont {Galbraith}}, \emph {et~al.},\ }\bibfield  {title}
  {\bibinfo {title} {Marine cloud brightening},\ }\href@noop {} {\bibfield
  {journal} {\bibinfo  {journal} {Philosophical Transactions of the Royal
  Society A: Mathematical, Physical and Engineering Sciences}\ }\textbf
  {\bibinfo {volume} {370}},\ \bibinfo {pages} {4217} (\bibinfo {year}
  {2012})}\BibitemShut {NoStop}%
\bibitem [{\citenamefont {{Cooper}}\ \emph {et~al.}(2014)\citenamefont
  {{Cooper}}, \citenamefont {{Foster}}, \citenamefont {{Galbraith}},
  \citenamefont {{Jain}}, \citenamefont {{Neukermans}},\ and\ \citenamefont
  {{Ormond}}}]{cooper2014}%
  \BibitemOpen
  \bibfield  {author} {\bibinfo {author} {\bibfnamefont {G.}~\bibnamefont
  {{Cooper}}}, \bibinfo {author} {\bibfnamefont {J.}~\bibnamefont {{Foster}}},
  \bibinfo {author} {\bibfnamefont {L.}~\bibnamefont {{Galbraith}}}, \bibinfo
  {author} {\bibfnamefont {S.}~\bibnamefont {{Jain}}}, \bibinfo {author}
  {\bibfnamefont {A.}~\bibnamefont {{Neukermans}}},\ and\ \bibinfo {author}
  {\bibfnamefont {B.}~\bibnamefont {{Ormond}}},\ }\bibfield  {title} {\bibinfo
  {title} {{Preliminary results for salt aerosol production intended for marine
  cloud brightening, using effervescent spray atomization}},\ }\href
  {https://doi.org/10.1098/rsta.2014.0055} {\bibfield  {journal} {\bibinfo
  {journal} {Philosophical Transactions of the Royal Society of London Series
  A}\ }\textbf {\bibinfo {volume} {372}},\ \bibinfo {pages} {20140055}
  (\bibinfo {year} {2014})}\BibitemShut {NoStop}%
\bibitem [{com()}]{comfre}%
  \BibitemOpen
  \href {https://comfre.iastate.edu/} {\bibinfo {title} {{Center for Multiphase
  Flow Research and Education (CoMFRE), Iowa State University,
  https://comfre.iastate.edu/}}}\BibitemShut {NoStop}%
\bibitem [{\citenamefont {Polin}\ \emph
  {et~al.}(2019{\natexlab{a}})\citenamefont {Polin}, \citenamefont {Carr},
  \citenamefont {Whitmer}, \citenamefont {Smith},\ and\ \citenamefont
  {Brown}}]{polin2019conventional}%
  \BibitemOpen
  \bibfield  {author} {\bibinfo {author} {\bibfnamefont {J.~P.}\ \bibnamefont
  {Polin}}, \bibinfo {author} {\bibfnamefont {H.~D.}\ \bibnamefont {Carr}},
  \bibinfo {author} {\bibfnamefont {L.~E.}\ \bibnamefont {Whitmer}}, \bibinfo
  {author} {\bibfnamefont {R.~G.}\ \bibnamefont {Smith}},\ and\ \bibinfo
  {author} {\bibfnamefont {R.~C.}\ \bibnamefont {Brown}},\ }\bibfield  {title}
  {\bibinfo {title} {Conventional and autothermal pyrolysis of corn stover:
  Overcoming the processing challenges of high-ash agricultural residues},\
  }\href@noop {} {\bibfield  {journal} {\bibinfo  {journal} {Journal of
  Analytical and Applied Pyrolysis}\ }\textbf {\bibinfo {volume} {143}},\
  \bibinfo {pages} {104679} (\bibinfo {year} {2019}{\natexlab{a}})}\BibitemShut
  {NoStop}%
\bibitem [{\citenamefont {Polin}\ \emph
  {et~al.}(2019{\natexlab{b}})\citenamefont {Polin}, \citenamefont {Peterson},
  \citenamefont {Whitmer}, \citenamefont {Smith},\ and\ \citenamefont
  {Brown}}]{polin2019process}%
  \BibitemOpen
  \bibfield  {author} {\bibinfo {author} {\bibfnamefont {J.~P.}\ \bibnamefont
  {Polin}}, \bibinfo {author} {\bibfnamefont {C.~A.}\ \bibnamefont {Peterson}},
  \bibinfo {author} {\bibfnamefont {L.~E.}\ \bibnamefont {Whitmer}}, \bibinfo
  {author} {\bibfnamefont {R.~G.}\ \bibnamefont {Smith}},\ and\ \bibinfo
  {author} {\bibfnamefont {R.~C.}\ \bibnamefont {Brown}},\ }\bibfield  {title}
  {\bibinfo {title} {Process intensification of biomass fast pyrolysis through
  autothermal operation of a fluidized bed reactor},\ }\href@noop {} {\bibfield
   {journal} {\bibinfo  {journal} {Applied Energy}\ }\textbf {\bibinfo {volume}
  {249}},\ \bibinfo {pages} {276} (\bibinfo {year}
  {2019}{\natexlab{b}})}\BibitemShut {NoStop}%
\bibitem [{\citenamefont {Polin}(2019)}]{polin2019thesis}%
  \BibitemOpen
  \bibfield  {author} {\bibinfo {author} {\bibfnamefont {J.~P.}\ \bibnamefont
  {Polin}},\ }\emph {\bibinfo {title} {Process intensification of biomass fast
  pyrolysis via autothermal operation}},\ \href@noop {} {Ph.D. thesis},\
  \bibinfo  {school} {Iowa State University} (\bibinfo {year}
  {2019})\BibitemShut {NoStop}%
\bibitem [{\citenamefont {Knowlton}\ \emph {et~al.}(2005)\citenamefont
  {Knowlton}, \citenamefont {Karri},\ and\ \citenamefont
  {Issangya}}]{knowlton2005scale}%
  \BibitemOpen
  \bibfield  {author} {\bibinfo {author} {\bibfnamefont {T.}~\bibnamefont
  {Knowlton}}, \bibinfo {author} {\bibfnamefont {S.}~\bibnamefont {Karri}},\
  and\ \bibinfo {author} {\bibfnamefont {A.}~\bibnamefont {Issangya}},\
  }\bibfield  {title} {\bibinfo {title} {Scale-up of fluidized-bed
  hydrodynamics},\ }\href@noop {} {\bibfield  {journal} {\bibinfo  {journal}
  {Powder Technology}\ }\textbf {\bibinfo {volume} {150}},\ \bibinfo {pages}
  {72} (\bibinfo {year} {2005})}\BibitemShut {NoStop}%
\bibitem [{\citenamefont {Breault}\ and\ \citenamefont
  {Guenther}(2009)}]{breault2009mass}%
  \BibitemOpen
  \bibfield  {author} {\bibinfo {author} {\bibfnamefont {R.~W.}\ \bibnamefont
  {Breault}}\ and\ \bibinfo {author} {\bibfnamefont {C.~P.}\ \bibnamefont
  {Guenther}},\ }\bibfield  {title} {\bibinfo {title} {Mass transfer in the
  core-annular and fast fluidization flow regimes of a cfb},\ }\href@noop {}
  {\bibfield  {journal} {\bibinfo  {journal} {Powder Technology}\ }\textbf
  {\bibinfo {volume} {190}},\ \bibinfo {pages} {385} (\bibinfo {year}
  {2009})}\BibitemShut {NoStop}%
\bibitem [{\citenamefont {Sun}\ \emph {et~al.}(2015)\citenamefont {Sun},
  \citenamefont {Tenneti},\ and\ \citenamefont
  {Subramaniam}}]{sun2015modeling}%
  \BibitemOpen
  \bibfield  {author} {\bibinfo {author} {\bibfnamefont {B.}~\bibnamefont
  {Sun}}, \bibinfo {author} {\bibfnamefont {S.}~\bibnamefont {Tenneti}},\ and\
  \bibinfo {author} {\bibfnamefont {S.}~\bibnamefont {Subramaniam}},\
  }\bibfield  {title} {\bibinfo {title} {Modeling average gas--solid heat
  transfer using particle-resolved direct numerical simulation},\ }\href@noop
  {} {\bibfield  {journal} {\bibinfo  {journal} {International Journal of Heat
  and Mass Transfer}\ }\textbf {\bibinfo {volume} {86}},\ \bibinfo {pages}
  {898} (\bibinfo {year} {2015})}\BibitemShut {NoStop}%
\bibitem [{\citenamefont {Shaffer}\ \emph {et~al.}(2013)\citenamefont
  {Shaffer}, \citenamefont {Gopalan}, \citenamefont {Breault}, \citenamefont
  {Cocco}, \citenamefont {Karri}, \citenamefont {Hays},\ and\ \citenamefont
  {Knowlton}}]{shaffer2013high}%
  \BibitemOpen
  \bibfield  {author} {\bibinfo {author} {\bibfnamefont {F.}~\bibnamefont
  {Shaffer}}, \bibinfo {author} {\bibfnamefont {B.}~\bibnamefont {Gopalan}},
  \bibinfo {author} {\bibfnamefont {R.~W.}\ \bibnamefont {Breault}}, \bibinfo
  {author} {\bibfnamefont {R.}~\bibnamefont {Cocco}}, \bibinfo {author}
  {\bibfnamefont {S.~R.}\ \bibnamefont {Karri}}, \bibinfo {author}
  {\bibfnamefont {R.}~\bibnamefont {Hays}},\ and\ \bibinfo {author}
  {\bibfnamefont {T.}~\bibnamefont {Knowlton}},\ }\bibfield  {title} {\bibinfo
  {title} {{High speed imaging of particle flow fields in CFB risers}},\
  }\href@noop {} {\bibfield  {journal} {\bibinfo  {journal} {Powder
  technology}\ }\textbf {\bibinfo {volume} {242}},\ \bibinfo {pages} {86}
  (\bibinfo {year} {2013})}\BibitemShut {NoStop}%
\bibitem [{\citenamefont {McMillan}\ \emph {et~al.}(2013)\citenamefont
  {McMillan}, \citenamefont {Shaffer}, \citenamefont {Gopalan}, \citenamefont
  {Chew}, \citenamefont {Hrenya}, \citenamefont {Hays}, \citenamefont {Karri},\
  and\ \citenamefont {Cocco}}]{mcmillan2013particle}%
  \BibitemOpen
  \bibfield  {author} {\bibinfo {author} {\bibfnamefont {J.}~\bibnamefont
  {McMillan}}, \bibinfo {author} {\bibfnamefont {F.}~\bibnamefont {Shaffer}},
  \bibinfo {author} {\bibfnamefont {B.}~\bibnamefont {Gopalan}}, \bibinfo
  {author} {\bibfnamefont {J.~W.}\ \bibnamefont {Chew}}, \bibinfo {author}
  {\bibfnamefont {C.}~\bibnamefont {Hrenya}}, \bibinfo {author} {\bibfnamefont
  {R.}~\bibnamefont {Hays}}, \bibinfo {author} {\bibfnamefont {S.~R.}\
  \bibnamefont {Karri}},\ and\ \bibinfo {author} {\bibfnamefont
  {R.}~\bibnamefont {Cocco}},\ }\bibfield  {title} {\bibinfo {title} {Particle
  cluster dynamics during fluidization},\ }\href@noop {} {\bibfield  {journal}
  {\bibinfo  {journal} {Chemical Engineering Science}\ }\textbf {\bibinfo
  {volume} {100}},\ \bibinfo {pages} {39} (\bibinfo {year} {2013})}\BibitemShut
  {NoStop}%
\bibitem [{\citenamefont {Jakobsen}(2008)}]{jakobsen2008chemical}%
  \BibitemOpen
  \bibfield  {author} {\bibinfo {author} {\bibfnamefont {H.~A.}\ \bibnamefont
  {Jakobsen}},\ }\bibfield  {title} {\bibinfo {title} {Chemical reactor
  modeling},\ }\href@noop {} {\bibfield  {journal} {\bibinfo  {journal}
  {Multiphase Reactive Flows}\ } (\bibinfo {year} {2008})}\BibitemShut
  {NoStop}%
\bibitem [{\citenamefont {Anglart}\ and\ \citenamefont
  {Nylund}(1996)}]{anglart1996cfd}%
  \BibitemOpen
  \bibfield  {author} {\bibinfo {author} {\bibfnamefont {H.}~\bibnamefont
  {Anglart}}\ and\ \bibinfo {author} {\bibfnamefont {O.}~\bibnamefont
  {Nylund}},\ }\bibfield  {title} {\bibinfo {title} {{CFD application to
  prediction of void distribution in two-phase bubbly flows in rod bundles}},\
  }\href@noop {} {\bibfield  {journal} {\bibinfo  {journal} {Nuclear
  Engineering and Design}\ }\textbf {\bibinfo {volume} {163}},\ \bibinfo
  {pages} {81} (\bibinfo {year} {1996})}\BibitemShut {NoStop}%
\bibitem [{\citenamefont {Serizawa}\ \emph {et~al.}(1997)\citenamefont
  {Serizawa}, \citenamefont {Huda}, \citenamefont {Yamada},\ and\ \citenamefont
  {Kataoka}}]{serizawa1997experiment}%
  \BibitemOpen
  \bibfield  {author} {\bibinfo {author} {\bibfnamefont {A.}~\bibnamefont
  {Serizawa}}, \bibinfo {author} {\bibfnamefont {K.}~\bibnamefont {Huda}},
  \bibinfo {author} {\bibfnamefont {Y.}~\bibnamefont {Yamada}},\ and\ \bibinfo
  {author} {\bibfnamefont {I.}~\bibnamefont {Kataoka}},\ }\bibfield  {title}
  {\bibinfo {title} {Experiment and numerical simulation of bubbly two-phase
  flow across horizontal and inclined rod bundles},\ }\href@noop {} {\bibfield
  {journal} {\bibinfo  {journal} {Nuclear engineering and design}\ }\textbf
  {\bibinfo {volume} {175}},\ \bibinfo {pages} {131} (\bibinfo {year}
  {1997})}\BibitemShut {NoStop}%
\bibitem [{\citenamefont {Kantarci}\ \emph {et~al.}(2005)\citenamefont
  {Kantarci}, \citenamefont {Borak},\ and\ \citenamefont
  {Ulgen}}]{kantarci2005bubble}%
  \BibitemOpen
  \bibfield  {author} {\bibinfo {author} {\bibfnamefont {N.}~\bibnamefont
  {Kantarci}}, \bibinfo {author} {\bibfnamefont {F.}~\bibnamefont {Borak}},\
  and\ \bibinfo {author} {\bibfnamefont {K.~O.}\ \bibnamefont {Ulgen}},\
  }\bibfield  {title} {\bibinfo {title} {Bubble column reactors},\ }\href@noop
  {} {\bibfield  {journal} {\bibinfo  {journal} {Process biochemistry}\
  }\textbf {\bibinfo {volume} {40}},\ \bibinfo {pages} {2263} (\bibinfo {year}
  {2005})}\BibitemShut {NoStop}%
\bibitem [{\citenamefont {Haghtalab}\ \emph {et~al.}(2012)\citenamefont
  {Haghtalab}, \citenamefont {Nabipoor},\ and\ \citenamefont
  {Farzad}}]{haghtalab2012kinetic}%
  \BibitemOpen
  \bibfield  {author} {\bibinfo {author} {\bibfnamefont {A.}~\bibnamefont
  {Haghtalab}}, \bibinfo {author} {\bibfnamefont {M.}~\bibnamefont
  {Nabipoor}},\ and\ \bibinfo {author} {\bibfnamefont {S.}~\bibnamefont
  {Farzad}},\ }\bibfield  {title} {\bibinfo {title} {Kinetic modeling of the
  fischer--tropsch synthesis in a slurry phase bubble column reactor using
  langmuir--freundlich isotherm},\ }\href@noop {} {\bibfield  {journal}
  {\bibinfo  {journal} {Fuel processing technology}\ }\textbf {\bibinfo
  {volume} {104}},\ \bibinfo {pages} {73} (\bibinfo {year} {2012})}\BibitemShut
  {NoStop}%
\bibitem [{\citenamefont {Storm}\ and\ \citenamefont
  {K{\"o}psel}(1992)}]{storm1992modelling}%
  \BibitemOpen
  \bibfield  {author} {\bibinfo {author} {\bibfnamefont {D.}~\bibnamefont
  {Storm}}\ and\ \bibinfo {author} {\bibfnamefont {R.}~\bibnamefont
  {K{\"o}psel}},\ }\bibfield  {title} {\bibinfo {title} {Modelling of catalytic
  coal hydrogenation in a bubble column reactor: 2. model calculations of
  bubble column cascades},\ }\href@noop {} {\bibfield  {journal} {\bibinfo
  {journal} {Fuel}\ }\textbf {\bibinfo {volume} {71}},\ \bibinfo {pages} {681}
  (\bibinfo {year} {1992})}\BibitemShut {NoStop}%
\bibitem [{\citenamefont {Winterbottom}(1993)}]{winterbottom1993bubble}%
  \BibitemOpen
  \bibfield  {author} {\bibinfo {author} {\bibfnamefont {J.}~\bibnamefont
  {Winterbottom}},\ }\bibfield  {title} {\bibinfo {title} {Bubble column
  reactors. by wolf-dieter deckwer. john wiley \& sons ltd, chichester, 1992,
  ix+ 533 pp., price: 110.00. isbn 0 471 91811 3},\ }\href@noop {} {\bibfield
  {journal} {\bibinfo  {journal} {Journal of Chemical Technology \&
  Biotechnology}\ }\textbf {\bibinfo {volume} {58}},\ \bibinfo {pages} {403}
  (\bibinfo {year} {1993})}\BibitemShut {NoStop}%
\bibitem [{\citenamefont {Ferrer}\ \emph {et~al.}(1985)\citenamefont {Ferrer},
  \citenamefont {David},\ and\ \citenamefont
  {Villermaux}}]{ferrer1985homogeneous}%
  \BibitemOpen
  \bibfield  {author} {\bibinfo {author} {\bibfnamefont {M.}~\bibnamefont
  {Ferrer}}, \bibinfo {author} {\bibfnamefont {R.}~\bibnamefont {David}},\ and\
  \bibinfo {author} {\bibfnamefont {J.}~\bibnamefont {Villermaux}},\ }\bibfield
   {title} {\bibinfo {title} {Homogeneous oxidation of n-butane in a
  self-stirred reactor},\ }\href@noop {} {\bibfield  {journal} {\bibinfo
  {journal} {Chemical engineering and processing}\ }\textbf {\bibinfo {volume}
  {19}},\ \bibinfo {pages} {119} (\bibinfo {year} {1985})}\BibitemShut
  {NoStop}%
\bibitem [{\citenamefont {Debellefontaine}\ and\ \citenamefont
  {Foussard}(2000)}]{debellefontaine2000wet}%
  \BibitemOpen
  \bibfield  {author} {\bibinfo {author} {\bibfnamefont {H.}~\bibnamefont
  {Debellefontaine}}\ and\ \bibinfo {author} {\bibfnamefont {J.~N.}\
  \bibnamefont {Foussard}},\ }\bibfield  {title} {\bibinfo {title} {Wet air
  oxidation for the treatment of industrial wastes. chemical aspects, reactor
  design and industrial applications in europe},\ }\href@noop {} {\bibfield
  {journal} {\bibinfo  {journal} {Waste management}\ }\textbf {\bibinfo
  {volume} {20}},\ \bibinfo {pages} {15} (\bibinfo {year} {2000})}\BibitemShut
  {NoStop}%
\bibitem [{\citenamefont {Ranjbar}\ \emph {et~al.}(2008)\citenamefont
  {Ranjbar}, \citenamefont {Inoue}, \citenamefont {Shiraishi}, \citenamefont
  {Katsuda},\ and\ \citenamefont {Katoh}}]{ranjbar2008high}%
  \BibitemOpen
  \bibfield  {author} {\bibinfo {author} {\bibfnamefont {R.}~\bibnamefont
  {Ranjbar}}, \bibinfo {author} {\bibfnamefont {R.}~\bibnamefont {Inoue}},
  \bibinfo {author} {\bibfnamefont {H.}~\bibnamefont {Shiraishi}}, \bibinfo
  {author} {\bibfnamefont {T.}~\bibnamefont {Katsuda}},\ and\ \bibinfo {author}
  {\bibfnamefont {S.}~\bibnamefont {Katoh}},\ }\bibfield  {title} {\bibinfo
  {title} {High efficiency production of astaxanthin by autotrophic cultivation
  of haematococcus pluvialis in a bubble column photobioreactor},\ }\href@noop
  {} {\bibfield  {journal} {\bibinfo  {journal} {Biochemical Engineering
  Journal}\ }\textbf {\bibinfo {volume} {39}},\ \bibinfo {pages} {575}
  (\bibinfo {year} {2008})}\BibitemShut {NoStop}%
\bibitem [{\citenamefont {Kosseva}\ \emph {et~al.}(1991)\citenamefont
  {Kosseva}, \citenamefont {Beschkov},\ and\ \citenamefont
  {Popov}}]{kosseva1991biotransformation}%
  \BibitemOpen
  \bibfield  {author} {\bibinfo {author} {\bibfnamefont {M.}~\bibnamefont
  {Kosseva}}, \bibinfo {author} {\bibfnamefont {V.}~\bibnamefont {Beschkov}},\
  and\ \bibinfo {author} {\bibfnamefont {R.}~\bibnamefont {Popov}},\ }\bibfield
   {title} {\bibinfo {title} {Biotransformation of d-sorbitol to l-sorbose by
  immobilized cells gluconobacter suboxydans in a bubble column},\ }\href@noop
  {} {\bibfield  {journal} {\bibinfo  {journal} {Journal of biotechnology}\
  }\textbf {\bibinfo {volume} {19}},\ \bibinfo {pages} {301} (\bibinfo {year}
  {1991})}\BibitemShut {NoStop}%
\bibitem [{\citenamefont {Nauha}\ and\ \citenamefont
  {Alopaeus}(2013)}]{nauha2013modeling}%
  \BibitemOpen
  \bibfield  {author} {\bibinfo {author} {\bibfnamefont {E.~K.}\ \bibnamefont
  {Nauha}}\ and\ \bibinfo {author} {\bibfnamefont {V.}~\bibnamefont
  {Alopaeus}},\ }\bibfield  {title} {\bibinfo {title} {Modeling method for
  combining fluid dynamics and algal growth in a bubble column
  photobioreactor},\ }\href@noop {} {\bibfield  {journal} {\bibinfo  {journal}
  {Chemical engineering journal}\ }\textbf {\bibinfo {volume} {229}},\ \bibinfo
  {pages} {559} (\bibinfo {year} {2013})}\BibitemShut {NoStop}%
\bibitem [{\citenamefont {Posten}(2009)}]{posten2009design}%
  \BibitemOpen
  \bibfield  {author} {\bibinfo {author} {\bibfnamefont {C.}~\bibnamefont
  {Posten}},\ }\bibfield  {title} {\bibinfo {title} {Design principles of
  photo-bioreactors for cultivation of microalgae},\ }\href@noop {} {\bibfield
  {journal} {\bibinfo  {journal} {Engineering in Life Sciences}\ }\textbf
  {\bibinfo {volume} {9}},\ \bibinfo {pages} {165} (\bibinfo {year}
  {2009})}\BibitemShut {NoStop}%
\bibitem [{\citenamefont {Mu{\~n}oz-Cobo}\ \emph {et~al.}(2012)\citenamefont
  {Mu{\~n}oz-Cobo}, \citenamefont {Chiva}, \citenamefont {Essa},\ and\
  \citenamefont {Mendes}}]{munoz2012simulation}%
  \BibitemOpen
  \bibfield  {author} {\bibinfo {author} {\bibfnamefont {J.~L.}\ \bibnamefont
  {Mu{\~n}oz-Cobo}}, \bibinfo {author} {\bibfnamefont {S.}~\bibnamefont
  {Chiva}}, \bibinfo {author} {\bibfnamefont {M.~A. A. E.~A.}\ \bibnamefont
  {Essa}},\ and\ \bibinfo {author} {\bibfnamefont {S.}~\bibnamefont {Mendes}},\
  }\bibfield  {title} {\bibinfo {title} {Simulation of bubbly flow in vertical
  pipes by coupling lagrangian and eulerian models with 3d random walks models:
  Validation with experimental data using multi-sensor conductivity probes and
  laser doppler anemometry},\ }\href@noop {} {\bibfield  {journal} {\bibinfo
  {journal} {Nuclear engineering and design}\ }\textbf {\bibinfo {volume}
  {242}},\ \bibinfo {pages} {285} (\bibinfo {year} {2012})}\BibitemShut
  {NoStop}%
\bibitem [{\citenamefont {Tryggvason}\ \emph {et~al.}(2013)\citenamefont
  {Tryggvason}, \citenamefont {Dabiri}, \citenamefont {Aboulhasanzadeh},\ and\
  \citenamefont {Lu}}]{tryggvason2013multiscale}%
  \BibitemOpen
  \bibfield  {author} {\bibinfo {author} {\bibfnamefont {G.}~\bibnamefont
  {Tryggvason}}, \bibinfo {author} {\bibfnamefont {S.}~\bibnamefont {Dabiri}},
  \bibinfo {author} {\bibfnamefont {B.}~\bibnamefont {Aboulhasanzadeh}},\ and\
  \bibinfo {author} {\bibfnamefont {J.}~\bibnamefont {Lu}},\ }\bibfield
  {title} {\bibinfo {title} {Multiscale considerations in direct numerical
  simulations of multiphase flows},\ }\href@noop {} {\bibfield  {journal}
  {\bibinfo  {journal} {Physics of Fluids}\ }\textbf {\bibinfo {volume} {25}},\
  \bibinfo {pages} {031302} (\bibinfo {year} {2013})}\BibitemShut {NoStop}%
\bibitem [{\citenamefont {Gvozdi{\'c}}\ \emph {et~al.}(2018)\citenamefont
  {Gvozdi{\'c}}, \citenamefont {Alm{\'e}ras}, \citenamefont {Mathai},
  \citenamefont {Zhu}, \citenamefont {van Gils}, \citenamefont {Verzicco},
  \citenamefont {Huisman}, \citenamefont {Sun},\ and\ \citenamefont
  {Lohse}}]{gvozdic2018experimental}%
  \BibitemOpen
  \bibfield  {author} {\bibinfo {author} {\bibfnamefont {B.}~\bibnamefont
  {Gvozdi{\'c}}}, \bibinfo {author} {\bibfnamefont {E.}~\bibnamefont
  {Alm{\'e}ras}}, \bibinfo {author} {\bibfnamefont {V.}~\bibnamefont {Mathai}},
  \bibinfo {author} {\bibfnamefont {X.}~\bibnamefont {Zhu}}, \bibinfo {author}
  {\bibfnamefont {D.~P.}\ \bibnamefont {van Gils}}, \bibinfo {author}
  {\bibfnamefont {R.}~\bibnamefont {Verzicco}}, \bibinfo {author}
  {\bibfnamefont {S.~G.}\ \bibnamefont {Huisman}}, \bibinfo {author}
  {\bibfnamefont {C.}~\bibnamefont {Sun}},\ and\ \bibinfo {author}
  {\bibfnamefont {D.}~\bibnamefont {Lohse}},\ }\bibfield  {title} {\bibinfo
  {title} {Experimental investigation of heat transport in homogeneous bubbly
  flow},\ }\href@noop {} {\bibfield  {journal} {\bibinfo  {journal} {Journal of
  fluid mechanics}\ }\textbf {\bibinfo {volume} {845}},\ \bibinfo {pages} {226}
  (\bibinfo {year} {2018})}\BibitemShut {NoStop}%
\bibitem [{\citenamefont {Wang}\ \emph {et~al.}(2019)\citenamefont {Wang},
  \citenamefont {Mathai},\ and\ \citenamefont {Sun}}]{wang2019self}%
  \BibitemOpen
  \bibfield  {author} {\bibinfo {author} {\bibfnamefont {Z.}~\bibnamefont
  {Wang}}, \bibinfo {author} {\bibfnamefont {V.}~\bibnamefont {Mathai}},\ and\
  \bibinfo {author} {\bibfnamefont {C.}~\bibnamefont {Sun}},\ }\bibfield
  {title} {\bibinfo {title} {Self-sustained biphasic catalytic particle
  turbulence},\ }\href@noop {} {\bibfield  {journal} {\bibinfo  {journal}
  {Nature communications}\ }\textbf {\bibinfo {volume} {10}},\ \bibinfo {pages}
  {1} (\bibinfo {year} {2019})}\BibitemShut {NoStop}%
\bibitem [{\citenamefont {Subramaniam}(2019)}]{subramaniam2019multiphase}%
  \BibitemOpen
  \bibfield  {author} {\bibinfo {author} {\bibfnamefont {S.}~\bibnamefont
  {Subramaniam}},\ }\bibfield  {title} {\bibinfo {title} {Multiphase flows:
  Rich physics, challenging theory, and big simulations},\ }\href@noop {}
  {\bibfield  {journal} {\bibinfo  {journal} {APS}\ ,\ \bibinfo {pages} {K01}}
  (\bibinfo {year} {2019})}\BibitemShut {NoStop}%
\bibitem [{\citenamefont {Balachandar}\ and\ \citenamefont
  {Eaton}(2010)}]{bala_eaton_2010}%
  \BibitemOpen
  \bibfield  {author} {\bibinfo {author} {\bibfnamefont {S.}~\bibnamefont
  {Balachandar}}\ and\ \bibinfo {author} {\bibfnamefont {J.~K.}\ \bibnamefont
  {Eaton}},\ }\bibfield  {title} {\bibinfo {title} {Turbulent dispersed
  multiphase flow},\ }\href@noop {} {\bibfield  {journal} {\bibinfo  {journal}
  {Annual review of fluid mechanics}\ }\textbf {\bibinfo {volume} {42}},\
  \bibinfo {pages} {111} (\bibinfo {year} {2010})}\BibitemShut {NoStop}%
\bibitem [{\citenamefont {Clift}\ \emph {et~al.}(1978)\citenamefont {Clift},
  \citenamefont {Grace},\ and\ \citenamefont {Weber}}]{clift}%
  \BibitemOpen
  \bibfield  {author} {\bibinfo {author} {\bibfnamefont {R.}~\bibnamefont
  {Clift}}, \bibinfo {author} {\bibfnamefont {J.~R.}\ \bibnamefont {Grace}},\
  and\ \bibinfo {author} {\bibfnamefont {M.~E.}\ \bibnamefont {Weber}},\
  }\href@noop {} {\emph {\bibinfo {title} {{Bubbles, Drops and Particles}}}}\
  (\bibinfo  {publisher} {Academic Press},\ \bibinfo {year} {1978})\BibitemShut
  {NoStop}%
\bibitem [{\citenamefont {Maxey}\ and\ \citenamefont
  {Riley}(1983)}]{maxey_riley_pf83}%
  \BibitemOpen
  \bibfield  {author} {\bibinfo {author} {\bibfnamefont {M.~R.}\ \bibnamefont
  {Maxey}}\ and\ \bibinfo {author} {\bibfnamefont {J.~J.}\ \bibnamefont
  {Riley}},\ }\bibfield  {title} {\bibinfo {title} {Equation of motion for a
  small rigid sphere in a nonuniform flow},\ }\href@noop {} {\bibfield
  {journal} {\bibinfo  {journal} {Phys. Fluids}\ }\textbf {\bibinfo {volume}
  {26}},\ \bibinfo {pages} {883} (\bibinfo {year} {1983})}\BibitemShut
  {NoStop}%
\bibitem [{\citenamefont {Marchisio}\ and\ \citenamefont
  {Fox}(2013)}]{fox_cup_book}%
  \BibitemOpen
  \bibfield  {author} {\bibinfo {author} {\bibfnamefont {D.~L.}\ \bibnamefont
  {Marchisio}}\ and\ \bibinfo {author} {\bibfnamefont {R.~O.}\ \bibnamefont
  {Fox}},\ }\href@noop {} {\emph {\bibinfo {title} {{Computational Models for
  Polydisperse Particulate and Multiphase Systems}}}}\ (\bibinfo  {publisher}
  {Cambridge University Press},\ \bibinfo {year} {2013})\BibitemShut {NoStop}%
\bibitem [{\citenamefont {Apte}\ \emph {et~al.}(2008)\citenamefont {Apte},
  \citenamefont {Mahesh},\ and\ \citenamefont {Lundgren}}]{apte2008accounting}%
  \BibitemOpen
  \bibfield  {author} {\bibinfo {author} {\bibfnamefont {S.}~\bibnamefont
  {Apte}}, \bibinfo {author} {\bibfnamefont {K.}~\bibnamefont {Mahesh}},\ and\
  \bibinfo {author} {\bibfnamefont {T.}~\bibnamefont {Lundgren}},\ }\bibfield
  {title} {\bibinfo {title} {Accounting for finite-size effects in simulations
  of disperse particle-laden flows},\ }\href@noop {} {\bibfield  {journal}
  {\bibinfo  {journal} {International Journal of Multiphase Flow}\ }\textbf
  {\bibinfo {volume} {34}},\ \bibinfo {pages} {260} (\bibinfo {year}
  {2008})}\BibitemShut {NoStop}%
\bibitem [{\citenamefont {Subramaniam}(2013)}]{shankar_pecs_2013}%
  \BibitemOpen
  \bibfield  {author} {\bibinfo {author} {\bibfnamefont {S.}~\bibnamefont
  {Subramaniam}},\ }\bibfield  {title} {\bibinfo {title}
  {Lagrangian--{Eulerian} methods for multiphase flows},\ }\href
  {https://doi.org/10.1016/j.pecs.2012.10.003} {\bibfield  {journal} {\bibinfo
  {journal} {Progress in Energy and Combustion Science}\ }\textbf {\bibinfo
  {volume} {39}},\ \bibinfo {pages} {215} (\bibinfo {year} {2013})}\BibitemShut
  {NoStop}%
\bibitem [{\citenamefont {Elghobashi}(1994)}]{elgho_asr_94}%
  \BibitemOpen
  \bibfield  {author} {\bibinfo {author} {\bibfnamefont {S.}~\bibnamefont
  {Elghobashi}},\ }\bibfield  {title} {\bibinfo {title} {On predicting
  particle--laden turbulent flows},\ }\href@noop {} {\bibfield  {journal}
  {\bibinfo  {journal} {Appl. Sci. Res.}\ }\textbf {\bibinfo {volume} {52}},\
  \bibinfo {pages} {309} (\bibinfo {year} {1994})}\BibitemShut {NoStop}%
\bibitem [{\citenamefont {McQuarrie}(2000)}]{mcquarrie2000statistical}%
  \BibitemOpen
  \bibfield  {author} {\bibinfo {author} {\bibfnamefont {D.}~\bibnamefont
  {McQuarrie}},\ }\href@noop {} {\emph {\bibinfo {title} {Statistical
  mechanics}}}\ (\bibinfo  {publisher} {University Science Books, Sausalito,
  CA},\ \bibinfo {year} {2000})\BibitemShut {NoStop}%
\bibitem [{\citenamefont {Liboff}(2003)}]{liboff_book}%
  \BibitemOpen
  \bibfield  {author} {\bibinfo {author} {\bibfnamefont {R.~L.}\ \bibnamefont
  {Liboff}},\ }\href@noop {} {\emph {\bibinfo {title} {Kinetic theory :
  {Classical, Quantum, and Relativistic descriptions}}}},\ \bibinfo {edition}
  {3rd}\ ed.\ (\bibinfo  {publisher} {Springer-Verlag},\ \bibinfo {address}
  {New York},\ \bibinfo {year} {2003})\BibitemShut {NoStop}%
\bibitem [{\citenamefont {Monin}\ and\ \citenamefont
  {Yaglom}(1975)}]{monin_yaglom2}%
  \BibitemOpen
  \bibfield  {author} {\bibinfo {author} {\bibfnamefont {A.~S.}\ \bibnamefont
  {Monin}}\ and\ \bibinfo {author} {\bibfnamefont {A.~M.}\ \bibnamefont
  {Yaglom}},\ }\href@noop {} {\emph {\bibinfo {title} {{Statistical Fluid
  Mechanics II}}}}\ (\bibinfo  {publisher} {MIT Press},\ \bibinfo {year}
  {1975})\BibitemShut {NoStop}%
\bibitem [{\citenamefont {Pope}(2000)}]{popebook}%
  \BibitemOpen
  \bibfield  {author} {\bibinfo {author} {\bibfnamefont {S.~B.}\ \bibnamefont
  {Pope}},\ }\href@noop {} {\emph {\bibinfo {title} {Turbulent Flows}}}\
  (\bibinfo  {publisher} {Cambridge University Press},\ \bibinfo {address}
  {Port Chester, NY},\ \bibinfo {year} {2000})\BibitemShut {NoStop}%
\bibitem [{\citenamefont {Brenner}(1972)}]{brenner1972suspension}%
  \BibitemOpen
  \bibfield  {author} {\bibinfo {author} {\bibfnamefont {H.}~\bibnamefont
  {Brenner}},\ }\bibfield  {title} {\bibinfo {title} {Suspension rheology},\
  }in\ \href@noop {} {\emph {\bibinfo {booktitle} {Progress in heat and mass
  transfer}}}\ (\bibinfo  {publisher} {Elsevier},\ \bibinfo {year} {1972})\
  pp.\ \bibinfo {pages} {89--129}\BibitemShut {NoStop}%
\bibitem [{\citenamefont {Hinch}(1977)}]{hinch1977averaged}%
  \BibitemOpen
  \bibfield  {author} {\bibinfo {author} {\bibfnamefont {E.}~\bibnamefont
  {Hinch}},\ }\bibfield  {title} {\bibinfo {title} {An averaged-equation
  approach to particle interactions in a fluid suspension},\ }\href@noop {}
  {\bibfield  {journal} {\bibinfo  {journal} {Journal of Fluid Mechanics}\
  }\textbf {\bibinfo {volume} {83}},\ \bibinfo {pages} {695} (\bibinfo {year}
  {1977})}\BibitemShut {NoStop}%
\bibitem [{\citenamefont {Koch}(1990)}]{koch_pof_1990}%
  \BibitemOpen
  \bibfield  {author} {\bibinfo {author} {\bibfnamefont {D.~L.}\ \bibnamefont
  {Koch}},\ }\bibfield  {title} {\bibinfo {title} {{Kinetic theory for a
  monodisperse gas--solid suspension}},\ }\href@noop {} {\bibfield  {journal}
  {\bibinfo  {journal} {Physics of Fluids}\ }\textbf {\bibinfo {volume} {A
  2(10)}},\ \bibinfo {pages} {1711} (\bibinfo {year} {1990})}\BibitemShut
  {NoStop}%
\bibitem [{\citenamefont {Tenneti}\ and\ \citenamefont
  {Subramaniam}(2014)}]{shankar2014arfm}%
  \BibitemOpen
  \bibfield  {author} {\bibinfo {author} {\bibfnamefont {S.}~\bibnamefont
  {Tenneti}}\ and\ \bibinfo {author} {\bibfnamefont {S.}~\bibnamefont
  {Subramaniam}},\ }\bibfield  {title} {\bibinfo {title} {Particle-resolved
  direct numerical simulation for gas-solid flow model development},\
  }\href@noop {} {\bibfield  {journal} {\bibinfo  {journal} {Annual review of
  fluid mechanics}\ }\textbf {\bibinfo {volume} {46}},\ \bibinfo {pages} {199}
  (\bibinfo {year} {2014})}\BibitemShut {NoStop}%
\bibitem [{\citenamefont {Jofre}\ \emph {et~al.}(2020)\citenamefont {Jofre},
  \citenamefont {del Rosario},\ and\ \citenamefont
  {Iaccarino}}]{jofre2020data}%
  \BibitemOpen
  \bibfield  {author} {\bibinfo {author} {\bibfnamefont {L.}~\bibnamefont
  {Jofre}}, \bibinfo {author} {\bibfnamefont {Z.~R.}\ \bibnamefont {del
  Rosario}},\ and\ \bibinfo {author} {\bibfnamefont {G.}~\bibnamefont
  {Iaccarino}},\ }\bibfield  {title} {\bibinfo {title} {Data-driven dimensional
  analysis of heat transfer in irradiated particle-laden turbulent flow},\
  }\href@noop {} {\bibfield  {journal} {\bibinfo  {journal} {International
  Journal of Multiphase Flow}\ }\textbf {\bibinfo {volume} {125}},\ \bibinfo
  {pages} {103198} (\bibinfo {year} {2020})}\BibitemShut {NoStop}%
\bibitem [{\citenamefont {Moore}\ \emph {et~al.}(2019)\citenamefont {Moore},
  \citenamefont {Balachandar},\ and\ \citenamefont {Akiki}}]{moore2019hybrid}%
  \BibitemOpen
  \bibfield  {author} {\bibinfo {author} {\bibfnamefont {W.}~\bibnamefont
  {Moore}}, \bibinfo {author} {\bibfnamefont {S.}~\bibnamefont {Balachandar}},\
  and\ \bibinfo {author} {\bibfnamefont {G.}~\bibnamefont {Akiki}},\ }\bibfield
   {title} {\bibinfo {title} {A hybrid point-particle force model that combines
  physical and data-driven approaches},\ }\href@noop {} {\bibfield  {journal}
  {\bibinfo  {journal} {Journal of Computational Physics}\ }\textbf {\bibinfo
  {volume} {385}},\ \bibinfo {pages} {187} (\bibinfo {year}
  {2019})}\BibitemShut {NoStop}%
\bibitem [{\citenamefont {Lumley}(1990)}]{lumley1990whither}%
  \BibitemOpen
  \bibfield  {author} {\bibinfo {author} {\bibfnamefont {J.~L.}\ \bibnamefont
  {Lumley}},\ }\href@noop {} {\emph {\bibinfo {title} {Whither turbulence?
  Turbulence at the crossroads}}},\ \bibinfo {series} {Lecture Notes in
  Physics}, Vol.\ \bibinfo {volume} {357}\ (\bibinfo  {publisher}
  {Springer--Verlag, New York},\ \bibinfo {year} {1990})\BibitemShut {NoStop}%
\bibitem [{\citenamefont {Adler}(1981)}]{adlerbook}%
  \BibitemOpen
  \bibfield  {author} {\bibinfo {author} {\bibfnamefont {R.~J.}\ \bibnamefont
  {Adler}},\ }\href@noop {} {\emph {\bibinfo {title} {The geometry of random
  fields}}}\ (\bibinfo  {publisher} {SIAM},\ \bibinfo {address} {Chichester,
  NY},\ \bibinfo {year} {1981})\BibitemShut {NoStop}%
\bibitem [{\citenamefont {Panchev}(1971)}]{panchev}%
  \BibitemOpen
  \bibfield  {author} {\bibinfo {author} {\bibfnamefont {S.}~\bibnamefont
  {Panchev}},\ }\href@noop {} {\emph {\bibinfo {title} {Random functions and
  turbulence}}}\ (\bibinfo  {publisher} {Pergamon Press},\ \bibinfo {address}
  {New York},\ \bibinfo {year} {1971})\BibitemShut {NoStop}%
\bibitem [{\citenamefont {Drew}(1983)}]{drew_arfm}%
  \BibitemOpen
  \bibfield  {author} {\bibinfo {author} {\bibfnamefont {D.~A.}\ \bibnamefont
  {Drew}},\ }\bibfield  {title} {\bibinfo {title} {Mathematical modeling of
  two--phase flow},\ }\href@noop {} {\bibfield  {journal} {\bibinfo  {journal}
  {Annu. Rev. Fluid Mech.}\ }\textbf {\bibinfo {volume} {15}},\ \bibinfo
  {pages} {261} (\bibinfo {year} {1983})}\BibitemShut {NoStop}%
\bibitem [{\citenamefont {Pai}\ and\ \citenamefont
  {Subramaniam}(2009)}]{pai_ss_jfm09}%
  \BibitemOpen
  \bibfield  {author} {\bibinfo {author} {\bibfnamefont {M.~G.}\ \bibnamefont
  {Pai}}\ and\ \bibinfo {author} {\bibfnamefont {S.}~\bibnamefont
  {Subramaniam}},\ }\bibfield  {title} {\bibinfo {title} {A comprehensive
  probability density function formalism for multiphase flows},\ }\href@noop {}
  {\bibfield  {journal} {\bibinfo  {journal} {Journal of Fluid Mechanics}\
  }\textbf {\bibinfo {volume} {628}},\ \bibinfo {pages} {181} (\bibinfo {year}
  {2009})}\BibitemShut {NoStop}%
\bibitem [{\citenamefont {Anderson}\ and\ \citenamefont
  {Jackson}(1967)}]{And_jack}%
  \BibitemOpen
  \bibfield  {author} {\bibinfo {author} {\bibfnamefont {T.~B.}\ \bibnamefont
  {Anderson}}\ and\ \bibinfo {author} {\bibfnamefont {R.}~\bibnamefont
  {Jackson}},\ }\bibfield  {title} {\bibinfo {title} {A fluid mechanical
  description of fluidized beds},\ }\href@noop {} {\bibfield  {journal}
  {\bibinfo  {journal} {Ind. Eng. Chem. Fundam.}\ }\textbf {\bibinfo {volume}
  {6}},\ \bibinfo {pages} {527} (\bibinfo {year} {1967})}\BibitemShut {NoStop}%
\bibitem [{\citenamefont {Ishii}(1975)}]{ishii_tft}%
  \BibitemOpen
  \bibfield  {author} {\bibinfo {author} {\bibfnamefont {M.}~\bibnamefont
  {Ishii}},\ }\href@noop {} {\emph {\bibinfo {title} {Thermo--fluid dynamic
  theory of two--phase flow}}}\ (\bibinfo  {publisher} {Eyrolles},\ \bibinfo
  {address} {Paris, France},\ \bibinfo {year} {1975})\BibitemShut {NoStop}%
\bibitem [{\citenamefont {Joseph}\ \emph {et~al.}(1990)\citenamefont {Joseph},
  \citenamefont {Lundgren}, \citenamefont {Jackson},\ and\ \citenamefont
  {Saville}}]{joseph_lundgren_etal_ijmf90}%
  \BibitemOpen
  \bibfield  {author} {\bibinfo {author} {\bibfnamefont {D.~D.}\ \bibnamefont
  {Joseph}}, \bibinfo {author} {\bibfnamefont {T.~S.}\ \bibnamefont
  {Lundgren}}, \bibinfo {author} {\bibfnamefont {R.}~\bibnamefont {Jackson}},\
  and\ \bibinfo {author} {\bibfnamefont {D.~A.}\ \bibnamefont {Saville}},\
  }\bibfield  {title} {\bibinfo {title} {Ensemble averaged and mixture theory
  equations for incompressible fluid--particle suspensions},\ }\href@noop {}
  {\bibfield  {journal} {\bibinfo  {journal} {Intl. J. Multiphase Flow}\
  }\textbf {\bibinfo {volume} {16}},\ \bibinfo {pages} {35} (\bibinfo {year}
  {1990})}\BibitemShut {NoStop}%
\bibitem [{\citenamefont {Smagorinsky}(1963)}]{smagorinsky1963general}%
  \BibitemOpen
  \bibfield  {author} {\bibinfo {author} {\bibfnamefont {J.}~\bibnamefont
  {Smagorinsky}},\ }\bibfield  {title} {\bibinfo {title} {General circulation
  experiments with the primitive equations: I. the basic experiment},\
  }\href@noop {} {\bibfield  {journal} {\bibinfo  {journal} {Monthly weather
  review}\ }\textbf {\bibinfo {volume} {91}},\ \bibinfo {pages} {99} (\bibinfo
  {year} {1963})}\BibitemShut {NoStop}%
\bibitem [{\citenamefont {Lilly}(1967)}]{lilly1967proceedings}%
  \BibitemOpen
  \bibfield  {author} {\bibinfo {author} {\bibfnamefont {D.}~\bibnamefont
  {Lilly}},\ }\bibfield  {title} {\bibinfo {title} {The representation of
  small--scale turbulence in numerical simulation experiments},\ }\bibinfo
  {organization} {IBM}\ (\bibinfo  {publisher} {IBM},\ \bibinfo {address}
  {Yorktown Heights, NY},\ \bibinfo {year} {1967})\ pp.\ \bibinfo {pages}
  {195--210}\BibitemShut {NoStop}%
\bibitem [{\citenamefont {Moin}\ and\ \citenamefont
  {Kim}(1982)}]{moin1982numerical}%
  \BibitemOpen
  \bibfield  {author} {\bibinfo {author} {\bibfnamefont {P.}~\bibnamefont
  {Moin}}\ and\ \bibinfo {author} {\bibfnamefont {J.}~\bibnamefont {Kim}},\
  }\bibfield  {title} {\bibinfo {title} {Numerical investigation of turbulent
  channel flow},\ }\href@noop {} {\bibfield  {journal} {\bibinfo  {journal}
  {Journal of fluid mechanics}\ }\textbf {\bibinfo {volume} {118}},\ \bibinfo
  {pages} {341} (\bibinfo {year} {1982})}\BibitemShut {NoStop}%
\bibitem [{\citenamefont {Smagorinsky}\ \emph {et~al.}(1993)\citenamefont
  {Smagorinsky}, \citenamefont {Galperin},\ and\ \citenamefont
  {Orszag}}]{smagorinsky1993large}%
  \BibitemOpen
  \bibfield  {author} {\bibinfo {author} {\bibfnamefont {J.}~\bibnamefont
  {Smagorinsky}}, \bibinfo {author} {\bibfnamefont {B.}~\bibnamefont
  {Galperin}},\ and\ \bibinfo {author} {\bibfnamefont {S.}~\bibnamefont
  {Orszag}},\ }\bibfield  {title} {\bibinfo {title} {Large eddy simulation of
  complex engineering and geophysical flows},\ }\href@noop {} {\bibfield
  {journal} {\bibinfo  {journal} {Evolution of physical oceanography}\ ,\
  \bibinfo {pages} {3}} (\bibinfo {year} {1993})}\BibitemShut {NoStop}%
\bibitem [{\citenamefont {Goc}\ \emph {et~al.}(2020)\citenamefont {Goc},
  \citenamefont {Bose},\ and\ \citenamefont {Moin}}]{goc2020wall}%
  \BibitemOpen
  \bibfield  {author} {\bibinfo {author} {\bibfnamefont {K.}~\bibnamefont
  {Goc}}, \bibinfo {author} {\bibfnamefont {S.}~\bibnamefont {Bose}},\ and\
  \bibinfo {author} {\bibfnamefont {P.}~\bibnamefont {Moin}},\ }\bibfield
  {title} {\bibinfo {title} {Wall-modeled large eddy simulation of an aircraft
  in landing configuration},\ }in\ \href@noop {} {\emph {\bibinfo {booktitle}
  {AIAA AVIATION 2020 FORUM}}}\ (\bibinfo {year} {2020})\ p.\ \bibinfo {pages}
  {3002}\BibitemShut {NoStop}%
\bibitem [{\citenamefont {Capecelatro}\ and\ \citenamefont
  {Desjardins}(2013)}]{olivier_les_jcp_2013}%
  \BibitemOpen
  \bibfield  {author} {\bibinfo {author} {\bibfnamefont {J.}~\bibnamefont
  {Capecelatro}}\ and\ \bibinfo {author} {\bibfnamefont {O.}~\bibnamefont
  {Desjardins}},\ }\bibfield  {title} {\bibinfo {title} {An {Euler}--{Lagrange}
  strategy for simulating particle-laden flows},\ }\href
  {https://doi.org/10.1016/j.jcp.2012.12.015} {\bibfield  {journal} {\bibinfo
  {journal} {Journal of Computational Physics}\ }\textbf {\bibinfo {volume}
  {238}},\ \bibinfo {pages} {1 } (\bibinfo {year} {2013})}\BibitemShut
  {NoStop}%
\bibitem [{\citenamefont {Subramaniam}(2000)}]{pprocess}%
  \BibitemOpen
  \bibfield  {author} {\bibinfo {author} {\bibfnamefont {S.}~\bibnamefont
  {Subramaniam}},\ }\bibfield  {title} {\bibinfo {title} {{Statistical
  representation of a spray as a point process}},\ }\href@noop {} {\bibfield
  {journal} {\bibinfo  {journal} {Phys. Fluids}\ }\textbf {\bibinfo {volume}
  {12}},\ \bibinfo {pages} {2413} (\bibinfo {year} {2000})}\BibitemShut
  {NoStop}%
\bibitem [{\citenamefont {Fox}(2012)}]{fox_arfm2011}%
  \BibitemOpen
  \bibfield  {author} {\bibinfo {author} {\bibfnamefont {R.~O.}\ \bibnamefont
  {Fox}},\ }\bibfield  {title} {\bibinfo {title} {{Large-eddy-simulation tools
  for multiphase flows}},\ }\href@noop {} {\bibfield  {journal} {\bibinfo
  {journal} {Annu. Rev. Fluid Mech.}\ }\textbf {\bibinfo {volume} {44}},\
  \bibinfo {pages} {47} (\bibinfo {year} {2012})}\BibitemShut {NoStop}%
\bibitem [{\citenamefont {Tenneti}\ \emph {et~al.}(2016)\citenamefont
  {Tenneti}, \citenamefont {Mehrabadi},\ and\ \citenamefont
  {Subramaniam}}]{tenneti2016stochastic}%
  \BibitemOpen
  \bibfield  {author} {\bibinfo {author} {\bibfnamefont {S.}~\bibnamefont
  {Tenneti}}, \bibinfo {author} {\bibfnamefont {M.}~\bibnamefont {Mehrabadi}},\
  and\ \bibinfo {author} {\bibfnamefont {S.}~\bibnamefont {Subramaniam}},\
  }\bibfield  {title} {\bibinfo {title} {Stochastic lagrangian model for
  hydrodynamic acceleration of inertial particles in gas--solid suspensions},\
  }\href@noop {} {\bibfield  {journal} {\bibinfo  {journal} {Journal of Fluid
  Mechanics}\ }\textbf {\bibinfo {volume} {788}},\ \bibinfo {pages} {695}
  (\bibinfo {year} {2016})}\BibitemShut {NoStop}%
\bibitem [{\citenamefont {Eaton}\ and\ \citenamefont
  {Fessler}(1994)}]{eaton_fessler_ijmf94}%
  \BibitemOpen
  \bibfield  {author} {\bibinfo {author} {\bibfnamefont {J.~K.}\ \bibnamefont
  {Eaton}}\ and\ \bibinfo {author} {\bibfnamefont {J.~R.}\ \bibnamefont
  {Fessler}},\ }\bibfield  {title} {\bibinfo {title} {{Preferential
  Concentration of Particles by Turbulence}},\ }\href@noop {} {\bibfield
  {journal} {\bibinfo  {journal} {Intl. J. Multiphase Flow}\ }\textbf {\bibinfo
  {volume} {20}},\ \bibinfo {pages} {169} (\bibinfo {year} {1994})}\BibitemShut
  {NoStop}%
\bibitem [{\citenamefont {Garzo}\ \emph {et~al.}(2012)\citenamefont {Garzo},
  \citenamefont {Tenneti}, \citenamefont {Subramaniam},\ and\ \citenamefont
  {Hrenya}}]{vgstsscmh_jfm_2012}%
  \BibitemOpen
  \bibfield  {author} {\bibinfo {author} {\bibfnamefont {V.}~\bibnamefont
  {Garzo}}, \bibinfo {author} {\bibfnamefont {S.}~\bibnamefont {Tenneti}},
  \bibinfo {author} {\bibfnamefont {S.}~\bibnamefont {Subramaniam}},\ and\
  \bibinfo {author} {\bibfnamefont {C.~M.}\ \bibnamefont {Hrenya}},\ }\bibfield
   {title} {\bibinfo {title} {Enskog kinetic theory for monodisperse gas--solid
  flows},\ }\href@noop {} {\bibfield  {journal} {\bibinfo  {journal} {Journal
  of Fluid Mechanics}\ }\textbf {\bibinfo {volume} {712}},\ \bibinfo {pages}
  {129} (\bibinfo {year} {2012})}\BibitemShut {NoStop}%
\bibitem [{\citenamefont {Kong}\ and\ \citenamefont
  {Fox}(2017)}]{kong2017solution}%
  \BibitemOpen
  \bibfield  {author} {\bibinfo {author} {\bibfnamefont {B.}~\bibnamefont
  {Kong}}\ and\ \bibinfo {author} {\bibfnamefont {R.~O.}\ \bibnamefont {Fox}},\
  }\bibfield  {title} {\bibinfo {title} {A solution algorithm for
  fluid--particle flows across all flow regimes},\ }\href@noop {} {\bibfield
  {journal} {\bibinfo  {journal} {Journal of Computational Physics}\ }\textbf
  {\bibinfo {volume} {344}},\ \bibinfo {pages} {575} (\bibinfo {year}
  {2017})}\BibitemShut {NoStop}%
\bibitem [{\citenamefont {Heylmun}\ \emph {et~al.}(2019)\citenamefont
  {Heylmun}, \citenamefont {Kong}, \citenamefont {Passalacqua},\ and\
  \citenamefont {Fox}}]{heylmun2019quadrature}%
  \BibitemOpen
  \bibfield  {author} {\bibinfo {author} {\bibfnamefont {J.~C.}\ \bibnamefont
  {Heylmun}}, \bibinfo {author} {\bibfnamefont {B.}~\bibnamefont {Kong}},
  \bibinfo {author} {\bibfnamefont {A.}~\bibnamefont {Passalacqua}},\ and\
  \bibinfo {author} {\bibfnamefont {R.~O.}\ \bibnamefont {Fox}},\ }\bibfield
  {title} {\bibinfo {title} {A quadrature-based moment method for polydisperse
  bubbly flows},\ }\href@noop {} {\bibfield  {journal} {\bibinfo  {journal}
  {Computer Physics Communications}\ }\textbf {\bibinfo {volume} {244}},\
  \bibinfo {pages} {187} (\bibinfo {year} {2019})}\BibitemShut {NoStop}%
\bibitem [{\citenamefont {Krall}\ and\ \citenamefont
  {Trivelpiece}(1973)}]{krall_trivelpiece}%
  \BibitemOpen
  \bibfield  {author} {\bibinfo {author} {\bibfnamefont {N.~A.}\ \bibnamefont
  {Krall}}\ and\ \bibinfo {author} {\bibfnamefont {A.~W.}\ \bibnamefont
  {Trivelpiece}},\ }\href@noop {} {\emph {\bibinfo {title} {Principles of
  plasma physics}}},\ International series in pure and applied physics\
  (\bibinfo  {publisher} {McGraw-Hill},\ \bibinfo {address} {New York},\
  \bibinfo {year} {1973})\BibitemShut {NoStop}%
\bibitem [{\citenamefont {Stoyan}\ and\ \citenamefont
  {Stoyan}(1995)}]{stoyan_stoyan}%
  \BibitemOpen
  \bibfield  {author} {\bibinfo {author} {\bibfnamefont {D.}~\bibnamefont
  {Stoyan}}\ and\ \bibinfo {author} {\bibfnamefont {H.}~\bibnamefont
  {Stoyan}},\ }\href@noop {} {\emph {\bibinfo {title} {Fractals, Random Shapes
  and Point Fields}}},\ Wiley Series in Probability and Mathematical
  Statistics\ (\bibinfo  {publisher} {John Wiley and Sons},\ \bibinfo {address}
  {New York},\ \bibinfo {year} {1995})\BibitemShut {NoStop}%
\bibitem [{\citenamefont {Stoyan}\ \emph {et~al.}(1995)\citenamefont {Stoyan},
  \citenamefont {Kendall},\ and\ \citenamefont {Mecke}}]{stoyan_kendall_mecke}%
  \BibitemOpen
  \bibfield  {author} {\bibinfo {author} {\bibfnamefont {D.}~\bibnamefont
  {Stoyan}}, \bibinfo {author} {\bibfnamefont {W.~S.}\ \bibnamefont
  {Kendall}},\ and\ \bibinfo {author} {\bibfnamefont {J.}~\bibnamefont
  {Mecke}},\ }\href@noop {} {\emph {\bibinfo {title} {Stochastic Geometry and
  its Applications}}},\ \bibinfo {edition} {2nd}\ ed.,\ Wiley Series in
  Probability and Mathematical Statistics\ (\bibinfo  {publisher} {John Wiley
  and Sons},\ \bibinfo {address} {New York},\ \bibinfo {year}
  {1995})\BibitemShut {NoStop}%
\bibitem [{\citenamefont {Sundaram}\ and\ \citenamefont
  {Collins}(1999)}]{sund_coll_jfm99}%
  \BibitemOpen
  \bibfield  {author} {\bibinfo {author} {\bibfnamefont {S.}~\bibnamefont
  {Sundaram}}\ and\ \bibinfo {author} {\bibfnamefont {L.~R.}\ \bibnamefont
  {Collins}},\ }\bibfield  {title} {\bibinfo {title} {A numerical study of the
  modulation of isotropic turbulence by suspended particles},\ }\href@noop {}
  {\bibfield  {journal} {\bibinfo  {journal} {J. Fluid Mech.}\ }\textbf
  {\bibinfo {volume} {379}},\ \bibinfo {pages} {105} (\bibinfo {year}
  {1999})}\BibitemShut {NoStop}%
\bibitem [{\citenamefont {Markutsya}\ \emph {et~al.}(2012)\citenamefont
  {Markutsya}, \citenamefont {Fox},\ and\ \citenamefont
  {Subramaniam}}]{markutsya2012coarse}%
  \BibitemOpen
  \bibfield  {author} {\bibinfo {author} {\bibfnamefont {S.}~\bibnamefont
  {Markutsya}}, \bibinfo {author} {\bibfnamefont {R.~O.}\ \bibnamefont {Fox}},\
  and\ \bibinfo {author} {\bibfnamefont {S.}~\bibnamefont {Subramaniam}},\
  }\bibfield  {title} {\bibinfo {title} {Coarse-graining approach to infer
  mesoscale interaction potentials from atomistic interactions for aggregating
  systems},\ }\href@noop {} {\bibfield  {journal} {\bibinfo  {journal}
  {Industrial \& engineering chemistry research}\ }\textbf {\bibinfo {volume}
  {51}},\ \bibinfo {pages} {16116} (\bibinfo {year} {2012})}\BibitemShut
  {NoStop}%
\bibitem [{\citenamefont {Rani}\ \emph {et~al.}(2014)\citenamefont {Rani},
  \citenamefont {Dhariwal},\ and\ \citenamefont {Koch}}]{rani2014stochastic}%
  \BibitemOpen
  \bibfield  {author} {\bibinfo {author} {\bibfnamefont {S.~L.}\ \bibnamefont
  {Rani}}, \bibinfo {author} {\bibfnamefont {R.}~\bibnamefont {Dhariwal}},\
  and\ \bibinfo {author} {\bibfnamefont {D.~L.}\ \bibnamefont {Koch}},\
  }\bibfield  {title} {\bibinfo {title} {A stochastic model for the relative
  motion of high stokes number particles in isotropic turbulence},\ }\href@noop
  {} {\bibfield  {journal} {\bibinfo  {journal} {Journal of fluid mechanics}\
  }\textbf {\bibinfo {volume} {756}},\ \bibinfo {pages} {870} (\bibinfo {year}
  {2014})}\BibitemShut {NoStop}%
\bibitem [{\citenamefont {Amsden}\ \emph {et~al.}(1989)\citenamefont {Amsden},
  \citenamefont {O'Rourke},\ and\ \citenamefont {Butler}}]{kiva2}%
  \BibitemOpen
  \bibfield  {author} {\bibinfo {author} {\bibfnamefont {A.~A.}\ \bibnamefont
  {Amsden}}, \bibinfo {author} {\bibfnamefont {P.~J.}\ \bibnamefont
  {O'Rourke}},\ and\ \bibinfo {author} {\bibfnamefont {T.~D.}\ \bibnamefont
  {Butler}},\ }\href@noop {} {\emph {\bibinfo {title} {{KIVA--II: A Computer
  Program for Chemically Reactive Flows with Sprays}}}},\ \bibinfo {type}
  {Tech. Rep.}\ \bibinfo {number} {LA--11560--MS}\ (\bibinfo  {institution}
  {Los Alamos National Laboratory},\ \bibinfo {year} {1989})\BibitemShut
  {NoStop}%
\bibitem [{\citenamefont {Garg}\ \emph {et~al.}(2007)\citenamefont {Garg},
  \citenamefont {Narayanan}, \citenamefont {Lakehal},\ and\ \citenamefont
  {Subramaniam}}]{garg2007accurate}%
  \BibitemOpen
  \bibfield  {author} {\bibinfo {author} {\bibfnamefont {R.}~\bibnamefont
  {Garg}}, \bibinfo {author} {\bibfnamefont {C.}~\bibnamefont {Narayanan}},
  \bibinfo {author} {\bibfnamefont {D.}~\bibnamefont {Lakehal}},\ and\ \bibinfo
  {author} {\bibfnamefont {S.}~\bibnamefont {Subramaniam}},\ }\bibfield
  {title} {\bibinfo {title} {Accurate numerical estimation of interphase
  momentum transfer in lagrangian--eulerian simulations of dispersed two-phase
  flows},\ }\href@noop {} {\bibfield  {journal} {\bibinfo  {journal}
  {International Journal of Multiphase Flow}\ }\textbf {\bibinfo {volume}
  {33}},\ \bibinfo {pages} {1337} (\bibinfo {year} {2007})}\BibitemShut
  {NoStop}%
\bibitem [{\citenamefont {Garg}\ \emph {et~al.}(2009)\citenamefont {Garg},
  \citenamefont {Narayanan},\ and\ \citenamefont
  {Subramaniam}}]{garg2009numerically}%
  \BibitemOpen
  \bibfield  {author} {\bibinfo {author} {\bibfnamefont {R.}~\bibnamefont
  {Garg}}, \bibinfo {author} {\bibfnamefont {C.}~\bibnamefont {Narayanan}},\
  and\ \bibinfo {author} {\bibfnamefont {S.}~\bibnamefont {Subramaniam}},\
  }\bibfield  {title} {\bibinfo {title} {A numerically convergent
  lagrangian--eulerian simulation method for dispersed two-phase flows},\
  }\href@noop {} {\bibfield  {journal} {\bibinfo  {journal} {International
  Journal of Multiphase Flow}\ }\textbf {\bibinfo {volume} {35}},\ \bibinfo
  {pages} {376} (\bibinfo {year} {2009})}\BibitemShut {NoStop}%
\bibitem [{\citenamefont {Subramaniam}(2001)}]{shankar_ddf}%
  \BibitemOpen
  \bibfield  {author} {\bibinfo {author} {\bibfnamefont {S.}~\bibnamefont
  {Subramaniam}},\ }\bibfield  {title} {\bibinfo {title} {{Statistical modeling
  of sprays using the droplet distribution function}},\ }\href@noop {}
  {\bibfield  {journal} {\bibinfo  {journal} {Phys. Fluids}\ }\textbf {\bibinfo
  {volume} {13}},\ \bibinfo {pages} {624} (\bibinfo {year} {2001})}\BibitemShut
  {NoStop}%
\bibitem [{\citenamefont {Drew}\ and\ \citenamefont
  {Passman}(1998)}]{drew_passman}%
  \BibitemOpen
  \bibfield  {author} {\bibinfo {author} {\bibfnamefont {D.~A.}\ \bibnamefont
  {Drew}}\ and\ \bibinfo {author} {\bibfnamefont {S.~L.}\ \bibnamefont
  {Passman}},\ }\href@noop {} {\emph {\bibinfo {title} {{Theory of
  Multicomponent Fluids}}}}\ (\bibinfo  {publisher} {Springer},\ \bibinfo
  {year} {1998})\BibitemShut {NoStop}%
\bibitem [{\citenamefont {Kataoka}\ and\ \citenamefont
  {Serizawa}(1989)}]{kataoka_ijmf}%
  \BibitemOpen
  \bibfield  {author} {\bibinfo {author} {\bibfnamefont {I.}~\bibnamefont
  {Kataoka}}\ and\ \bibinfo {author} {\bibfnamefont {A.}~\bibnamefont
  {Serizawa}},\ }\bibfield  {title} {\bibinfo {title} {Basic equations of
  turbulence in gas--liquid two--phase flow},\ }\href@noop {} {\bibfield
  {journal} {\bibinfo  {journal} {Intl. J. Multiphase Flow}\ }\textbf {\bibinfo
  {volume} {15}},\ \bibinfo {pages} {843} (\bibinfo {year} {1989})}\BibitemShut
  {NoStop}%
\bibitem [{\citenamefont {Jackson}(1997)}]{jackson1997locally}%
  \BibitemOpen
  \bibfield  {author} {\bibinfo {author} {\bibfnamefont {R.}~\bibnamefont
  {Jackson}},\ }\bibfield  {title} {\bibinfo {title} {Locally averaged
  equations of motion for a mixture of identical spherical particles and a
  newtonian fluid},\ }\href@noop {} {\bibfield  {journal} {\bibinfo  {journal}
  {Chemical Engineering Science}\ }\textbf {\bibinfo {volume} {52}},\ \bibinfo
  {pages} {2457} (\bibinfo {year} {1997})}\BibitemShut {NoStop}%
\bibitem [{\citenamefont {Hrenya}\ and\ \citenamefont
  {Sinclair}(1997)}]{Hrenya_aiche97}%
  \BibitemOpen
  \bibfield  {author} {\bibinfo {author} {\bibfnamefont {C.}~\bibnamefont
  {Hrenya}}\ and\ \bibinfo {author} {\bibfnamefont {J.}~\bibnamefont
  {Sinclair}},\ }\bibfield  {title} {\bibinfo {title} {Effects of
  particle-phase turbulence in gas-solid flows},\ }\href@noop {} {\bibfield
  {journal} {\bibinfo  {journal} {AIChE Journal}\ }\textbf {\bibinfo {volume}
  {43}},\ \bibinfo {pages} {853 } (\bibinfo {year} {1997})},\ \bibinfo {note}
  {gas-solid flow;Gas-solid suspensions;Inelastic collisions;Particles
  segregation patterns;}\BibitemShut {NoStop}%
\bibitem [{\citenamefont {Simonin}(1995)}]{simonin_95}%
  \BibitemOpen
  \bibfield  {author} {\bibinfo {author} {\bibfnamefont {O.}~\bibnamefont
  {Simonin}},\ }\bibfield  {title} {\bibinfo {title} {Two--fluid model approach
  for turbulent reactive two--phase flows.},\ }\href@noop {} {\bibfield
  {journal} {\bibinfo  {journal} {Summer school on numerical modelling and
  prediction of dispersed two-phase flows. IMVU, Meserburg, Germany}\ }
  (\bibinfo {year} {1995})}\BibitemShut {NoStop}%
\bibitem [{\citenamefont {Ahmadi}\ and\ \citenamefont
  {Ma}(1990)}]{ahmadi_ma_1990a}%
  \BibitemOpen
  \bibfield  {author} {\bibinfo {author} {\bibfnamefont {G.}~\bibnamefont
  {Ahmadi}}\ and\ \bibinfo {author} {\bibfnamefont {D.}~\bibnamefont {Ma}},\
  }\bibfield  {title} {\bibinfo {title} {A thermodynamical formulation for
  dispersed multiphase turbulent flows--1: {Basic} theory},\ }\href@noop {}
  {\bibfield  {journal} {\bibinfo  {journal} {Intl. J. Multiphase Flow}\
  }\textbf {\bibinfo {volume} {16}},\ \bibinfo {pages} {323} (\bibinfo {year}
  {1990})}\BibitemShut {NoStop}%
\bibitem [{\citenamefont {Xu}\ and\ \citenamefont
  {Subramaniam}(2006{\natexlab{a}})}]{xu2006multiscale}%
  \BibitemOpen
  \bibfield  {author} {\bibinfo {author} {\bibfnamefont {Y.}~\bibnamefont
  {Xu}}\ and\ \bibinfo {author} {\bibfnamefont {S.}~\bibnamefont
  {Subramaniam}},\ }\bibfield  {title} {\bibinfo {title} {A multiscale model
  for dilute turbulent gas-particle flows based on the equilibration of energy
  concept},\ }\href@noop {} {\bibfield  {journal} {\bibinfo  {journal} {Physics
  of Fluids}\ }\textbf {\bibinfo {volume} {18}},\ \bibinfo {pages} {033301}
  (\bibinfo {year} {2006}{\natexlab{a}})}\BibitemShut {NoStop}%
\bibitem [{\citenamefont {Fox}(2014)}]{fox2014multiphase}%
  \BibitemOpen
  \bibfield  {author} {\bibinfo {author} {\bibfnamefont {R.~O.}\ \bibnamefont
  {Fox}},\ }\bibfield  {title} {\bibinfo {title} {On multiphase turbulence
  models for collisional fluid-particle flows},\ }\href@noop {} {\bibfield
  {journal} {\bibinfo  {journal} {Journal of Fluid Mechanics}\ }\textbf
  {\bibinfo {volume} {742}},\ \bibinfo {pages} {368} (\bibinfo {year}
  {2014})}\BibitemShut {NoStop}%
\bibitem [{\citenamefont {Xu}\ and\ \citenamefont
  {Subramaniam}(2006{\natexlab{b}})}]{xu_ss_pf06}%
  \BibitemOpen
  \bibfield  {author} {\bibinfo {author} {\bibfnamefont {Y.}~\bibnamefont
  {Xu}}\ and\ \bibinfo {author} {\bibfnamefont {S.}~\bibnamefont
  {Subramaniam}},\ }\bibfield  {title} {\bibinfo {title} {A multiscale model
  for dilute turbulent gas-particle flows based on the equilibration of energy
  concept},\ }\href@noop {} {\bibfield  {journal} {\bibinfo  {journal} {{Phys.
  Fluids}}\ }\textbf {\bibinfo {volume} {18}},\ \bibinfo {pages} {033301}
  (\bibinfo {year} {2006}{\natexlab{b}})}\BibitemShut {NoStop}%
\bibitem [{\citenamefont {Sundaram}\ and\ \citenamefont
  {Collins}(1994{\natexlab{a}})}]{sund_coll_ijmf1}%
  \BibitemOpen
  \bibfield  {author} {\bibinfo {author} {\bibfnamefont {S.}~\bibnamefont
  {Sundaram}}\ and\ \bibinfo {author} {\bibfnamefont {L.~R.}\ \bibnamefont
  {Collins}},\ }\bibfield  {title} {\bibinfo {title} {{Spectrum of density
  fluctuations in a particle--fluid system---I. Monodisperse spheres}},\
  }\href@noop {} {\bibfield  {journal} {\bibinfo  {journal} {Intl. J.
  Multiphase Flow}\ }\textbf {\bibinfo {volume} {20}},\ \bibinfo {pages} {1021}
  (\bibinfo {year} {1994}{\natexlab{a}})}\BibitemShut {NoStop}%
\bibitem [{\citenamefont {Torquato}\ and\ \citenamefont
  {Stell}(1982)}]{torquato1982microstructure}%
  \BibitemOpen
  \bibfield  {author} {\bibinfo {author} {\bibfnamefont {S.}~\bibnamefont
  {Torquato}}\ and\ \bibinfo {author} {\bibfnamefont {G.}~\bibnamefont
  {Stell}},\ }\bibfield  {title} {\bibinfo {title} {Microstructure of two-phase
  random media. i. the n-point probability functions},\ }\href@noop {}
  {\bibfield  {journal} {\bibinfo  {journal} {The Journal of Chemical Physics}\
  }\textbf {\bibinfo {volume} {77}},\ \bibinfo {pages} {2071} (\bibinfo {year}
  {1982})}\BibitemShut {NoStop}%
\bibitem [{\citenamefont {Sundaram}\ and\ \citenamefont
  {Collins}(1994{\natexlab{b}})}]{sund_coll_ijmf2}%
  \BibitemOpen
  \bibfield  {author} {\bibinfo {author} {\bibfnamefont {S.}~\bibnamefont
  {Sundaram}}\ and\ \bibinfo {author} {\bibfnamefont {L.~R.}\ \bibnamefont
  {Collins}},\ }\bibfield  {title} {\bibinfo {title} {Spectrum of density
  fluctuations in a particle--fluid system---ii. polydisperse spheres},\
  }\href@noop {} {\bibfield  {journal} {\bibinfo  {journal} {Intl. J.
  Multiphase Flow}\ }\textbf {\bibinfo {volume} {20}},\ \bibinfo {pages} {1039}
  (\bibinfo {year} {1994}{\natexlab{b}})}\BibitemShut {NoStop}%
\bibitem [{\citenamefont {Lu}\ and\ \citenamefont
  {Torquato}(1990)}]{lu1990local}%
  \BibitemOpen
  \bibfield  {author} {\bibinfo {author} {\bibfnamefont {B.}~\bibnamefont
  {Lu}}\ and\ \bibinfo {author} {\bibfnamefont {S.}~\bibnamefont {Torquato}},\
  }\bibfield  {title} {\bibinfo {title} {Local volume fraction fluctuations in
  heterogeneous media},\ }\href@noop {} {\bibfield  {journal} {\bibinfo
  {journal} {The Journal of chemical physics}\ }\textbf {\bibinfo {volume}
  {93}},\ \bibinfo {pages} {3452} (\bibinfo {year} {1990})}\BibitemShut
  {NoStop}%
\bibitem [{\citenamefont {Murphy}(2017)}]{murphy2017analysis}%
  \BibitemOpen
  \bibfield  {author} {\bibinfo {author} {\bibfnamefont {E.}~\bibnamefont
  {Murphy}},\ }\emph {\bibinfo {title} {Analysis and Modeling of Structure
  Formation in Granular and Fluid-Solid Flows}},\ \href@noop {} {Ph.D.
  thesis},\ \bibinfo  {school} {Iowa State University} (\bibinfo {year}
  {2017})\BibitemShut {NoStop}%
\bibitem [{\citenamefont {Capecelatro}\ \emph {et~al.}(2014)\citenamefont
  {Capecelatro}, \citenamefont {Desjardins},\ and\ \citenamefont
  {Fox}}]{capecelatro2014numerical}%
  \BibitemOpen
  \bibfield  {author} {\bibinfo {author} {\bibfnamefont {J.}~\bibnamefont
  {Capecelatro}}, \bibinfo {author} {\bibfnamefont {O.}~\bibnamefont
  {Desjardins}},\ and\ \bibinfo {author} {\bibfnamefont {R.~O.}\ \bibnamefont
  {Fox}},\ }\bibfield  {title} {\bibinfo {title} {Numerical study of
  collisional particle dynamics in cluster-induced turbulence},\ }\href@noop {}
  {\bibfield  {journal} {\bibinfo  {journal} {Journal of Fluid Mechanics}\
  }\textbf {\bibinfo {volume} {747}} (\bibinfo {year} {2014})}\BibitemShut
  {NoStop}%
\bibitem [{\citenamefont {Syamlal}\ \emph {et~al.}(1993)\citenamefont
  {Syamlal}, \citenamefont {Rogers},\ and\ \citenamefont
  {O'Brien}}]{mfix_theory}%
  \BibitemOpen
  \bibfield  {author} {\bibinfo {author} {\bibfnamefont {M.}~\bibnamefont
  {Syamlal}}, \bibinfo {author} {\bibfnamefont {W.}~\bibnamefont {Rogers}},\
  and\ \bibinfo {author} {\bibfnamefont {T.~J.}\ \bibnamefont {O'Brien}},\
  }\href@noop {} {\emph {\bibinfo {title} {{MFIX Documentation: Theory
  Guide}}}},\ \bibinfo {type} {Tech. Rep.}\ \bibinfo {number}
  {{DOE/METC-95/1013, NTIS/DE95000031}}\ (\bibinfo  {institution} {National
  Energy Technology Laboratory, Department of Energy},\ \bibinfo {year}
  {1993})\ \bibinfo {note} {see also URL http://www.mfix.org}\BibitemShut
  {NoStop}%
\bibitem [{\citenamefont {Kumaran}(2003)}]{kumaran2003stability}%
  \BibitemOpen
  \bibfield  {author} {\bibinfo {author} {\bibfnamefont {V.}~\bibnamefont
  {Kumaran}},\ }\bibfield  {title} {\bibinfo {title} {Stability of a sheared
  particle suspension},\ }\href@noop {} {\bibfield  {journal} {\bibinfo
  {journal} {Physics of Fluids}\ }\textbf {\bibinfo {volume} {15}},\ \bibinfo
  {pages} {3625} (\bibinfo {year} {2003})}\BibitemShut {NoStop}%
\bibitem [{\citenamefont {Mehrabadi}\ \emph {et~al.}(2015)\citenamefont
  {Mehrabadi}, \citenamefont {Tenneti}, \citenamefont {Garg},\ and\
  \citenamefont {Subramaniam}}]{mehrabadi2015pseudo}%
  \BibitemOpen
  \bibfield  {author} {\bibinfo {author} {\bibfnamefont {M.}~\bibnamefont
  {Mehrabadi}}, \bibinfo {author} {\bibfnamefont {S.}~\bibnamefont {Tenneti}},
  \bibinfo {author} {\bibfnamefont {R.}~\bibnamefont {Garg}},\ and\ \bibinfo
  {author} {\bibfnamefont {S.}~\bibnamefont {Subramaniam}},\ }\bibfield
  {title} {\bibinfo {title} {Pseudo-turbulent gas-phase velocity fluctuations
  in homogeneous gas-solid flow: fixed particle assemblies and freely evolving
  suspensions},\ }\href@noop {} {\bibfield  {journal} {\bibinfo  {journal}
  {Journal of Fluid Mechanics}\ }\textbf {\bibinfo {volume} {770}},\ \bibinfo
  {pages} {210} (\bibinfo {year} {2015})}\BibitemShut {NoStop}%
\bibitem [{\citenamefont {Sun}\ \emph {et~al.}(2016)\citenamefont {Sun},
  \citenamefont {Tenneti}, \citenamefont {Subramaniam},\ and\ \citenamefont
  {Koch}}]{sun2016pseudo}%
  \BibitemOpen
  \bibfield  {author} {\bibinfo {author} {\bibfnamefont {B.}~\bibnamefont
  {Sun}}, \bibinfo {author} {\bibfnamefont {S.}~\bibnamefont {Tenneti}},
  \bibinfo {author} {\bibfnamefont {S.}~\bibnamefont {Subramaniam}},\ and\
  \bibinfo {author} {\bibfnamefont {D.~L.}\ \bibnamefont {Koch}},\ }\bibfield
  {title} {\bibinfo {title} {Pseudo-turbulent heat flux and average gas--phase
  conduction during gas--solid heat transfer: flow past random fixed particle
  assemblies},\ }\href@noop {} {\bibfield  {journal} {\bibinfo  {journal}
  {Journal of Fluid Mechanics}\ }\textbf {\bibinfo {volume} {798}},\ \bibinfo
  {pages} {299} (\bibinfo {year} {2016})}\BibitemShut {NoStop}%
\bibitem [{\citenamefont {Akiki}\ \emph {et~al.}(2016)\citenamefont {Akiki},
  \citenamefont {Jackson},\ and\ \citenamefont {Balachandar}}]{akiki2016force}%
  \BibitemOpen
  \bibfield  {author} {\bibinfo {author} {\bibfnamefont {G.}~\bibnamefont
  {Akiki}}, \bibinfo {author} {\bibfnamefont {T.}~\bibnamefont {Jackson}},\
  and\ \bibinfo {author} {\bibfnamefont {S.}~\bibnamefont {Balachandar}},\
  }\bibfield  {title} {\bibinfo {title} {Force variation within arrays of
  monodisperse spherical particles},\ }\href@noop {} {\bibfield  {journal}
  {\bibinfo  {journal} {Physical Review Fluids}\ }\textbf {\bibinfo {volume}
  {1}},\ \bibinfo {pages} {044202} (\bibinfo {year} {2016})}\BibitemShut
  {NoStop}%
\bibitem [{\citenamefont {Hu}\ \emph {et~al.}(2001)\citenamefont {Hu},
  \citenamefont {Patankar},\ and\ \citenamefont {Zhu}}]{hu_jcp2001}%
  \BibitemOpen
  \bibfield  {author} {\bibinfo {author} {\bibfnamefont {H.~H.}\ \bibnamefont
  {Hu}}, \bibinfo {author} {\bibfnamefont {N.~A.}\ \bibnamefont {Patankar}},\
  and\ \bibinfo {author} {\bibfnamefont {M.~Y.}\ \bibnamefont {Zhu}},\
  }\bibfield  {title} {\bibinfo {title} {Direct numerical simulations of
  fluid-solid systems using the arbitrary {Lagrangian}--{Eulerian} technique},\
  }\href@noop {} {\bibfield  {journal} {\bibinfo  {journal} {J. Comp. Phys.}\
  }\textbf {\bibinfo {volume} {169}},\ \bibinfo {pages} {427} (\bibinfo {year}
  {2001})}\BibitemShut {NoStop}%
\bibitem [{\citenamefont {Nomura}\ and\ \citenamefont
  {Hughes}(1992)}]{nomura_hughes}%
  \BibitemOpen
  \bibfield  {author} {\bibinfo {author} {\bibfnamefont {T.}~\bibnamefont
  {Nomura}}\ and\ \bibinfo {author} {\bibfnamefont {T.~J.~R.}\ \bibnamefont
  {Hughes}},\ }\bibfield  {title} {\bibinfo {title} {An arbitrary
  {Lagrangian}--{Eulerian} finite element method for interaction of fluid and a
  rigid body},\ }\href@noop {} {\bibfield  {journal} {\bibinfo  {journal}
  {Comput. Meth. Appl. Mech. Eng.}\ }\textbf {\bibinfo {volume} {95}},\
  \bibinfo {pages} {115} (\bibinfo {year} {1992})}\BibitemShut {NoStop}%
\bibitem [{\citenamefont {Bagchi}\ and\ \citenamefont
  {Balachandar}(2003)}]{bagchi_balachandar_03}%
  \BibitemOpen
  \bibfield  {author} {\bibinfo {author} {\bibfnamefont {P.}~\bibnamefont
  {Bagchi}}\ and\ \bibinfo {author} {\bibfnamefont {S.}~\bibnamefont
  {Balachandar}},\ }\bibfield  {title} {\bibinfo {title} {Effect of turbulence
  on the drag and lift of a particle},\ }\href@noop {} {\bibfield  {journal}
  {\bibinfo  {journal} {Phys. Fluids}\ }\textbf {\bibinfo {volume} {15}},\
  \bibinfo {pages} {3496} (\bibinfo {year} {2003})}\BibitemShut {NoStop}%
\bibitem [{\citenamefont {Bagchi}\ and\ \citenamefont
  {Balachandar}(2004)}]{bagchi_balachandar_04}%
  \BibitemOpen
  \bibfield  {author} {\bibinfo {author} {\bibfnamefont {P.}~\bibnamefont
  {Bagchi}}\ and\ \bibinfo {author} {\bibfnamefont {S.}~\bibnamefont
  {Balachandar}},\ }\bibfield  {title} {\bibinfo {title} {Response of the wake
  of an isolated particle to an isotropic turbulent flow},\ }\href@noop {}
  {\bibfield  {journal} {\bibinfo  {journal} {J. Fluid Mech.}\ }\textbf
  {\bibinfo {volume} {518}},\ \bibinfo {pages} {95} (\bibinfo {year}
  {2004})}\BibitemShut {NoStop}%
\bibitem [{\citenamefont {Burton}\ and\ \citenamefont
  {Eaton}(2005)}]{burton_eaton}%
  \BibitemOpen
  \bibfield  {author} {\bibinfo {author} {\bibfnamefont {T.~M.}\ \bibnamefont
  {Burton}}\ and\ \bibinfo {author} {\bibfnamefont {J.~K.}\ \bibnamefont
  {Eaton}},\ }\bibfield  {title} {\bibinfo {title} {Fully resolved simulations
  of particle-turbulence interaction},\ }\href@noop {} {\bibfield  {journal}
  {\bibinfo  {journal} {J. Fluid Mech.}\ }\textbf {\bibinfo {volume} {545}},\
  \bibinfo {pages} {67} (\bibinfo {year} {2005})}\BibitemShut {NoStop}%
\bibitem [{\citenamefont {Patankar}\ \emph {et~al.}(2000)\citenamefont
  {Patankar}, \citenamefont {Singh}, \citenamefont {Joseph}, \citenamefont
  {Glowinski},\ and\ \citenamefont {Pan}}]{patankaretal_ijmf2000}%
  \BibitemOpen
  \bibfield  {author} {\bibinfo {author} {\bibfnamefont {N.}~\bibnamefont
  {Patankar}}, \bibinfo {author} {\bibfnamefont {P.}~\bibnamefont {Singh}},
  \bibinfo {author} {\bibfnamefont {D.~D.}\ \bibnamefont {Joseph}}, \bibinfo
  {author} {\bibfnamefont {R.}~\bibnamefont {Glowinski}},\ and\ \bibinfo
  {author} {\bibfnamefont {T.~W.}\ \bibnamefont {Pan}},\ }\bibfield  {title}
  {\bibinfo {title} {{A new formulation of the distributed Lagrange
  multipliers/fictitious domain method for particulate flow}},\ }\href@noop {}
  {\bibfield  {journal} {\bibinfo  {journal} {Intl. J. Multiphase Flow}\
  }\textbf {\bibinfo {volume} {26}},\ \bibinfo {pages} {1509} (\bibinfo {year}
  {2000})}\BibitemShut {NoStop}%
\bibitem [{\citenamefont {Glowinski}\ \emph {et~al.}(2001)\citenamefont
  {Glowinski}, \citenamefont {Pan}, \citenamefont {Hesla}, \citenamefont
  {Joseph},\ and\ \citenamefont {P\'{e}riaux}}]{joseph_jcp2001}%
  \BibitemOpen
  \bibfield  {author} {\bibinfo {author} {\bibfnamefont {R.}~\bibnamefont
  {Glowinski}}, \bibinfo {author} {\bibfnamefont {T.~W.}\ \bibnamefont {Pan}},
  \bibinfo {author} {\bibfnamefont {T.~I.}\ \bibnamefont {Hesla}}, \bibinfo
  {author} {\bibfnamefont {D.~D.}\ \bibnamefont {Joseph}},\ and\ \bibinfo
  {author} {\bibfnamefont {J.}~\bibnamefont {P\'{e}riaux}},\ }\bibfield
  {title} {\bibinfo {title} {A fictitious domain approach to the direct
  numerical simulation of incompressible viscous flow past moving rigid bodies:
  Application to particulate flow},\ }\href@noop {} {\bibfield  {journal}
  {\bibinfo  {journal} {J. Comp. Phys.}\ }\textbf {\bibinfo {volume} {169}},\
  \bibinfo {pages} {363} (\bibinfo {year} {2001})}\BibitemShut {NoStop}%
\bibitem [{\citenamefont {Ladd}\ and\ \citenamefont
  {Verberg}(2001)}]{ladd_verberg_lbm}%
  \BibitemOpen
  \bibfield  {author} {\bibinfo {author} {\bibfnamefont {A.~J.~C.}\
  \bibnamefont {Ladd}}\ and\ \bibinfo {author} {\bibfnamefont {R.}~\bibnamefont
  {Verberg}},\ }\bibfield  {title} {\bibinfo {title} {Lattice--{Boltzmann}
  simulations of particle-fluid suspensions},\ }\href@noop {} {\bibfield
  {journal} {\bibinfo  {journal} {Journal of Statistical Physics}\ }\textbf
  {\bibinfo {volume} {104}},\ \bibinfo {pages} {1191} (\bibinfo {year}
  {2001})}\BibitemShut {NoStop}%
\bibitem [{\citenamefont {Sharma}\ and\ \citenamefont
  {Patankar}(2005)}]{sharma_patankar_05}%
  \BibitemOpen
  \bibfield  {author} {\bibinfo {author} {\bibfnamefont {N.}~\bibnamefont
  {Sharma}}\ and\ \bibinfo {author} {\bibfnamefont {N.}~\bibnamefont
  {Patankar}},\ }\bibfield  {title} {\bibinfo {title} {A fast computation
  technique for the direct numerical simulation of rigid particulate flows},\
  }\href@noop {} {\bibfield  {journal} {\bibinfo  {journal} {J. Comp. Phys.}\
  }\textbf {\bibinfo {volume} {205}},\ \bibinfo {pages} {439} (\bibinfo {year}
  {2005})}\BibitemShut {NoStop}%
\bibitem [{\citenamefont {Apte}\ \emph {et~al.}(2009)\citenamefont {Apte},
  \citenamefont {Martin},\ and\ \citenamefont {Patankar}}]{apte_etal}%
  \BibitemOpen
  \bibfield  {author} {\bibinfo {author} {\bibfnamefont {S.}~\bibnamefont
  {Apte}}, \bibinfo {author} {\bibfnamefont {M.}~\bibnamefont {Martin}},\ and\
  \bibinfo {author} {\bibfnamefont {N.}~\bibnamefont {Patankar}},\ }\bibfield
  {title} {\bibinfo {title} {A numerical method for fully resolved simulation
  ({FRS}) of rigid particle--flow interactions in complex flows},\ }\href@noop
  {} {\bibfield  {journal} {\bibinfo  {journal} {J. Comp. Phys.}\ }\textbf
  {\bibinfo {volume} {228}},\ \bibinfo {pages} {2712} (\bibinfo {year}
  {2009})}\BibitemShut {NoStop}%
\bibitem [{\citenamefont {Peskin}(1981)}]{peskin_81}%
  \BibitemOpen
  \bibfield  {author} {\bibinfo {author} {\bibfnamefont {C.~S.}\ \bibnamefont
  {Peskin}},\ }\bibfield  {title} {\bibinfo {title} {The fluid dynamics of
  heart valves: experimental, theoretical, and computational methods},\
  }\href@noop {} {\bibfield  {journal} {\bibinfo  {journal} {Annu. Rev. Fluid
  Mech.}\ }\textbf {\bibinfo {volume} {14}},\ \bibinfo {pages} {235} (\bibinfo
  {year} {1981})}\BibitemShut {NoStop}%
\bibitem [{\citenamefont {Mohd-Yusof}(1996)}]{jamalphd}%
  \BibitemOpen
  \bibfield  {author} {\bibinfo {author} {\bibfnamefont {J.}~\bibnamefont
  {Mohd-Yusof}},\ }\emph {\bibinfo {title} {Interaction of Massive Particles
  with Turbulence}},\ \href@noop {} {Ph.D. thesis},\ \bibinfo  {school}
  {Cornell University} (\bibinfo {year} {1996})\BibitemShut {NoStop}%
\bibitem [{\citenamefont {Taira}\ and\ \citenamefont
  {Colonius}(2007)}]{taira_col_ibm_jcp2007}%
  \BibitemOpen
  \bibfield  {author} {\bibinfo {author} {\bibfnamefont {K.}~\bibnamefont
  {Taira}}\ and\ \bibinfo {author} {\bibfnamefont {T.}~\bibnamefont
  {Colonius}},\ }\bibfield  {title} {\bibinfo {title} {The immersed boundary
  method: A projection approach},\ }\href
  {https://doi.org/10.1016/j.jcp.2007.03.005} {\bibfield  {journal} {\bibinfo
  {journal} {Journal of Computational Physics}\ }\textbf {\bibinfo {volume}
  {225}},\ \bibinfo {pages} {2118 } (\bibinfo {year} {2007})}\BibitemShut
  {NoStop}%
\bibitem [{\citenamefont {Oguz}\ and\ \citenamefont
  {Prosperetti}(2001)}]{prosperetti_oguz_jcp01}%
  \BibitemOpen
  \bibfield  {author} {\bibinfo {author} {\bibfnamefont {H.}~\bibnamefont
  {Oguz}}\ and\ \bibinfo {author} {\bibfnamefont {A.}~\bibnamefont
  {Prosperetti}},\ }\bibfield  {title} {\bibinfo {title} {{PHYSALIS}: A new
  o({N}) method for the numerical simulation of disperse systems. {Part I}:
  {Potential} flow of spheres},\ }\href@noop {} {\bibfield  {journal} {\bibinfo
   {journal} {J. Comp. Phys.}\ }\textbf {\bibinfo {volume} {167}},\ \bibinfo
  {pages} {196} (\bibinfo {year} {2001})}\BibitemShut {NoStop}%
\bibitem [{\citenamefont {Uhlmann}(2005)}]{Uhlmann:2005aa}%
  \BibitemOpen
  \bibfield  {author} {\bibinfo {author} {\bibfnamefont {M.}~\bibnamefont
  {Uhlmann}},\ }\bibfield  {title} {\bibinfo {title} {An immersed boundary
  method with direct forcing for the simulation of particulate flows},\ }\href
  {https://doi.org/DOI 10.1016/j.jcp.2005.03.017} {\bibfield  {journal}
  {\bibinfo  {journal} {J. Comp. Phys.}\ }\textbf {\bibinfo {volume} {209}},\
  \bibinfo {pages} {448} (\bibinfo {year} {2005})}\BibitemShut {NoStop}%
\bibitem [{\citenamefont {Garg}(2009)}]{rgarg_thesis}%
  \BibitemOpen
  \bibfield  {author} {\bibinfo {author} {\bibfnamefont {R.}~\bibnamefont
  {Garg}},\ }\emph {\bibinfo {title} {Modeling and simulation of two-phase
  flows}},\ \href@noop {} {Ph.D. thesis},\ \bibinfo  {school} {Iowa State
  University} (\bibinfo {year} {2009})\BibitemShut {NoStop}%
\bibitem [{\citenamefont {Kim}\ and\ \citenamefont {Choi}(2006)}]{kim_choi_06}%
  \BibitemOpen
  \bibfield  {author} {\bibinfo {author} {\bibfnamefont {D.}~\bibnamefont
  {Kim}}\ and\ \bibinfo {author} {\bibfnamefont {H.}~\bibnamefont {Choi}},\
  }\bibfield  {title} {\bibinfo {title} {Immersed boundary method for flow
  around an arbitrarily moving body},\ }\href@noop {} {\bibfield  {journal}
  {\bibinfo  {journal} {J. Comp. Phys.}\ }\textbf {\bibinfo {volume} {21}},\
  \bibinfo {pages} {662} (\bibinfo {year} {2006})}\BibitemShut {NoStop}%
\bibitem [{\citenamefont {Lucci}\ \emph {et~al.}(2010)\citenamefont {Lucci},
  \citenamefont {Ferrante},\ and\ \citenamefont {Elgobashi}}]{lucci_etal_2010}%
  \BibitemOpen
  \bibfield  {author} {\bibinfo {author} {\bibfnamefont {F.}~\bibnamefont
  {Lucci}}, \bibinfo {author} {\bibfnamefont {A.}~\bibnamefont {Ferrante}},\
  and\ \bibinfo {author} {\bibfnamefont {S.}~\bibnamefont {Elgobashi}},\
  }\bibfield  {title} {\bibinfo {title} {Modulation of isotropic turbulence by
  particles of {Taylor} length-scale size},\ }\href@noop {} {\bibfield
  {journal} {\bibinfo  {journal} {J. Fluid Mech.}\ }\textbf {\bibinfo {volume}
  {650}} (\bibinfo {year} {2010})}\BibitemShut {NoStop}%
\bibitem [{\citenamefont {Scardovelli}\ and\ \citenamefont
  {Zaleski}(1999)}]{zaleski_whole_domain}%
  \BibitemOpen
  \bibfield  {author} {\bibinfo {author} {\bibfnamefont {R.}~\bibnamefont
  {Scardovelli}}\ and\ \bibinfo {author} {\bibfnamefont {S.}~\bibnamefont
  {Zaleski}},\ }\bibfield  {title} {\bibinfo {title} {Direct numerical
  simulation of free--surface and interfacial flow},\ }\href@noop {} {\bibfield
   {journal} {\bibinfo  {journal} {Annu. Rev. Fluid Mech.}\ }\textbf {\bibinfo
  {volume} {31}},\ \bibinfo {pages} {567} (\bibinfo {year} {1999})}\BibitemShut
  {NoStop}%
\bibitem [{\citenamefont {Wylie}\ \emph {et~al.}(2003)\citenamefont {Wylie},
  \citenamefont {Koch},\ and\ \citenamefont {Ladd}}]{wylie_koch_jfm_03}%
  \BibitemOpen
  \bibfield  {author} {\bibinfo {author} {\bibfnamefont {J.~J.}\ \bibnamefont
  {Wylie}}, \bibinfo {author} {\bibfnamefont {D.~L.}\ \bibnamefont {Koch}},\
  and\ \bibinfo {author} {\bibfnamefont {A.~J.}\ \bibnamefont {Ladd}},\
  }\bibfield  {title} {\bibinfo {title} {Rheology of suspensions with high
  particle inertia and moderate fluid inertia},\ }\href@noop {} {\bibfield
  {journal} {\bibinfo  {journal} {Journal of Fluid Mechanics}\ }\textbf
  {\bibinfo {volume} {480}},\ \bibinfo {pages} {95 } (\bibinfo {year}
  {2003})}\BibitemShut {NoStop}%
\bibitem [{\citenamefont {Tenneti}\ \emph {et~al.}(2010)\citenamefont
  {Tenneti}, \citenamefont {Garg}, \citenamefont {Hrenya}, \citenamefont
  {Fox},\ and\ \citenamefont {Subramaniam}}]{tenneti_pt_2010}%
  \BibitemOpen
  \bibfield  {author} {\bibinfo {author} {\bibfnamefont {S.}~\bibnamefont
  {Tenneti}}, \bibinfo {author} {\bibfnamefont {R.}~\bibnamefont {Garg}},
  \bibinfo {author} {\bibfnamefont {C.~M.}\ \bibnamefont {Hrenya}}, \bibinfo
  {author} {\bibfnamefont {R.~O.}\ \bibnamefont {Fox}},\ and\ \bibinfo {author}
  {\bibfnamefont {S.}~\bibnamefont {Subramaniam}},\ }\bibfield  {title}
  {\bibinfo {title} {Direct numerical simulation of gas--solid suspensions at
  moderate {Reynolds} number: {Quantifying} the coupling between hydrodynamic
  forces and particle velocity fluctuations},\ }\href@noop {} {\bibfield
  {journal} {\bibinfo  {journal} {Powder Technology}\ }\textbf {\bibinfo
  {volume} {203}},\ \bibinfo {pages} {57} (\bibinfo {year} {2010})}\BibitemShut
  {NoStop}%
\bibitem [{\citenamefont {Kriebitzsch}\ \emph {et~al.}(2013)\citenamefont
  {Kriebitzsch}, \citenamefont {van~der Hoef},\ and\ \citenamefont
  {Kuipers}}]{kriebitzsch_2013}%
  \BibitemOpen
  \bibfield  {author} {\bibinfo {author} {\bibfnamefont {S.}~\bibnamefont
  {Kriebitzsch}}, \bibinfo {author} {\bibfnamefont {M.}~\bibnamefont {van~der
  Hoef}},\ and\ \bibinfo {author} {\bibfnamefont {J.}~\bibnamefont {Kuipers}},\
  }\bibfield  {title} {\bibinfo {title} {Fully resolved simulation of a
  gas-fluidized bed: A critical test of {DEM} models},\ }\href
  {https://doi.org/10.1016/j.ces.2012.12.038} {\bibfield  {journal} {\bibinfo
  {journal} {Chem. Eng. Sci.}\ }\textbf {\bibinfo {volume} {91}},\ \bibinfo
  {pages} {1} (\bibinfo {year} {2013})}\BibitemShut {NoStop}%
\bibitem [{\citenamefont {Zhou}\ \emph {et~al.}(2014)\citenamefont {Zhou},
  \citenamefont {Xiong}, \citenamefont {Wang}, \citenamefont {Wang},
  \citenamefont {Ren},\ and\ \citenamefont {Ge}}]{zhou_2014}%
  \BibitemOpen
  \bibfield  {author} {\bibinfo {author} {\bibfnamefont {G.}~\bibnamefont
  {Zhou}}, \bibinfo {author} {\bibfnamefont {Q.}~\bibnamefont {Xiong}},
  \bibinfo {author} {\bibfnamefont {L.}~\bibnamefont {Wang}}, \bibinfo {author}
  {\bibfnamefont {X.}~\bibnamefont {Wang}}, \bibinfo {author} {\bibfnamefont
  {X.}~\bibnamefont {Ren}},\ and\ \bibinfo {author} {\bibfnamefont
  {W.}~\bibnamefont {Ge}},\ }\bibfield  {title} {\bibinfo {title}
  {Structure-dependent drag in gas--solid flows studied with direct numerical
  simulation},\ }\href {https://doi.org/10.1016/j.ces.2014.04.025} {\bibfield
  {journal} {\bibinfo  {journal} {Chem. Eng. Sci.}\ }\textbf {\bibinfo {volume}
  {116}},\ \bibinfo {pages} {9} (\bibinfo {year} {2014})}\BibitemShut {NoStop}%
\bibitem [{\citenamefont {Luo}\ \emph {et~al.}(2016)\citenamefont {Luo},
  \citenamefont {Tan}, \citenamefont {Wang},\ and\ \citenamefont
  {Fan}}]{luo_2016}%
  \BibitemOpen
  \bibfield  {author} {\bibinfo {author} {\bibfnamefont {K.}~\bibnamefont
  {Luo}}, \bibinfo {author} {\bibfnamefont {J.}~\bibnamefont {Tan}}, \bibinfo
  {author} {\bibfnamefont {Z.}~\bibnamefont {Wang}},\ and\ \bibinfo {author}
  {\bibfnamefont {J.}~\bibnamefont {Fan}},\ }\bibfield  {title} {\bibinfo
  {title} {Particle-resolved direct numerical simulation of gas--solid dynamics
  in experimental fluidized beds},\ }\href {https://doi.org/10.1002/aic.15186}
  {\bibfield  {journal} {\bibinfo  {journal} {AIChEJ.}\ }\textbf {\bibinfo
  {volume} {62}},\ \bibinfo {pages} {1917} (\bibinfo {year}
  {2016})}\BibitemShut {NoStop}%
\bibitem [{\citenamefont {Tang}\ \emph {et~al.}(2016)\citenamefont {Tang},
  \citenamefont {Peters},\ and\ \citenamefont {Kuipers}}]{tang_free_2016}%
  \BibitemOpen
  \bibfield  {author} {\bibinfo {author} {\bibfnamefont {Y.}~\bibnamefont
  {Tang}}, \bibinfo {author} {\bibfnamefont {E.~A. J.~F.}\ \bibnamefont
  {Peters}},\ and\ \bibinfo {author} {\bibfnamefont {J.~A.~M.}\ \bibnamefont
  {Kuipers}},\ }\bibfield  {title} {\bibinfo {title} {Direct numerical
  simulations of dynamic gas--solid suspensions},\ }\href
  {https://doi.org/10.1002/aic.15197} {\bibfield  {journal} {\bibinfo
  {journal} {AIChEJ.}\ }\textbf {\bibinfo {volume} {62}},\ \bibinfo {pages}
  {1958} (\bibinfo {year} {2016})}\BibitemShut {NoStop}%
\bibitem [{\citenamefont {Rubinstein}\ \emph {et~al.}(2016)\citenamefont
  {Rubinstein}, \citenamefont {Derksen},\ and\ \citenamefont
  {Sundaresan}}]{rubinstein_2016}%
  \BibitemOpen
  \bibfield  {author} {\bibinfo {author} {\bibfnamefont {G.~J.}\ \bibnamefont
  {Rubinstein}}, \bibinfo {author} {\bibfnamefont {J.~J.}\ \bibnamefont
  {Derksen}},\ and\ \bibinfo {author} {\bibfnamefont {S.}~\bibnamefont
  {Sundaresan}},\ }\bibfield  {title} {\bibinfo {title} {Lattice {B}oltzmann
  simulations of low-{Reynolds}-number flow past fluidized spheres: effect of
  {S}tokes number on drag force},\ }\href
  {https://doi.org/10.1017/jfm.2015.679} {\bibfield  {journal} {\bibinfo
  {journal} {J. Fluid Mech.}\ }\textbf {\bibinfo {volume} {788}},\ \bibinfo
  {pages} {576} (\bibinfo {year} {2016})}\BibitemShut {NoStop}%
\bibitem [{\citenamefont {Rubinstein}\ \emph {et~al.}(2017)\citenamefont
  {Rubinstein}, \citenamefont {Ozel}, \citenamefont {Yin}, \citenamefont
  {Derksen},\ and\ \citenamefont {Sundaresan}}]{rubinstein_2017}%
  \BibitemOpen
  \bibfield  {author} {\bibinfo {author} {\bibfnamefont {G.~J.}\ \bibnamefont
  {Rubinstein}}, \bibinfo {author} {\bibfnamefont {A.}~\bibnamefont {Ozel}},
  \bibinfo {author} {\bibfnamefont {X.}~\bibnamefont {Yin}}, \bibinfo {author}
  {\bibfnamefont {J.~J.}\ \bibnamefont {Derksen}},\ and\ \bibinfo {author}
  {\bibfnamefont {S.}~\bibnamefont {Sundaresan}},\ }\bibfield  {title}
  {\bibinfo {title} {Lattice {B}oltzmann simulations of low-{R}eynolds-number
  flows past fluidized spheres: effect of inhomogeneities on the drag force},\
  }\href {https://doi.org/10.1017/jfm.2017.705} {\bibfield  {journal} {\bibinfo
   {journal} {J. Fluid Mech.}\ }\textbf {\bibinfo {volume} {833}},\ \bibinfo
  {pages} {599} (\bibinfo {year} {2017})}\BibitemShut {NoStop}%
\bibitem [{\citenamefont {Zaidi}(2018)}]{zaidi_2018}%
  \BibitemOpen
  \bibfield  {author} {\bibinfo {author} {\bibfnamefont {A.~A.}\ \bibnamefont
  {Zaidi}},\ }\bibfield  {title} {\bibinfo {title} {Study of particle inertia
  effects on drag force of finite sized particles in settling process},\ }\href
  {https://doi.org/10.1016/j.cherd.2018.02.013} {\bibfield  {journal} {\bibinfo
   {journal} {Chem. Eng. Res. Des.}\ }\textbf {\bibinfo {volume} {132}},\
  \bibinfo {pages} {714} (\bibinfo {year} {2018})}\BibitemShut {NoStop}%
\bibitem [{\citenamefont {Garg}\ \emph {et~al.}(2011)\citenamefont {Garg},
  \citenamefont {Tenneti}, \citenamefont {Mohd-Yusof},\ and\ \citenamefont
  {Subramaniam}}]{garg_book}%
  \BibitemOpen
  \bibfield  {author} {\bibinfo {author} {\bibfnamefont {R.}~\bibnamefont
  {Garg}}, \bibinfo {author} {\bibfnamefont {S.}~\bibnamefont {Tenneti}},
  \bibinfo {author} {\bibfnamefont {J.}~\bibnamefont {Mohd-Yusof}},\ and\
  \bibinfo {author} {\bibfnamefont {S.}~\bibnamefont {Subramaniam}},\
  }\bibfield  {title} {\bibinfo {title} {Direct numerical simulation of
  gas-solids flow based on the immersed boundary method},\ }in\ \href@noop {}
  {\emph {\bibinfo {booktitle} {Computational Gas-Solids Flows and Reacting
  Systems: Theory, Methods and Practice}}},\ \bibinfo {editor} {edited by\
  \bibinfo {editor} {\bibfnamefont {S.}~\bibnamefont {Pannala}}, \bibinfo
  {editor} {\bibfnamefont {M.}~\bibnamefont {Syamlal}},\ and\ \bibinfo {editor}
  {\bibfnamefont {T.~J.}\ \bibnamefont {O'Brien}}}\ (\bibinfo  {publisher} {IGI
  Global},\ \bibinfo {year} {2011})\ pp.\ \bibinfo {pages}
  {245--276}\BibitemShut {NoStop}%
\bibitem [{\citenamefont {Tenneti}\ \emph {et~al.}(2011)\citenamefont
  {Tenneti}, \citenamefont {Garg},\ and\ \citenamefont
  {Subramaniam}}]{tenneti_ijmf_2011}%
  \BibitemOpen
  \bibfield  {author} {\bibinfo {author} {\bibfnamefont {S.}~\bibnamefont
  {Tenneti}}, \bibinfo {author} {\bibfnamefont {R.}~\bibnamefont {Garg}},\ and\
  \bibinfo {author} {\bibfnamefont {S.}~\bibnamefont {Subramaniam}},\
  }\bibfield  {title} {\bibinfo {title} {Drag law for monodisperse gas--solid
  systems using particle--resolved direct numerical simulation of flow past
  fixed assemblies of spheres},\ }\href@noop {} {\bibfield  {journal} {\bibinfo
   {journal} {Intl. J. Multiphase Flow}\ }\textbf {\bibinfo {volume} {37}},\
  \bibinfo {pages} {1072} (\bibinfo {year} {2011})}\BibitemShut {NoStop}%
\bibitem [{\citenamefont {Prosperetti}\ and\ \citenamefont
  {Tryggvason}(2007)}]{prosperetti_trygg_book_2007}%
  \BibitemOpen
  \bibfield  {author} {\bibinfo {author} {\bibfnamefont {A.}~\bibnamefont
  {Prosperetti}}\ and\ \bibinfo {author} {\bibfnamefont {G.}~\bibnamefont
  {Tryggvason}},\ }\href@noop {} {\emph {\bibinfo {title} {Computational
  methods for multiphase flow}}}\ (\bibinfo  {publisher} {Cambridge University
  Press},\ \bibinfo {year} {2007})\BibitemShut {NoStop}%
\bibitem [{\citenamefont {Xu}\ and\ \citenamefont
  {Subramaniam}(2010)}]{xu2010effect}%
  \BibitemOpen
  \bibfield  {author} {\bibinfo {author} {\bibfnamefont {Y.}~\bibnamefont
  {Xu}}\ and\ \bibinfo {author} {\bibfnamefont {S.}~\bibnamefont
  {Subramaniam}},\ }\bibfield  {title} {\bibinfo {title} {Effect of particle
  clusters on carrier flow turbulence: A direct numerical simulation study},\
  }\href@noop {} {\bibfield  {journal} {\bibinfo  {journal} {Flow, Turbulence
  and Combustion}\ }\textbf {\bibinfo {volume} {85}},\ \bibinfo {pages} {735}
  (\bibinfo {year} {2010})}\BibitemShut {NoStop}%
\bibitem [{\citenamefont {Sun}(2016)}]{sun2016modeling}%
  \BibitemOpen
  \bibfield  {author} {\bibinfo {author} {\bibfnamefont {B.}~\bibnamefont
  {Sun}},\ }\emph {\bibinfo {title} {Modeling heat and mass transfer in
  reacting gas-solid flow using particle-resolved direct numerical
  simulation}},\ \href@noop {} {Ph.D. thesis},\ \bibinfo  {school} {Iowa State
  University} (\bibinfo {year} {2016})\BibitemShut {NoStop}%
\bibitem [{\citenamefont {Allen}\ and\ \citenamefont
  {Tildesley}(1989)}]{allen_tildesley_cslbook}%
  \BibitemOpen
  \bibfield  {author} {\bibinfo {author} {\bibfnamefont {M.~P.}\ \bibnamefont
  {Allen}}\ and\ \bibinfo {author} {\bibfnamefont {D.~J.}\ \bibnamefont
  {Tildesley}},\ }\href@noop {} {\emph {\bibinfo {title} {Computer Simulation
  of Liquids}}},\ molecular dynamics\ (\bibinfo  {publisher} {Oxford University
  Press},\ \bibinfo {address} {Oxford, United Kingdom},\ \bibinfo {year}
  {1989})\BibitemShut {NoStop}%
\bibitem [{\citenamefont {Zhou}\ and\ \citenamefont
  {Balachandar}(2018)}]{zhou2018investigation}%
  \BibitemOpen
  \bibfield  {author} {\bibinfo {author} {\bibfnamefont {K.}~\bibnamefont
  {Zhou}}\ and\ \bibinfo {author} {\bibfnamefont {S.}~\bibnamefont
  {Balachandar}},\ }\bibfield  {title} {\bibinfo {title} {Investigation of
  direct forcing immersed boundary method}\ }(\bibinfo  {publisher} {APS},\
  \bibinfo {year} {2018})\BibitemShut {NoStop}%
\bibitem [{\citenamefont {Weller}\ \emph {et~al.}(2018)\citenamefont {Weller},
  \citenamefont {Greenshields},\ and\ \citenamefont
  {Santos}}]{weller2018openfoam}%
  \BibitemOpen
  \bibfield  {author} {\bibinfo {author} {\bibfnamefont {H.}~\bibnamefont
  {Weller}}, \bibinfo {author} {\bibfnamefont {C.}~\bibnamefont
  {Greenshields}},\ and\ \bibinfo {author} {\bibfnamefont {B.}~\bibnamefont
  {Santos}},\ }\href@noop {} {\bibinfo {title} {Openfoam}} (\bibinfo {year}
  {2018}),\ \bibinfo {note} {uRL http://www.openfoam.org}\BibitemShut {NoStop}%
\bibitem [{\citenamefont {Igci}\ \emph {et~al.}(2008)\citenamefont {Igci},
  \citenamefont {Andrews~IV}, \citenamefont {Sundaresan}, \citenamefont
  {Pannala},\ and\ \citenamefont {O'Brien}}]{igci2008filtered}%
  \BibitemOpen
  \bibfield  {author} {\bibinfo {author} {\bibfnamefont {Y.}~\bibnamefont
  {Igci}}, \bibinfo {author} {\bibfnamefont {A.~T.}\ \bibnamefont
  {Andrews~IV}}, \bibinfo {author} {\bibfnamefont {S.}~\bibnamefont
  {Sundaresan}}, \bibinfo {author} {\bibfnamefont {S.}~\bibnamefont
  {Pannala}},\ and\ \bibinfo {author} {\bibfnamefont {T.}~\bibnamefont
  {O'Brien}},\ }\bibfield  {title} {\bibinfo {title} {Filtered two-fluid models
  for fluidized gas-particle suspensions},\ }\href@noop {} {\bibfield
  {journal} {\bibinfo  {journal} {AIChE Journal}\ }\textbf {\bibinfo {volume}
  {54}},\ \bibinfo {pages} {1431} (\bibinfo {year} {2008})}\BibitemShut
  {NoStop}%
\bibitem [{\citenamefont {Ozarkar}\ \emph {et~al.}(2015)\citenamefont
  {Ozarkar}, \citenamefont {Yan}, \citenamefont {Wang}, \citenamefont
  {Milioli}, \citenamefont {Milioli},\ and\ \citenamefont
  {Sundaresan}}]{ozarkar2015validation}%
  \BibitemOpen
  \bibfield  {author} {\bibinfo {author} {\bibfnamefont {S.~S.}\ \bibnamefont
  {Ozarkar}}, \bibinfo {author} {\bibfnamefont {X.}~\bibnamefont {Yan}},
  \bibinfo {author} {\bibfnamefont {S.}~\bibnamefont {Wang}}, \bibinfo {author}
  {\bibfnamefont {C.~C.}\ \bibnamefont {Milioli}}, \bibinfo {author}
  {\bibfnamefont {F.~E.}\ \bibnamefont {Milioli}},\ and\ \bibinfo {author}
  {\bibfnamefont {S.}~\bibnamefont {Sundaresan}},\ }\bibfield  {title}
  {\bibinfo {title} {Validation of filtered two-fluid models for gas--particle
  flows against experimental data from bubbling fluidized bed},\ }\href@noop {}
  {\bibfield  {journal} {\bibinfo  {journal} {Powder Technology}\ }\textbf
  {\bibinfo {volume} {284}},\ \bibinfo {pages} {159} (\bibinfo {year}
  {2015})}\BibitemShut {NoStop}%
\bibitem [{\citenamefont {Xiu}\ and\ \citenamefont
  {Karniadakis}(2003)}]{xiu2003modeling}%
  \BibitemOpen
  \bibfield  {author} {\bibinfo {author} {\bibfnamefont {D.}~\bibnamefont
  {Xiu}}\ and\ \bibinfo {author} {\bibfnamefont {G.~E.}\ \bibnamefont
  {Karniadakis}},\ }\bibfield  {title} {\bibinfo {title} {Modeling uncertainty
  in flow simulations via generalized polynomial chaos},\ }\href@noop {}
  {\bibfield  {journal} {\bibinfo  {journal} {Journal of computational
  physics}\ }\textbf {\bibinfo {volume} {187}},\ \bibinfo {pages} {137}
  (\bibinfo {year} {2003})}\BibitemShut {NoStop}%
\bibitem [{\citenamefont {Soize}\ and\ \citenamefont
  {Ghanem}(2009)}]{soize2009reduced}%
  \BibitemOpen
  \bibfield  {author} {\bibinfo {author} {\bibfnamefont {C.}~\bibnamefont
  {Soize}}\ and\ \bibinfo {author} {\bibfnamefont {R.~G.}\ \bibnamefont
  {Ghanem}},\ }\bibfield  {title} {\bibinfo {title} {Reduced chaos
  decomposition with random coefficients of vector-valued random variables and
  random fields},\ }\href@noop {} {\bibfield  {journal} {\bibinfo  {journal}
  {Computer Methods in Applied Mechanics and Engineering}\ }\textbf {\bibinfo
  {volume} {198}},\ \bibinfo {pages} {1926} (\bibinfo {year}
  {2009})}\BibitemShut {NoStop}%
\bibitem [{\citenamefont {{\"O}ztireli}\ and\ \citenamefont
  {Gross}(2012)}]{oztireli2012analysis}%
  \BibitemOpen
  \bibfield  {author} {\bibinfo {author} {\bibfnamefont {A.~C.}\ \bibnamefont
  {{\"O}ztireli}}\ and\ \bibinfo {author} {\bibfnamefont {M.}~\bibnamefont
  {Gross}},\ }\bibfield  {title} {\bibinfo {title} {Analysis and synthesis of
  point distributions based on pair correlation},\ }\href@noop {} {\bibfield
  {journal} {\bibinfo  {journal} {ACM Transactions on Graphics (TOG)}\ }\textbf
  {\bibinfo {volume} {31}},\ \bibinfo {pages} {1} (\bibinfo {year}
  {2012})}\BibitemShut {NoStop}%
\bibitem [{\citenamefont {Colucci}\ \emph {et~al.}(1998)\citenamefont
  {Colucci}, \citenamefont {Jaberi}, \citenamefont {Givi},\ and\ \citenamefont
  {Pope}}]{colucci1998filtered}%
  \BibitemOpen
  \bibfield  {author} {\bibinfo {author} {\bibfnamefont {P.}~\bibnamefont
  {Colucci}}, \bibinfo {author} {\bibfnamefont {F.}~\bibnamefont {Jaberi}},
  \bibinfo {author} {\bibfnamefont {P.}~\bibnamefont {Givi}},\ and\ \bibinfo
  {author} {\bibfnamefont {S.}~\bibnamefont {Pope}},\ }\bibfield  {title}
  {\bibinfo {title} {Filtered density function for large eddy simulation of
  turbulent reacting flows},\ }\href@noop {} {\bibfield  {journal} {\bibinfo
  {journal} {Physics of Fluids}\ }\textbf {\bibinfo {volume} {10}},\ \bibinfo
  {pages} {499} (\bibinfo {year} {1998})}\BibitemShut {NoStop}%
\bibitem [{\citenamefont {Fox}(2003)}]{fox_book}%
  \BibitemOpen
  \bibfield  {author} {\bibinfo {author} {\bibfnamefont {R.~O.}\ \bibnamefont
  {Fox}},\ }\href@noop {} {\emph {\bibinfo {title} {{Computational Models for
  Turbulent Reacting Flows}}}}\ (\bibinfo  {publisher} {Cambridge University
  Press},\ \bibinfo {year} {2003})\BibitemShut {NoStop}%
\bibitem [{\citenamefont {Klimontovich}(2012)}]{klimontovich_2012}%
  \BibitemOpen
  \bibfield  {author} {\bibinfo {author} {\bibfnamefont {Y.}~\bibnamefont
  {Klimontovich}},\ }\href {https://books.google.com/books?id=WjlqCQAAQBAJ}
  {\emph {\bibinfo {title} {Statistical Theory of Open Systems: Volume 1: A
  Unified Approach to Kinetic Description of Processes in Active Systems}}},\
  Fundamental Theories of Physics\ (\bibinfo  {publisher} {Springer
  Netherlands},\ \bibinfo {year} {2012})\BibitemShut {NoStop}%
\bibitem [{\citenamefont {Van~Noije}\ \emph {et~al.}(1997)\citenamefont
  {Van~Noije}, \citenamefont {Ernst}, \citenamefont {Brito},\ and\
  \citenamefont {Orza}}]{van1997mesoscopic}%
  \BibitemOpen
  \bibfield  {author} {\bibinfo {author} {\bibfnamefont {T.}~\bibnamefont
  {Van~Noije}}, \bibinfo {author} {\bibfnamefont {M.}~\bibnamefont {Ernst}},
  \bibinfo {author} {\bibfnamefont {R.}~\bibnamefont {Brito}},\ and\ \bibinfo
  {author} {\bibfnamefont {J.}~\bibnamefont {Orza}},\ }\bibfield  {title}
  {\bibinfo {title} {Mesoscopic theory of granular fluids},\ }\href@noop {}
  {\bibfield  {journal} {\bibinfo  {journal} {Physical review letters}\
  }\textbf {\bibinfo {volume} {79}},\ \bibinfo {pages} {411} (\bibinfo {year}
  {1997})}\BibitemShut {NoStop}%
\bibitem [{\citenamefont {Nicholson}(1992)}]{plasma_theory}%
  \BibitemOpen
  \bibfield  {author} {\bibinfo {author} {\bibfnamefont {D.~R.}\ \bibnamefont
  {Nicholson}},\ }\href@noop {} {\emph {\bibinfo {title} {{Introduction to
  Plasma Theory}}}}\ (\bibinfo  {publisher} {Krieger Publishing Co.},\ \bibinfo
  {address} {Malabar, FL, USA},\ \bibinfo {year} {1992})\ \bibinfo {note} {now
  out of print}\BibitemShut {NoStop}%
\bibitem [{\citenamefont {Klimontovich}(1986)}]{klimontovich_1986}%
  \BibitemOpen
  \bibfield  {author} {\bibinfo {author} {\bibfnamefont {Y.}~\bibnamefont
  {Klimontovich}},\ }\href {https://books.google.com/books?id=mznSwAEACAAJ}
  {\emph {\bibinfo {title} {Statistical Physics}}}\ (\bibinfo  {publisher}
  {Taylor \& Francis},\ \bibinfo {year} {1986})\BibitemShut {NoStop}%
\bibitem [{\citenamefont {Quintanilla}\ and\ \citenamefont
  {Torquato}(1997)}]{quintanilla1997local}%
  \BibitemOpen
  \bibfield  {author} {\bibinfo {author} {\bibfnamefont {J.}~\bibnamefont
  {Quintanilla}}\ and\ \bibinfo {author} {\bibfnamefont {S.}~\bibnamefont
  {Torquato}},\ }\bibfield  {title} {\bibinfo {title} {Local volume fraction
  fluctuations in random media},\ }\href@noop {} {\bibfield  {journal}
  {\bibinfo  {journal} {The Journal of chemical physics}\ }\textbf {\bibinfo
  {volume} {106}},\ \bibinfo {pages} {2741} (\bibinfo {year}
  {1997})}\BibitemShut {NoStop}%
\end{thebibliography}%

\end{document}